\documentclass{aa}
\usepackage{natbib}
\bibpunct{(}{)}{;}{a}{}{,}    
\usepackage{tabularx,hhline,amsmath,amssymb}
\usepackage{graphics}
\usepackage[small]{subfigure}
\newcommand{\lns}{Log(N)--Log(S)}

\begin{document}

\setcounter{table}{0}
\setcounter{figure}{0}

\sloppy
\title{The Milky Way in X-rays for an outside observer}
\subtitle{Log(N)-Log(S) and Luminosity Function of X-ray binaries from
RXTE/ASM data}
\author{H.-J. Grimm\inst{1} \and M. Gilfanov\inst{1,2} \and
R. Sunyaev\inst{1,2}}
\institute{Max-Planck-Institut f\"ur Astrophysik,Garching, Germany
\and Space Research Institute, Moscow, Russia}
\offprints{Hans-Jakob Grimm, \email{grimm@mpa-garching.mpg.de}}
\mail{Max-Planck-Institut f\"ur Astrophysik,
Karl-Schwarzschild-Str. 1, 85748 Garching, Germany}
\date{Received 18-09-01 / Accepted 03-06-02}

\abstract{We study the \lns~and X-ray luminosity function in the 2--10
keV energy band, and the spatial (3-D) distribution of bright, $L_X
\ge 10^{34} - 10^{35}$ erg s$^{-1}$, X-ray binaries in the Milky
Way. In agreement with theoretical expectations and earlier results we
found significant differences between the spatial distributions of low
(LMXB) and high (HMXB) mass X-ray binaries. The volume density of LMXB
sources peaks strongly at the Galactic Bulge whereas HMXBs tend to
avoid the inner $\sim 3-4$ kpc of the Galaxy. In addition HMXBs are
more concentrated towards the Galactic Plane (scale heights of
$\approx 150$ and $\approx 410$ pc for HMXB and LMXB correspondingly)
and show clear signatures of the spiral structure in their spatial
distribution. The \lns~distributions and the X-ray luminosity
functions are also noticeably different. LMXB sources have a flatter
\lns~distribution and luminosity function. The integrated 2-10 keV
luminosities of all X-ray binaries in the Galaxy, {\em averaged} over
1996--2000, are $\sim 2-3 \cdot  10^{39}$ (LMXB) and $\sim 2-3 \cdot
10^{38}$ (HMXB) erg s$^{-1}$. Normalised to the stellar mass and the
star formation rate, respectively, these correspond to $\sim 5 \cdot
10^{28}$ erg s$^{-1}$ M$^{-1}_{\odot}$ for LMXBs and $\sim 5\cdot
10^{37}$ erg s$^{-1}$/(M$_{\odot}$ yr$^{-1}$) for HMXBs. Due to the
shallow slopes of the luminosity functions the integrated emission of
X-ray binaries is dominated by the $\sim$ 5--10 most luminous sources
which determine the appearance of the Milky Way in the standard X-ray
band for an outside observer. In particular variability of individual
sources or an outburst of a bright transient source can increase the
integrated luminosity of the Milky  Way by as much as a factor of
$\sim 2$. Although the average LMXB luminosity function shows a break
near the Eddington luminosity for a 1.4 M$_{\odot}$ neutron star, at
least 12 sources showed episodes of super-Eddington luminosity during
ASM observations. We provide  the maps of distribution of X-ray
binaries in the Milky Way in various projections, which can be
compared to images of nearby galaxies taken by CHANDRA and
XMM-Newton.
\keywords{X-rays: binaries -- X-rays: galaxies -- Galaxy: general --
Galaxy: structure -- Galaxies: spiral -- Stars: luminosity function}}

\maketitle

\section{Introduction}
Recently the CHANDRA  X-ray observatory studied the
distributions and luminosity functions of X-ray binaries in at
least 7 spiral, e.g \object{M 81}~\citep{tennant:01}, 2 elliptical,
e.g. \object{NGC 4697} \citep{sarazin:00}, and 2 starburst galaxies,
\object{M 82} \citep{zezas:01} and Antennae \citep{fabbiano:01}. The
main discovery of these CHANDRA observations was the existence of
numerous point-like sources with luminosities in the CHANDRA spectral
band considerably higher than the Eddington luminosity of a 1.4
M$_{\odot}$ neutron star. Nearby galaxies observed by CHANDRA
have a great advantage compared to observations of X-ray sources in
our Galaxy: All objects observed in a particular galaxy are
equidistant and therefore it is straightforward to construct the
luminosity function in the CHANDRA band. However, even with the
angular resolution and sensitivity of CHANDRA we are restricted to
nearby galaxies ($d \la 50$ Mpc) and we are able to observe only
the high luminosity end of the luminosity function.

Observations of compact sources inside our Galaxy thus open the unique
possibility to construct a luminosity function in a much broader range
of luminosities and this might be important to construct the
synthesised spectrum of the LMXB and HMXB populations of the Galaxy in
a broad spectral range from 0.1--500 keV using data from all existing
spacecraft.

In this paper we use data of the All-Sky Monitor
(ASM)~\citep{levine:96} aboard the Rossi X-ray Timing
Explorer~\citep{brandt:96} to investigate the following topics.
\begin{itemize}
  \item Using ASM data, existing information about the source
  distances and a model of the mass distribution in the 
  Milky Way we constructed the luminosity function of high and low
  mass X-ray binaries in our Galaxy.

  \item Distribution and number of high mass X-ray binaries
  are expected to trace the location and reflect the rate of star
  formation. The X-ray luminosity of starburst galaxies might become
  an additional source of information about the star formation rate in
  these galaxies. Moreover knowledge of the luminosity of HMXBs versus
  star formation rate opens the way to find the volume emissivity of
  our universe at different redshifts in hard X-rays due to starburst
  and young galaxies.

  \item The luminosity of the LMXB component is proportional to the
  total mass of the old stellar population of the Milky
  Way. Concerning the volume emissivity of galaxies these data will
  provide information about the contribution of elliptical galaxies
  and old star populations in spiral galaxies only at sufficiently
  low redshifts ($z < 0.4-0.5$). To obtain the volume emissivity due
  to old star populations at higher redshift we need to know a model
  of the mass exchange rate evolution.

  \item Our Galaxy should become an important point in the future
  calibration curves of $L_{\text{HMXB}}/SFR$ and
  $L_{\text{LMXB}}/M_{\text{galaxy}}$.

  \item Our analysis of ASM data permits us to show how our Galaxy
  would look from outside in different projections. This will allow us
  to compare data about our Galaxy with new CHANDRA observations.

  \item Surprisingly enough, just the comparatively few most luminous
  Galactic X-ray binaries practically dominate the X-ray luminosity of
  our Galaxy. The majority of the brightest X-ray binaries are
  extremely variable on all time scales from milliseconds to
  years--tens of years. Therefore the luminosity of our Galaxy as a
  whole would also be subject to strong variability. This is important
  because with a powerful X-ray telescope such as XEUS it will be
  possible to detect X-ray flux from distant galaxies on the level of
  $L \sim 10^{40}$ erg s$^{-1}$ but only short time scale variability
  would permit to distinguish the collective emission of X-ray
  binaries from the low luminosity, AGN-type activity of the
  nucleus. Black holes are unable to produce strong variability with
  characteristic times significantly shorter than a few 0.01s
  $\frac{M_{\text{BH}}}{M_{\odot}}$ \citep{sunyaev:00}. For
  super-massive black holes the characteristic time is of order or
  above $\sim 10^{3}$s.

  \item Our analysis of ASM data and data from other spacecraft
  shows that at least for 17 X-ray sources in our Galaxy ASM or other
  spacecraft detected flux reaching or exceeding the level
  corresponding to the Eddington critical luminosity for a 1.4
  M$_{\odot}$ neutron star, see Table \ref{tab:edd_source}. Maximal
  fluxes detected were up to 10 times higher than the Eddington value
  for a neutron star. In at least 7 sources the compact object has
  been identified as a neutron star based on the detection of X-ray
  pulsations or X-ray bursts, therefore we know with certainty that
  the peak luminosity exceeded the Eddington limit. Moreover, the
  total number of super-Eddington sources might be higher because we
  know from the broad band observations that the bulk of the
  luminosity can be emitted outside the ASM sensitivity band.
\end{itemize}

\begin{table*}[tb]
\caption{Persistent, transient and extragalactic X-ray binaries for
which episodes of near- or super-Eddington flux were detected. The
maximum luminosities from ASM observations refer to the dwell-time
light curves (90 s observations every 90 minutes). Thus the values
might differ from the luminosities given in
Tables \ref{tab:bright_l} and \ref{tab:bright_h}.}
  \begin{tabularx}{1.01\linewidth}{|l|l|c|cc|c|l|c|l|}
    \hline
    Source  & type         & M$_1$   &\multicolumn{2}{c|}{Luminosity
[$10^{38}$erg s$^{-1}$]}  & Energy range & Ref.$^{(b)}$& distance &
Ref.$^{(c)}$\\
          & & [M$_{\odot}$]& average$^{(a)}$ & peak&  [keV] & & [kpc]&\\
    \hline
    \multicolumn{9}{|c|}{Persistent sources}\\
    \hline
    \object{Cir X-1}      & LMXB & NS  & 4.4   & 12  & 2--10 & (1)& 10.9 & (i)\\
    \object{GRS 1915+105} & LMXB &14-30& 3.7   & 15  & 2--10 & (1)& 12.5 & (ii)\\
    \object{Sco X-1}      & LMXB & NS  & 2.7   & 9.4 & 2--10 & (1)& 2.8  & (iii)\\
    \object{Cyg X-2}      & LMXB & NS  & 1.8   & 4.2 & 2--10 & (1)& 11.3 & (i),(iv),(v),(vi),(vii)\\
    \object{GX 349+2}     & LMXB & NS  & 1.6   & 3.2 & 2--10 & (1)&  9.2 & (i),(vii),(viii)\\
    \object{GX 17+2}      & LMXB & NS  & 1.5   & 3.0 & 2--10 & (1)&  9.5 & (i),(vii),(ix),(x)\\
    \object{GX 5-1}       & LMXB & NS  & 1.4   & 2.2 & 2--10 & (1)&  7.2 & (i),(vii)\\
    \object{GX 340+0}     & LMXB & NS  & 1.3   & 2.2 & 2--10 & (1)& 11.0 & (i),(vii)\\
    \object{Cyg X-3}      & HMXB & NS(?)& 0.5  & 2.1 & 2--10 & (1)&  9.0 & (xi)\\
    \object{X 1624-490}   & LMXB & NS  & 0.24  & 3.3 & 2--10 & (1)& 13.5 & (ix)\\
    \object{GRO J1744-28} & LMXB & NS  & 0.15  & 4   & 8--20 & (2)&8.5 & (xii)\\
    \hline
    \multicolumn{9}{|c|}{Transient sources}\\
    \hline
    \object{V4641 Sgr}    & HMXB & 9.6  &    & 33  & 2--10 & (1) & 9.9 & (xiii) \\
    \object{GS 2023+338}  & LMXB & 12   &    & 11  & 1--40 & (3) & 4.3 & (xiv)\\
    \object{4U 1608-52}   & LMXB & NS   &    & 9.2 & 2--20 & (4) & 4.0 & (i),(x),(xv)\\
    \object{N Musc 91}    & LMXB & 7    &    & 6.1 & 1--6  & (5) & 5.5 & (xvi),(xvii),(xviii)\\
    \object{XTE J1550-564}& LMXB & 10.5 &    & 5.3 & 2--10 & (1) & 5.3 & (xix)\\
    \object{N Oph 77}     & LMXB & 5    &    & 5.4 & 2--18 & (6) & 7.0 & (xviii),(xx)\\
    \object{GS 2000+251}  & LMXB & 6    &    & 2.2 & 1--6  & (7) & 2.7 & (xviii),(xxi)\\
    \hline
    \multicolumn{9}{|c|}{Magellanic Clouds sources}\\
    \hline
    \object{SMC X-1} & HMXB & NS  & 2.0   & 17  & 2--10 & (1) & 60$^{(d)}$ & \\
    \object{LMC X-1} &HMXB& 4.7$\sqrt{\cos i}$ & 1.5 & 13 & 2--10 & (1) & 50$^{(d)}$&\\
    \object{LMC X-2} & LMXB & NS  & 1.5   & 17  & 2--10 & (1) & 50$^{(d)}$ & \\
    \object{LMC X-3} & HMXB &$>5.8$ & 1.5 & 17  & 2--10 & (1) & 50$^{(d)}$ & \\
    \object{LMC X-4} & HMXB & NS  & 0.38  & 15  & 2--10 & (1) & 50$^{(d)}$ & \\
    \hline
  \end{tabularx}
  \label{tab:edd_source}
  $^{(a)}$ average luminosity observed by ASM.\\
  $^{(b)}$ Reference for the peak luminosity.\\
  (1) ASM (this paper), (2) \citet{sazonov:97},
  (3) \citet{tanaka:92}, (4) \citet{nakamura:89},
  (5) \citet{kitamoto:92}, (6) \citet{watson:78},
  (7) \citet{tsunemi:89}.\\
  $^{(c)}$ Reference(s) for the distance:
(i) \citet{vanparadijs:95}, (ii) \citet{mirabel:94},
(iii) \citet{bradshaw:99}, (iv) \citet{orosz:99},
(v) \citet{cowley:79}, (vi) \citet{smale:98},
(vii) \citet{penninx:89}, (viii) \citet{wachter:96},
(ix) \citet{christian:97}, (x) \citet{ebisuzaki:84},
(xi) \citet{predehl:00}, (xii) \citet{nishiuchi:99},
(xiii) \citet{orosz:00}, (xiv) \citet{king:93},
(xv) \citet{nakamura:89}, (xvi) \citet{greiner:94},
(xvii) \citet{orosz:96}, (xviii) \citet{barret:97},
(xix) \citet{orosz:02}, (xx) \citet{martin:95},
(xxi) \citet{chevalier:90}\\
  $^{(d)}$ assuming a distance of 50 kpc for LMC and 60 kpc for SMC.\\
\end{table*}

In terms of the spatial distribution of X-ray binaries this paper
elaborates on works done earlier that also distinguished between low
and high mass systems but used substantially smaller samples.

Previously \citet{white:80}, \citet{lamb:80}, \citet{nagase:89} and
\citet{verbunt:96} noted the correlation of the positions of accreting
X-ray pulsars with high mass companions with the location of spiral
arm features of the Milky Way. Based on a larger sample of HMXBs with
measured distances we show that indeed the spatial distribution of
HMXBs follows the spiral structure of the Galaxy.

Using distance estimates and angular distribution of LMXBs
\citet{vanparadijs:95} and \citet{white:96} investigated
the spatial distribution of LMXBs and BHC in our Galaxy,
particularly in the Galactic disk. They estimated values for the
vertical (290 pc and 710 pc for BHC and NS binaries) and radial
scales (4.5 kpc for NS binaries) of the disk. These values are in
general agreement with those obtained in this paper, that are based
on a considerably larger number of sources. \citet{grebenev:96} found
good agreement between the source distribution  observed by
ART-P/GRANAT in the Galactic Centre region and the stellar mass
distribution in the Galactic Bulge. We thus have a reasonably good
knowledge about the distribution of LMXBs in the Galaxy.

%
%
\section{RXTE All-Sky Monitor Data}
\begin{figure*}[htb]
  \resizebox{\hsize}{!}{\includegraphics{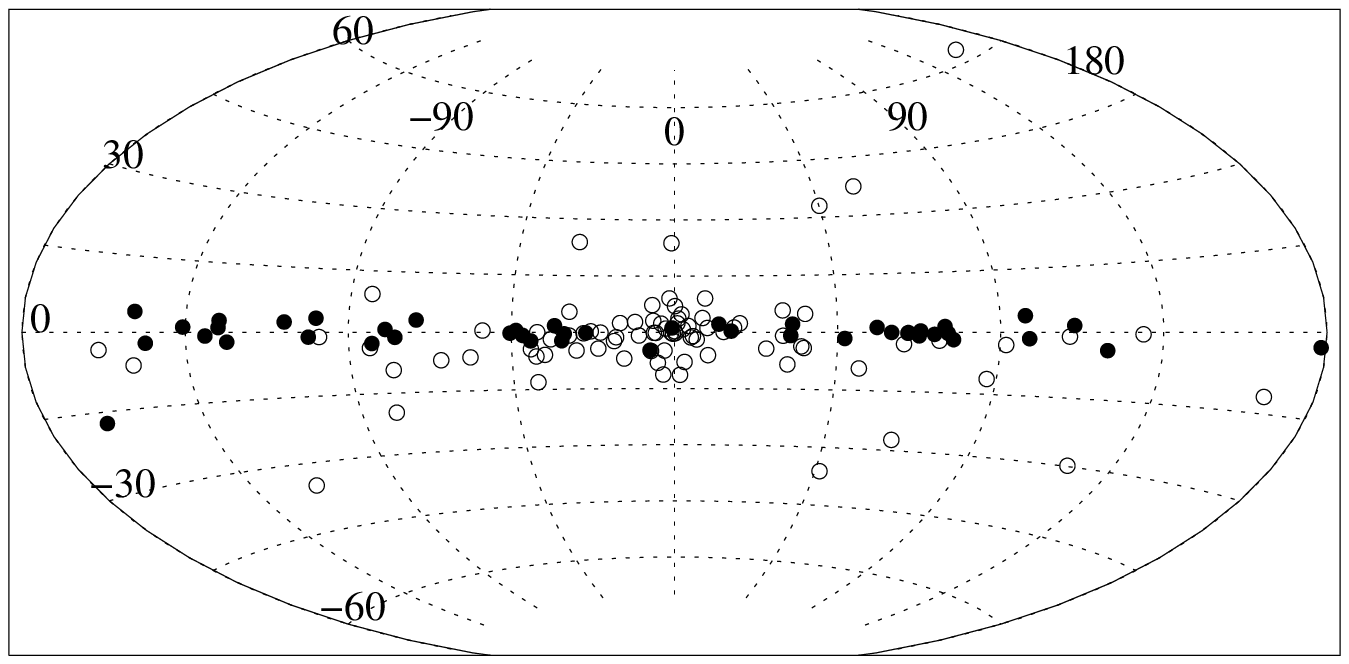}}
  \caption{Distribution of LMXBs (open circles) and HMXBs (filled
  circles) in the Galaxy. In total 86  LMXBs and 52 HMXBs are
  shown. Note the significant concentration of HMXBs 
  towards the Galactic Plane and the clustering of LMXBs in the
  Galactic Bulge.}
\label{fig:asm_sky}
\end{figure*}
In order to construct the \lns~distributions and luminosity functions
we used the publicly available data of ASM. The ASM instrument is
sensitive in the 2-10 keV energy band which is divided into 3 broad
energy channels and provides 80\% sky coverage for every satellite
orbit ($\sim$ 90 minutes). Due to its all-sky nature and long
operational time, $\sim$ 5 years, the ASM instrument is ideally suited
for studying time averaged properties of sources. The light curves are
obtained by RXTE GOF~\citep{levine:96} for a preselected set of
sources from the ASM catalogue. The catalogue consists of sources
which have reached an intensity of more than 5 mCrab at any time
\citep{xte-asm}, and as of June 2000 included 340 sources of which 217
are galactic and 112 extragalactic, and 10 unidentified. The
distribution of galactic sources on the sky is shown in
Fig. \ref{fig:asm_sky}. For a detailed description of selection
criteria and a list of sources see \citet{xte-asm}. The 1 day
sensitivity of ASM is $\approx$ 10 mCrab corresponding to a count rate
of 0.75 cnts s$^{-1}$. The ASM count rate has been converted to energy
flux assuming a Crab-like spectrum and using the observed Crab count
rate:
\begin{equation}
  F[\text{erg s$^{-1}$ cm$^{-2}$}] = 3.2 \cdot 10^{-10} \cdot
  R[\text{cnts s$^{-1}$}].
\end{equation}
The 1-dwell ASM light curves have been retrieved from the RXTE public
archive
\footnote{ftp://legacy.gsfc.nasa.gov/xte/data/archive/ASMProducts/\
definitive\_1dwell/}
at HEASARC and cover a time period from the start of the mission
through 27/04/00. In order to construct \lns~the light curves have
been averaged over the entire period of available data which might
differ for different sources. We did not account in any way for
orbital variations or eclipses, as e.g. in \object{Cen X-3}.

Important for the analysis presented below are the questions of
systematic errors in the light curves and of the completeness limit of
the ASM catalogue.

\subsection{Systematic errors}
\label{sec:syserr}
\begin{figure}[ht]
  \resizebox{\hsize}{!}{\includegraphics{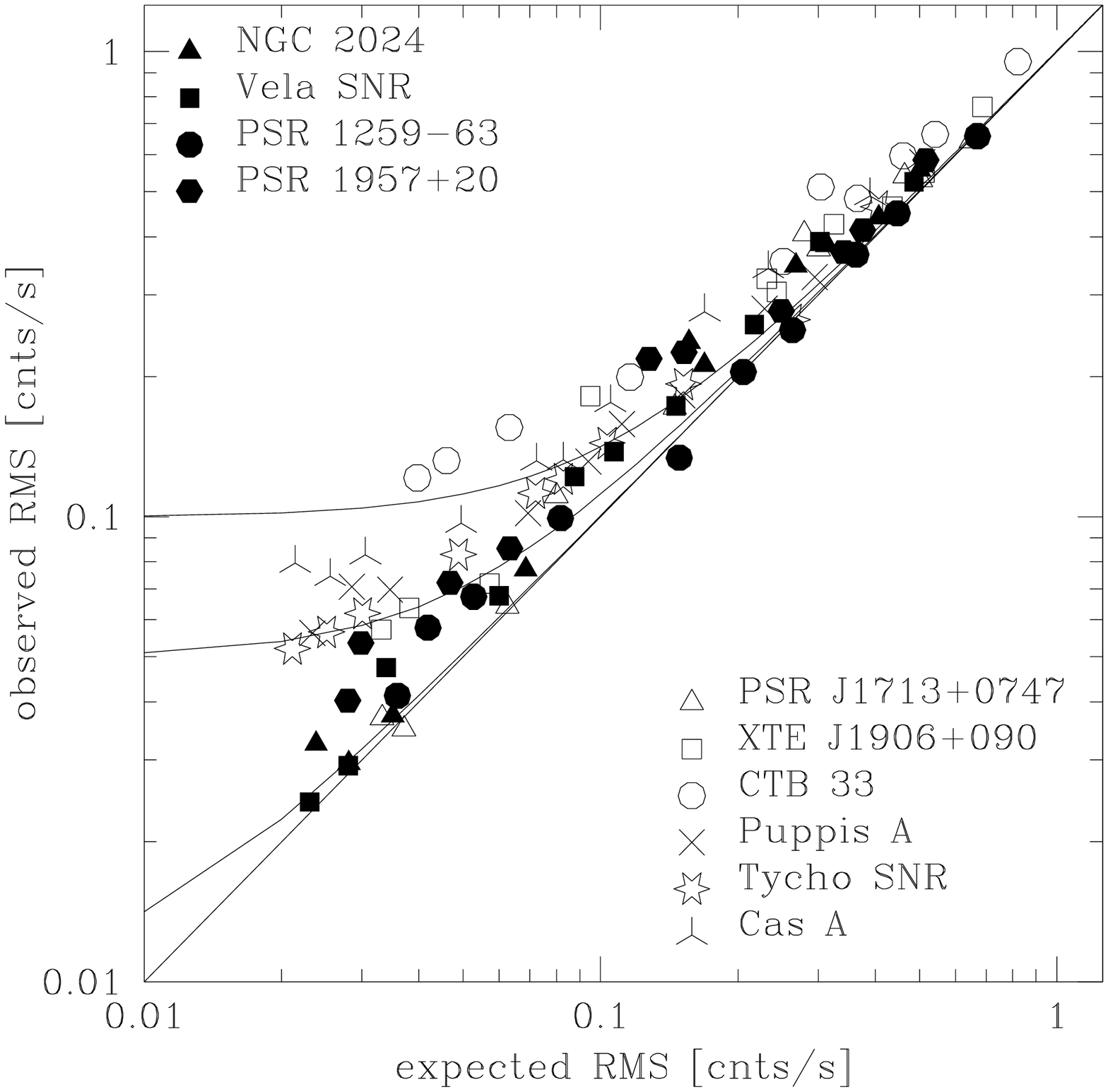}}
  \caption{Observed versus expected RMS for 10 different sources and
  for different time binnings. The bin duration varies from dwell time
  scale, i.e. $\sim 90$ seconds (upper right corner), to 200 days
  (lower left corner). Although there is considerable spread, the
  observed RMS is generally higher than expected, especially at large
  bin durations exceeding 50 days (expected RMS $< 0.1$ cnts
  s$^{-1}$). Assuming that systematic and statistical errors are
  independent the systematic error may be added to the statistical
  error in quadrature. This is shown by the solid curves for three
  different values of the systematic error: 0.01, 0.05 and 0.1 cnts
  s$^{-1}$.}
  \label{fig:rms}
\end{figure}
The ASM light curves are assumed to have a systematic error at the
level of $\sim$ 3\% which is added in quadrature to the statistical
errors in the light curves provided by the RXTE GOF. The systematic
error has been estimated using \object{Crab} data and refers to the
$\sim$ dwell--day time scales. The formal errors for the average
fluxes calculated from the entire ASM light curves are very small
$\sim 0.1-0.2$ mCrab ($\sim 1-2 \cdot 10^{-2}$ cnts s$^{-1}$).
In the presence of systematic errors this might not correctly
characterise the accuracy of the average flux estimate, especially for
weak sources. The contribution of systematic errors to the average
flux estimate depends on their statistical properties, in particular
their correlation time scale. In order to investigate these properties
we selected several sources believed to have constant X-ray flux, like
SNRs or rotation powered pulsars, see Table \ref{tab:rms}, and
rebinned their light curves with different bin durations ranging from
1 to 200 days.
\begin{table}[tb]
\caption{List of the sources used to estimate systematic errors.}
\begin{tabularx}{\linewidth}{l|cc}
\hline
Source & average flux$^{(a)}$ & excess RMS$^{(b)}$\\
 & [cnts s$^{-1}$] & [cnts s$^{-1}$]\\
\hline
\object{Cas A}          & 4.9  $\pm$ 0.007   & $\sim$ 0.08\\
\object{Tycho SNR}      & 1.3  $\pm$ 0.007   & $\sim$ 0.04\\
\object{Puppis A}       & 0.84 $\pm$ 0.008   & $\sim$ 0.05\\
\object{Vela pulsar}    & 0.75 $\pm$ 0.008   & $\sim$ 0.01\\
\object{CTB 33}         & 0.35 $\pm$ 0.014   & $\sim$ 0.07\\
\object{PSR 1259-63}    & 0.18 $\pm$ 0.012   & $\sim$ 0.01\\
\object{NGC 2024}       & 0.09 $\pm$ 0.008   & $\sim$ 0.02\\
\object{PSR J1713+0747} & 0.07 $\pm$ 0.015   & $\sim$ 0.01\\
\object{PSR 1957+20}    & 0.06 $\pm$ 0.012   & $\sim$ 0.02\\
\object{XTE J1906+090}  & 0.04 $\pm$ 0.011   & $\sim$ 0.03\\
\hline
\end{tabularx}
\label{tab:rms}
$^{(a)}$the errors are formally calculated using the errors in
the light curves.\\
$^{(b)}$upper limit on the unaccounted contribution of the
systematic errors to the averaged flux, estimated from
Fig. \ref{fig:rms}.
\end{table}
For each binned light curve we computed the expected RMS from the
errors given with the light curves and compared it with the observed
RMS. The results are shown in Fig. \ref{fig:rms}. Ideally there should
be a one-to-one correspondence between expected and observed RMS
(straight line in Fig. \ref{fig:rms}). As can be seen from
Fig. \ref{fig:rms} this is not the case. The observed RMS somewhat
exceeds the expected value, the discrepancy increasing towards large
bin durations ($\sim$ 50-200 days). The excess variance at large bin
durations (lower-left part in Fig. \ref{fig:rms}) gives an upper limit
on the unaccounted systematic error in the averaged flux estimate. As
can be seen from Fig. \ref{fig:rms} the particular value of the
systematic error, though varying from source to source, is in the
range of 0.01--0.1 cnts s$^{-1}$. We assumed a value of 0.05 cnts
s$^{-1}$ (to be added in quadrature to the statistical error). We
further verified that our conclusions are not sensitive to the value
of the systematic error.

For 15 sources we obtain statistically significant, $\ge 3 \sigma$,
negative average count rates. The majority of these sources, namely
14, are located in the Small and Large Magellanic Cloud and their
negative average flux is apparently caused by source interference in
these crowded regions. The remaining source also appears to suffer
from interference with nearby sources. In particular, we have noticed
that some of the light curves show a clear drop below zero count rate
coincident in time with addition of new sources located nearby to the
ASM catalogue. All these sources are excluded from our analysis.

%
%
\begin{figure}[t]
  \resizebox{\hsize}{!}{\includegraphics{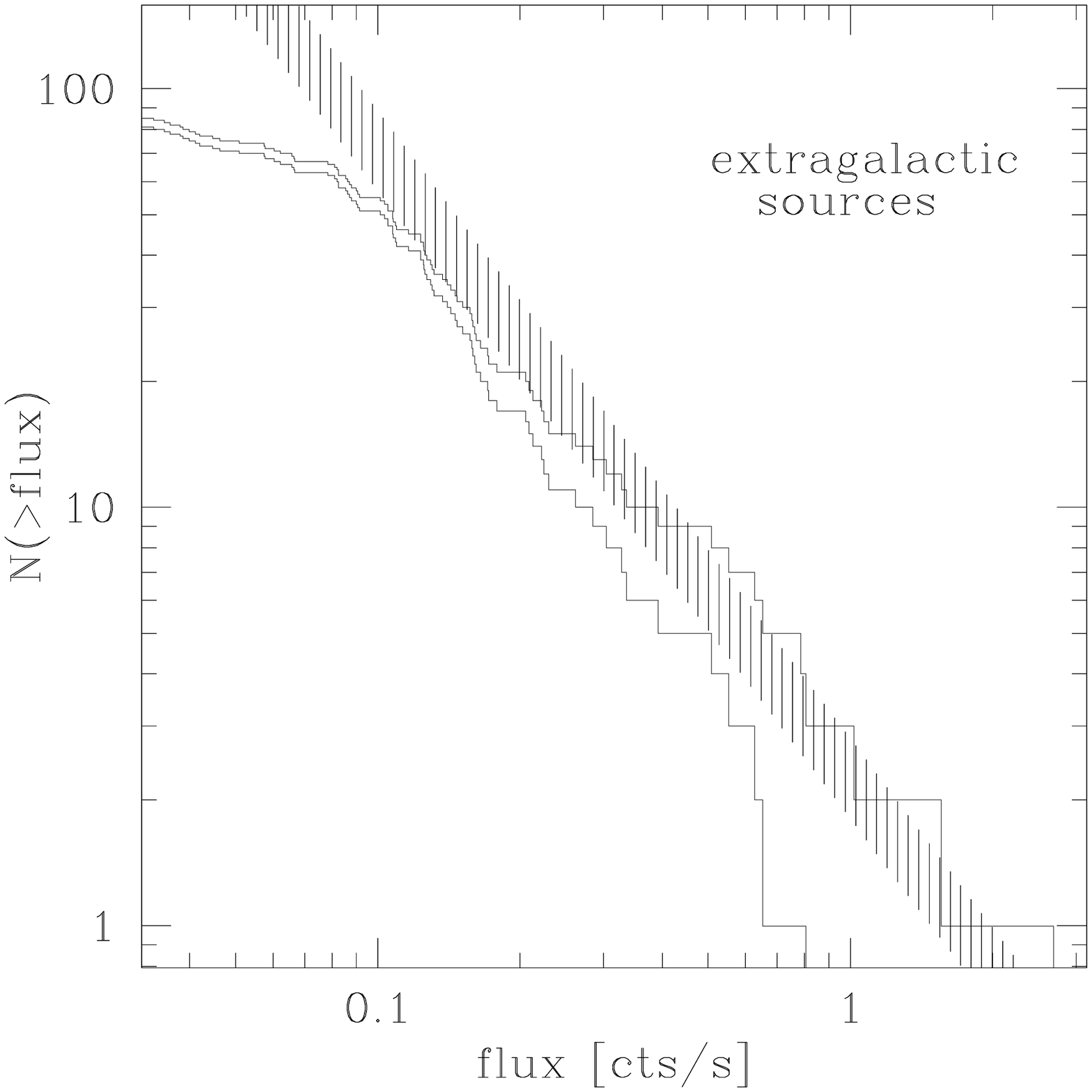}}
  \caption{\lns~distribution of extragalactic sources. Magellanic
  Cloud sources have been omitted. The upper histogram contains all
  extragalactic sources, the lower histogram excludes 4 nearby galaxy
  clusters (\object{Perseus}, \object{Virgo}/\object{M 87},
  \object{Coma} and \object{Centaurus}). The shaded region shows the
  \lns~obtained by HEAO-1 A-2 for high latitude ($|b| > 20^{\circ}$)
  sources \citep{piccinotti:82}. The width of the shaded region
  roughly accounts for the uncertainty of the RXTE/ASM and HEAO-1 A-2
  calibration.} 
  \label{fig:extragal_logn}
\end{figure}

\begin{figure}[t]
  \resizebox{\hsize}{!}{\includegraphics{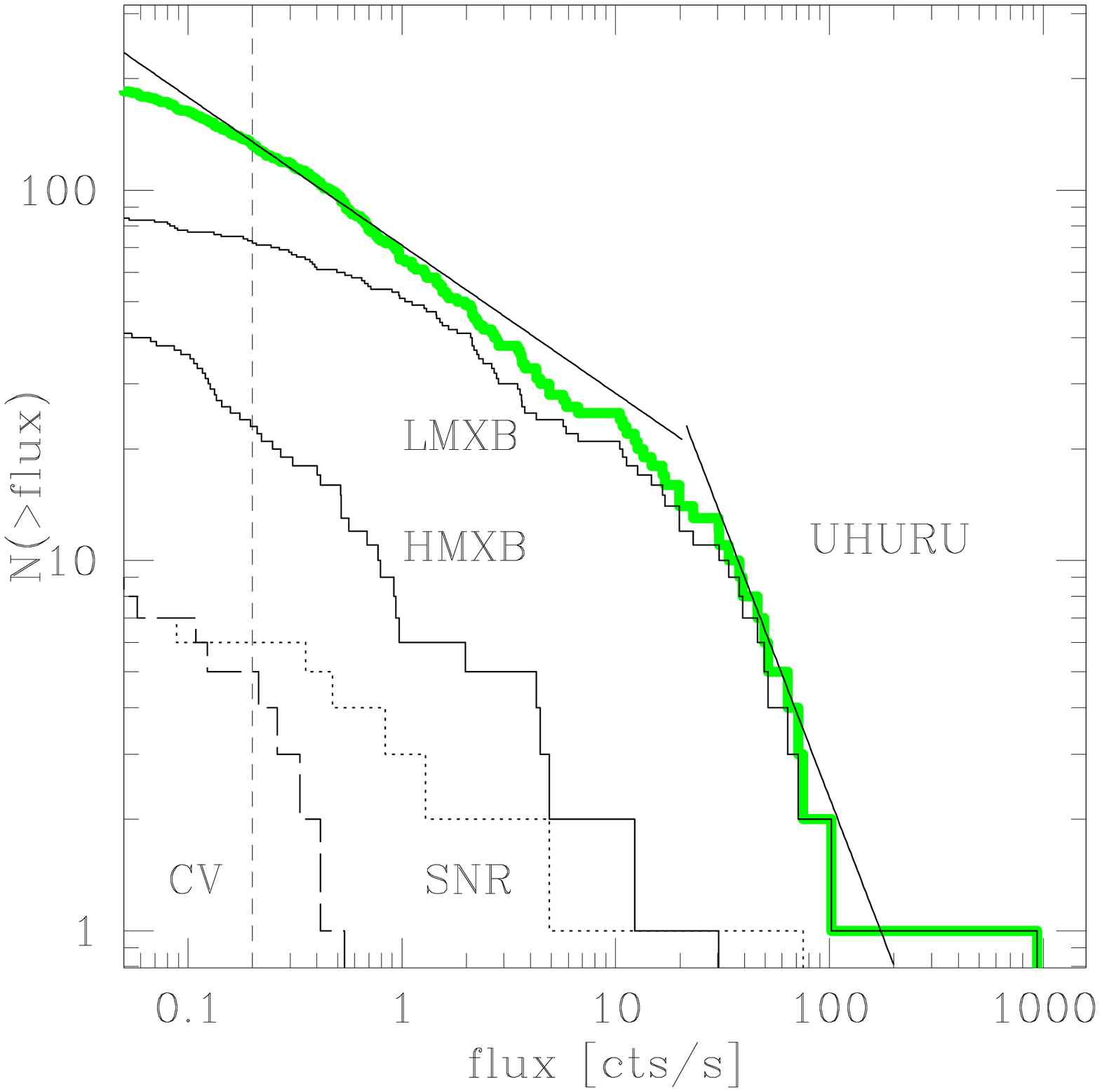}}
  \caption{Number--flux relation for all galactic sources derived from
  the entire ASM sample. The broken solid line shows schematically the
  number--flux relation for the low--latitude $|b|<20^{\circ}$ sources
  obtained by UHURU \citep{matilsky:73}. The vertical dashed line shows
  approximate completeness limit of the ASM sample. The thick grey
  histogram shows the \lns~for all Galactic sources observed by
  ASM. The four lower histograms show the contributions of different
  classes of sources to the total galactic \lns.}
  \label{fig:gal_logn}
\end{figure}

\begin{figure*}[tb]
  \resizebox{0.5\hsize}{!}{\includegraphics{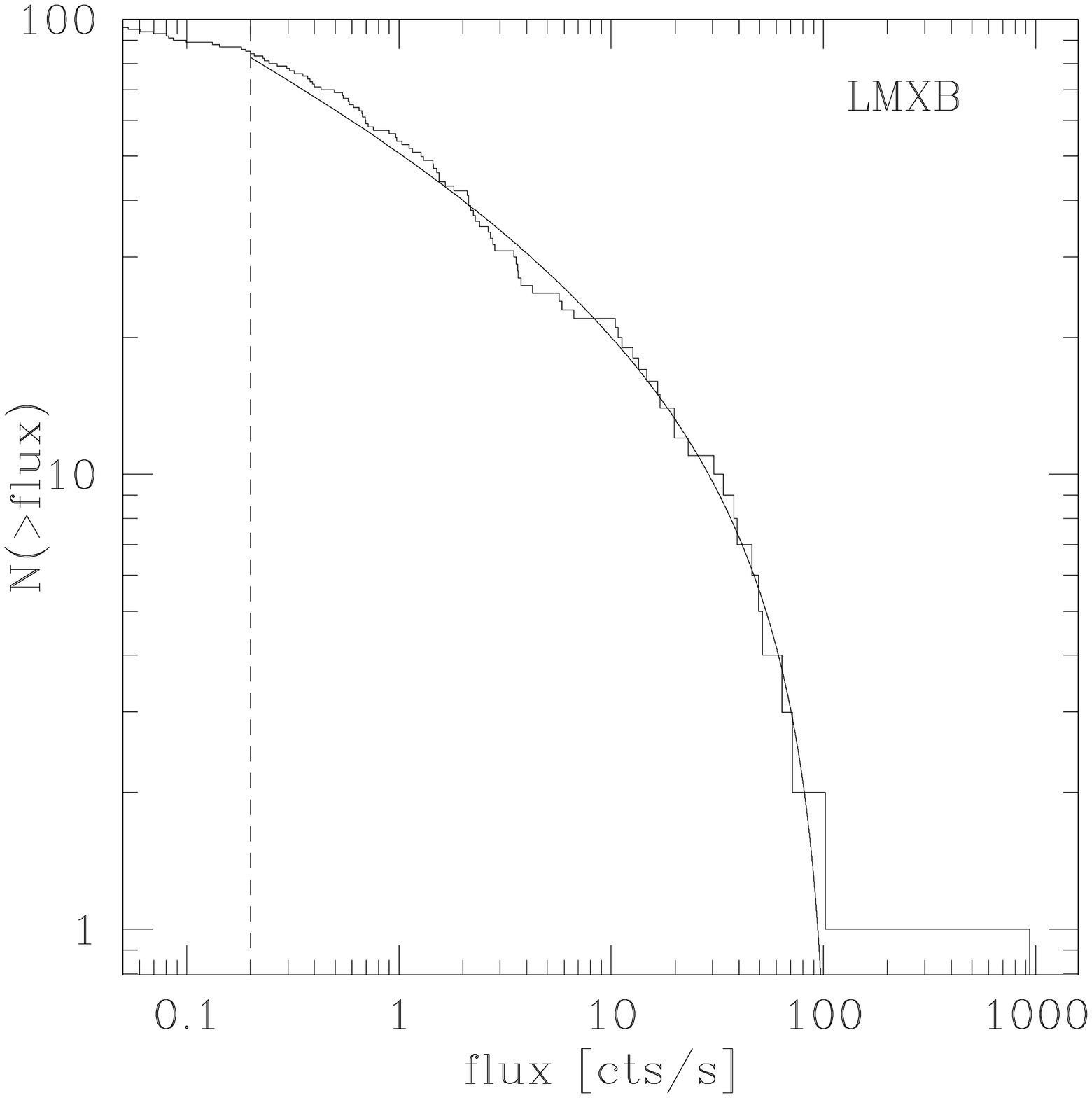}}
  \resizebox{0.5\hsize}{!}{\includegraphics{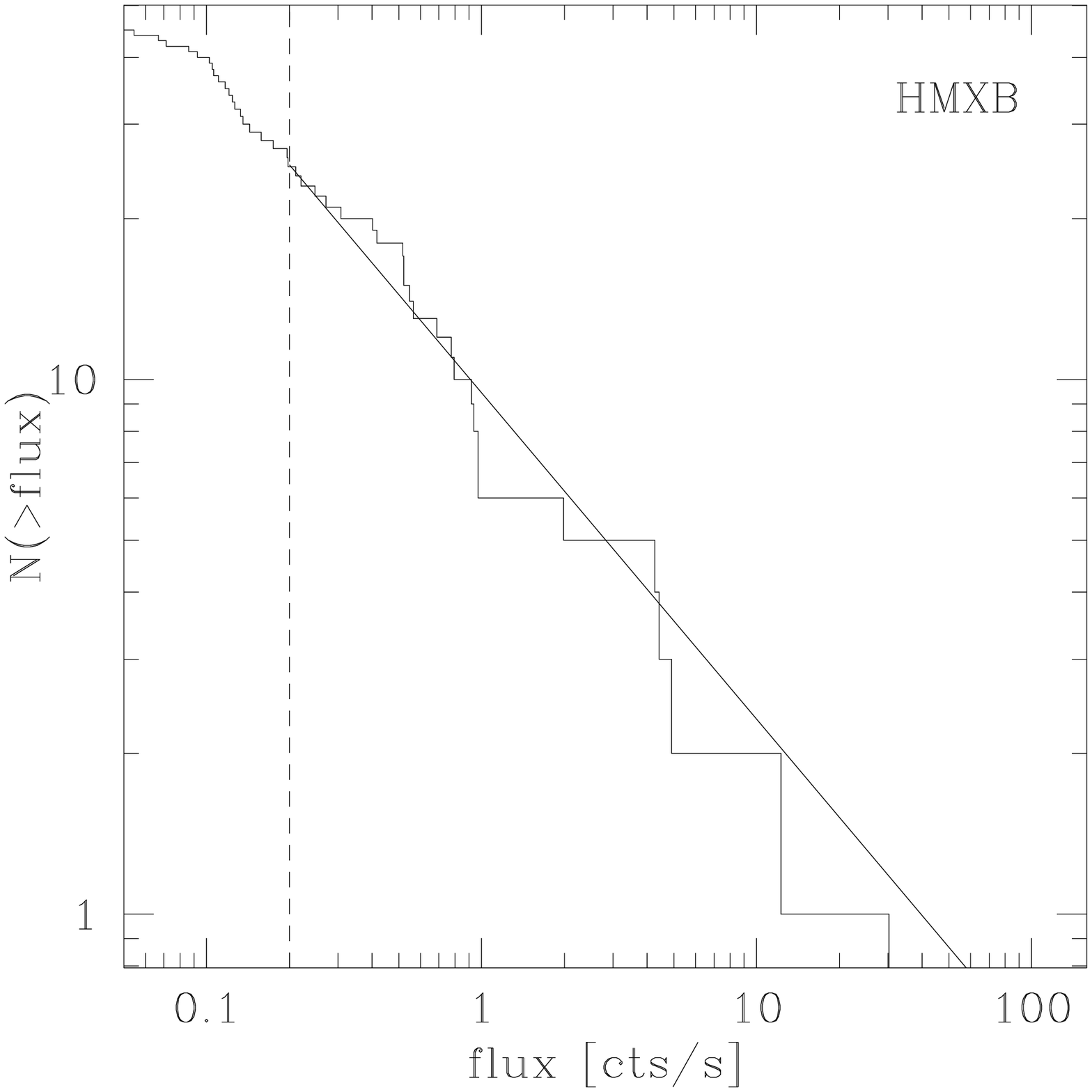}}
  \caption{Number--flux relation for galactic X-ray binaries. The
  vertical dashed line corresponds to our completeness limit of 0.2
  cnts s$^{-1}$. The solid lines are the best fit models to the ASM
  data -- a power law for HMXBs and a power law with cutoff in the
  differential \lns~ distributions at 110 cnts s$^{-1}$ for LMXBs (see
  Eqs.(\ref{eq:lnls_h}) and (\ref{eq:cut})).} 
  \label{fig:xb_logn}
\end{figure*}

\begin{table*}[ht]
\caption{The best fit values for the number--flux relation for
different classes of galactic sources from the ASM catalogue.}
\newcolumntype{Y}{>{\centering\arraybackslash}X}
\begin{tabularx}{\linewidth}{|Y|r|Y|Y|Y|Y|}
\hline
Subsample & no. of sources$^{(1)}$/all sources & cutoff [cnts s$^{-1}$]
&norma\-lisation & slope & quality of fit (K-S test)\\
\hline
all galactic & 131/217$^{(2)}$ & 110 & 88  & $0.34\pm 0.05$         & 92\% \\
             & 132/217 &  --  & 72  & $0.41\pm 0.04$         & 51\% \\
LMXB         & 83/105$^{(2)}$ & 110 & 83 & $0.2 \pm 0.06$    & 71\% \\
             & 84/105  &  --  & 56  & $0.3\pm 0.05$          & 0.5\% \\
HMXB         &  25/51  &  --  & 9.4 & $0.61_{-0.12}^{+0.14}$ & 46\% \\
SNR          &   6/7   &  --  & 4.8 & $0.36_{-0.19}^{+0.22}$ & 98\% \\
CV           &   5/10  &  --  & 0.5 & $1.68 \pm 0.61$        & 98\% \\
\hline
\end{tabularx}
$^{(1)}$ Number of sources above the completeness limit of 0.2 cnts s$^{-1}$.\\
$^{(2)}$ For fits with a cutoff the brightest source, \object{Sco X-1},
was excluded.
\label{tab:logn-logs}
\end{table*}

\subsection{Completeness}

Important for the analysis presented below are two aspects of
completeness: 
\begin{enumerate}
\item 
completeness flux limit of the ASM sample of the X-ray sources 
\item 
completeness of the sample of galactic X-ray
binaries which are  optically identified and for which distance
measurements are available
\end{enumerate}
The first problem arises for example in studying \lns~ distribution
of all galactic sources and is addressed below. The second problem is
important in analysing \lns~ distributions of various types of 
galactic X-ray sources and especially their luminosity functions. It
is discussed in Sect. \ref{sec:spat}.

Due to the present method of construction of the ASM catalogue its
completeness limit is difficult to assess in any straightforward
way. By definition the ASM sample includes all sources, galactic and
extragalactic, which have reached an intensity of 5 mCrab at any time,
which corresponds to a completeness limit of $\sim$ 0.37 cnts
s$^{-1}$. On the other hand we know from the same ASM light curves
that non-transient Galactic X-ray binaries have typical values of the
ratio of maximum flux (on the time scale of dwell--\-$\sim$ day) to
average flux of the order of few. Therefore, in terms of long term
average values the ASM catalogue might be complete down to lower
fluxes.

In order to indirectly probe the completeness limit of the ASM sample
we use the fact that the \lns~relation for extragalactic sources is
well known and follows a power law with index
$-$3/2~\citep{forman:78}, down to $\sim 3.8 \cdot 10^{-14}$ erg
s$^{-1}$ cm$^{-2}$~\citep{ogasaka:98} which corresponds to ASM count
rate of $1.2 \cdot 10^{-4}$ cnts s$^{-1}$. The \lns~relation for
extragalactic sources based on ASM data is compared with HEAO A-1 and
ASCA results in Fig. \ref{fig:extragal_logn}. One can see that
flattening of the source counts caused by incompleteness of the sample
begins at a count rate of $\sim 0.1$ cnts s$^{-1}$.

Therefore we set, somewhat arbitrarily, the completness limit of the
ASM sample of the X-ray sources at 0.2 cnts s$^{-1}$. We
verified that our conclusions are not sensitive to the exact value. 

\begin{figure*}[tb]
  \resizebox{0.5\hsize}{!}{\includegraphics{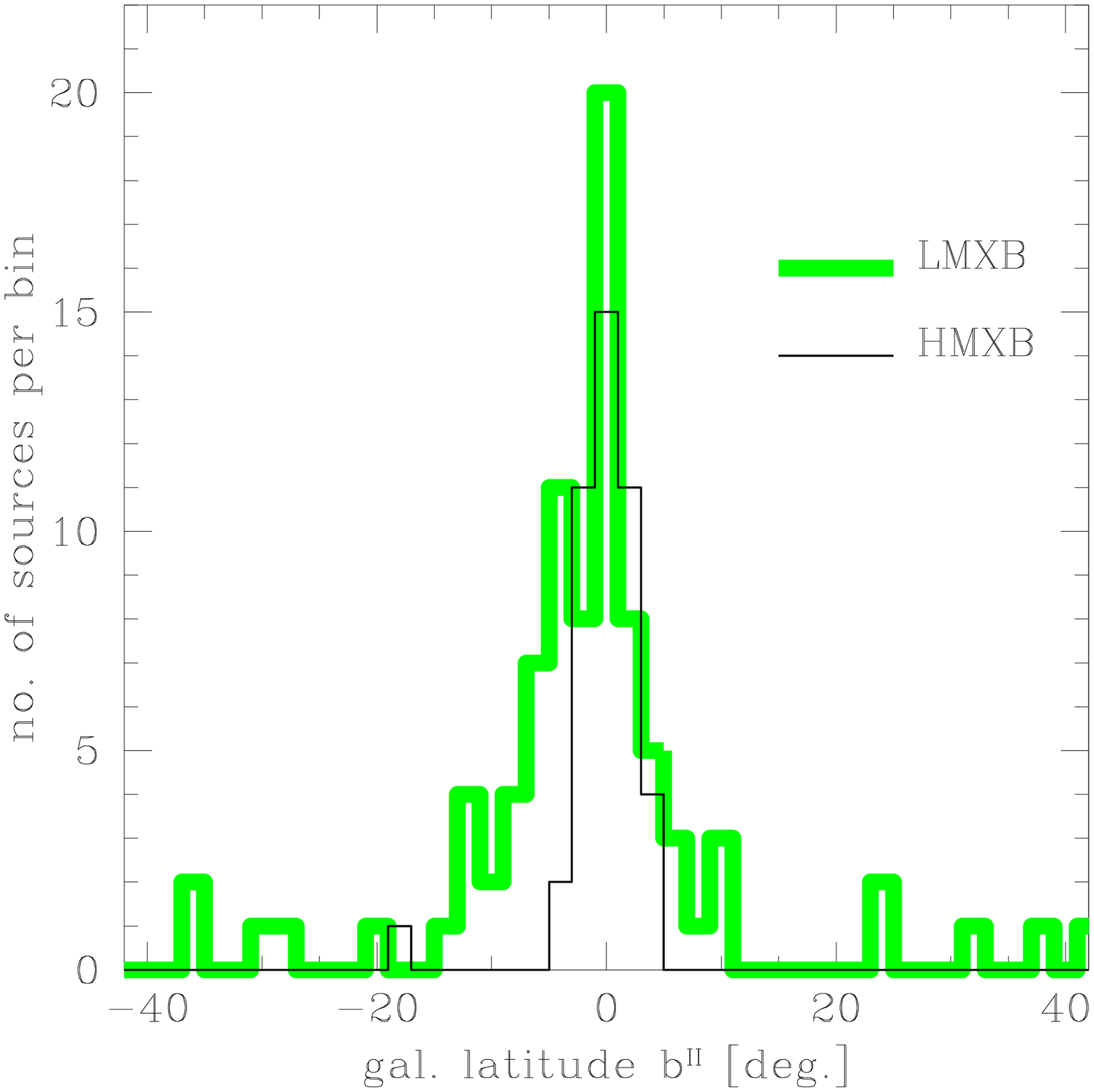}}
  \resizebox{0.5\hsize}{!}{\includegraphics{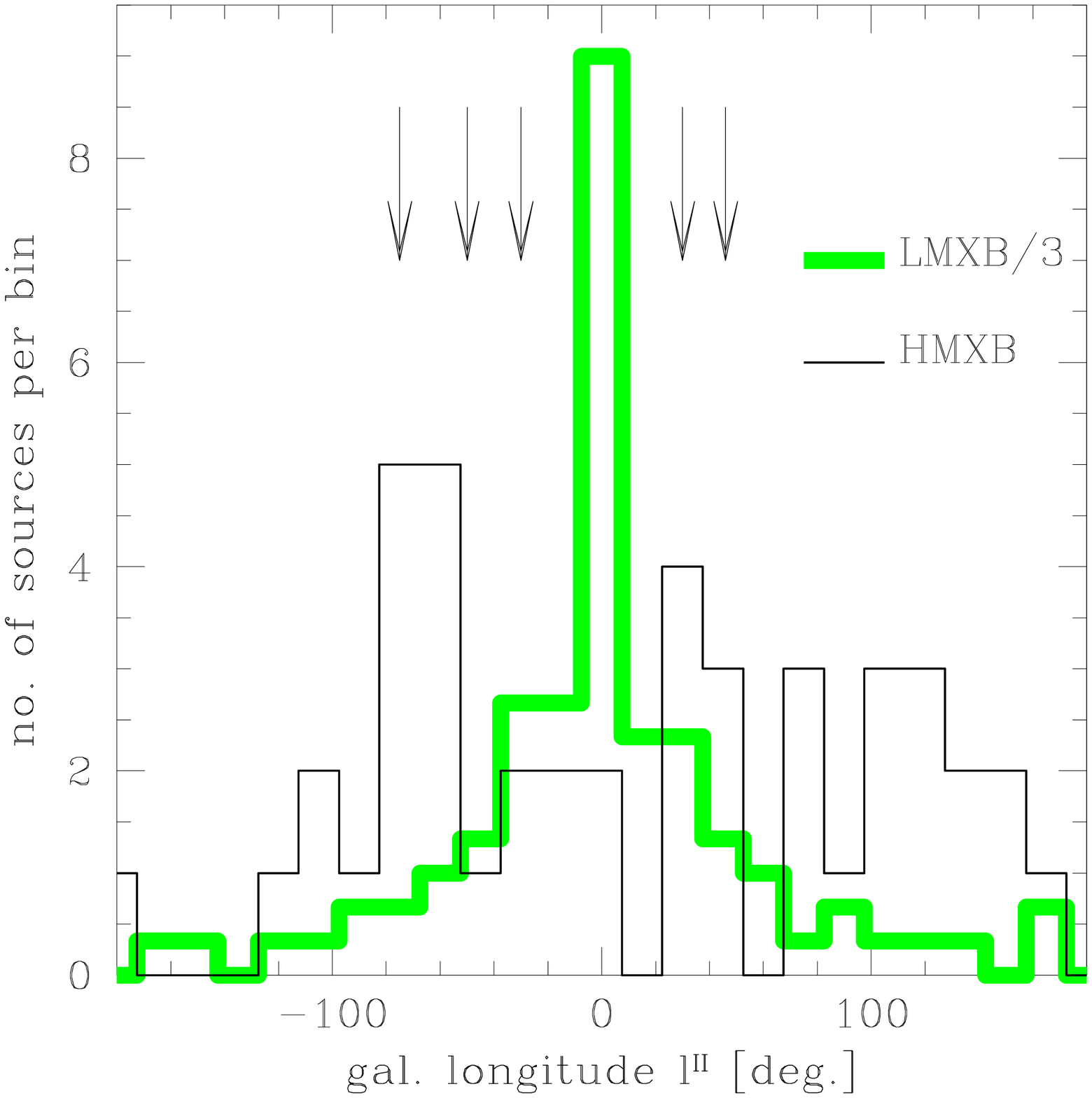}}
  \caption{The distribution of Galactic HMXBs (solid lines) and LMXBs
  (thick grey lines) against Galactic latitude $b^{II}$ (left panel)
  and longitude $l^{II}$ (right panel). The distribution against
  $b^{II}$ of HMXBs shows a stronger concentration towards the
  Galactic plane compared to LMXBs. Along $l^{II}$ LMXBs show a strong
  concentration in the direction towards the Galactic centre. The
  arrows in the right panel mark the positions of the tangential
  points of spiral arms. The broad hump in the HMXB distribution at
  $l^{II} = 100^{\circ}-160^{\circ}$ is mostly composed of relatively
  low luminosity sources in the Perseus and Cygnus arms. Note that on
  the right panel the number of LMXBs is divided by 3.}
  \label{fig:gal_ang}
\end{figure*}

%
%
\section{The \lns~distributions}
In order to calculate the number--flux relations the ASM light curves
were averaged over the entire time span of available data for each
source. The resulting \lns~relation for galactic
sources is shown in Fig. \ref{fig:gal_logn}. The
differentiation between galactic and extragalactic sources was done
using SIMBAD database. The overall shape and normalisation of the
\lns~relation of Galactic sources is similar to that obtained by UHURU
\citep{forman:78} and ARIEL V \citep{warwick:81a}. The UHURU
result~\citep{matilsky:73} is schematically shown in
Fig. \ref{fig:gal_logn} by the solid line. The \lns~relation for
different types of Galactic sources is also shown in
Fig. \ref{fig:gal_logn}.

We further selected X-ray binaries from the sample and divided them
into low mass (LMXB) and high mass (HMXB) binaries according to the
mass of the optical companion, using the mass of the secondary, M$_2$,
of 2.5 M$_{\odot}$ to separate high and low mass systems. The precise
value of this boundary affects classification of only few X-ray
binaries (\object{Her X-1}, \object{GX 1+4}, \object{GRO J1655-40}
etc.). In doing so we used SIMBAD  database, the Catalogue of X-ray
Binaries \citep{paradijs:94}, the Catalogue of CV, LMXB and related
objects \citep{ritter:98}, the catalogues of low-mass X-ray binaries
\citep{liu:01} and high-mass X-ray binaries \citep{liu:00} and in some
cases publications on individual sources.
Recently the donor star in \object{GRS 1915+105} was identified to be
a K or M giant \citep{greiner:01} so this source is classified as an
LMXB. Of 115 galactic X-ray binaries with average ASM flux exceeding
our completeness limit of 0.2 cnts s$^{-1}$ only 6 sources were left 
unclassified. The fraction of unclassified sources is $\sim 5\%$ and
they have fluxes in the $3\cdot10^{-1}$--13 cnts s$^{-1}$ range and
therefore should not affect our conclusions in any significant way. 
The compilation of galactic X-ray binaries with type, optical
companion, average flux and, if available, distance and average
luminosity is available in electronic form at
http://www.mpa-garching.mpg.de/$\sim$grimm/.
The resulting \lns~relations for LMXBs and  HMXBs are shown in
Fig. \ref{fig:xb_logn}.

To fit the observed \lns~distributions we used the usual power law in
the form:
\begin{eqnarray}
    N(>S) = k \cdot S^{-a}
\label{eq:lnls_h}
\end{eqnarray}
where $N(>S)$ is the number of sources with fluxes higher than $S$,
$a$ is the slope, and $k$ the normalisation. $S$ is measured in ASM
cnts s$^{-1}$. In order to calculate the best fit values of the
parameters we use a Maximum-Likelihood method in the form suggested by
\citet{murdoch:73}. This implementation of the M-L method takes into
account the errors associated with the flux. Since the systematic
error dominates the averaged flux error we used the value of 0.05 cnts
s$^{-1}$ from Sect. \ref{sec:syserr} as an estimate of the error. The
error is assumed to be Gaussian. Only sources with an averaged flux
above 0.2 cnts s$^{-1}$ were used in the fit. The best fit values for
different types of Galactic sources are given in Table
\ref{tab:logn-logs}. The errors given are an estimate of the 1$\sigma$
errors for one parameter of interest derived from the
Maximum-Likelihood method. In order to characterise the quality of the
fit we used the Kolmogorov-Smirnov test.

As is obvious from Fig. \ref{fig:xb_logn} and the results of the K-S
test (Table \ref{tab:logn-logs}) a simple power law distribution does
not describe the observed \lns~relation for LMXBs. A gradual
steepening of the \lns~relation occurs towards higher fluxes. Similar
behaviour was also found by UHURU \citep{matilsky:73} and OSO-7
\citep{markert:79}. We therefore modified the simple power law in the
form:
\begin{eqnarray}
  N(>S) = k \cdot (S^{-a} - S_{max}^{-a})
  \label{eq:cut}
\end{eqnarray}
This corresponds to a cutoff in the differential \lns~relation at flux
$S = S_{max}$. The value of the cutoff was chosen to $S_{max}$ =
110 cnts s$^{-1}$. The results, however, are not very sensitive to the
actual value of $S_{max}$. The above value of $S_{max}$ corresponds to
the ASM flux from a 1.4 M$_{\odot}$ neutron star located at a distance
of 6.5 kpc (average distance of LMXBs from the Sun) and radiating at
Eddington luminosity. For fitting the \lns~ of all galactic and LMXB
sources with cutoff we excluded the brightest source, \object{Sco X-1},
from the sample since its flux is far higher than the cutoff. As can
be seen from Table \ref{tab:logn-logs} and Fig. \ref{fig:xb_logn}
introduction of the cutoff significantly improves the quality of the
fit for LMXBs. On the other hand it does not change significantly the
results for other types of Galactic sources, especially HMXBs. Note
that the steepening of the \lns~for LMXBs is not an artifact of the
incompleteness of the source sample at low fluxes. The numbers do not
change qualitatively if we increase the low flux limit by a factor of
2 -- the values of K-S probability are 6\% and 68\% for a single
power law and a power law with cutoff in the form of Eq. (\ref{eq:cut}),
respectively.

%
%
\section{Spatial distribution of X-ray binaries}
\label{sec:spat}

Progress in the number of distance determinations and
identifications of secondary stars in X-ray binaries in the last
decade opens the opportunity to study the 3-D distribution of XRBs in
more detail than was previously possible. Notwithstanding the still
relatively small number of X-ray sources and the sometimes poor
accuracy of distance determinations it is now possible to compare the
observed distribution of XRBs with theoretical expectations.
This is not only interesting in itself but, because of the flux
limited nature of the ASM sample, knowledge of the spatial
distribution is required in order to derive the luminosity
function. Due to the above mentioned uncertainties and the flux
limitation of the sample it is still not possible to unambiguously
determine shape and parameters of the XRB distribution. We therefore
adopted an approach in which we use the standard model of the stellar
mass distribution in the Galaxy as a starting point and adjust,
whenever possible, its parameters to fit observed distributions of low
and high mass X-ray binaries. As the luminosity function depends
somewhat on the assumed spatial distribution, we verify that
variations of the parameters, which can not be determined from the
data do not affect derived luminosity functions significantly.

\subsection{Angular distribution of X-ray binaries}

The all-sky map shown in Fig. \ref{fig:asm_sky} demonstrates vividly
that the angular distributions of high and low mass X-ray binaries
over the sky differ significantly. This fact is further illustrated
by the angular distributions against Galactic latitude and longitude
shown in Fig. \ref{fig:gal_ang}. The figures illustrate the well-known
fact that HMXBs are strongly concentrated towards the Galactic
plane. In addition drastic difference in the longitude distributions
of HMXBs and LMXBs can be noticed, with the latter significantly
concentrated towards the Galactic Centre/Bulge and the former
distributed in clumps approximately coinciding with the location of
tangential points of the spiral arms,see
e.g. \citet{englmaier:99,simonson:76}.

\subsection{Source distances and 3-D distribution of X-ray binaries}
\begin{figure}[t]
  \resizebox{\hsize}{!}{\includegraphics{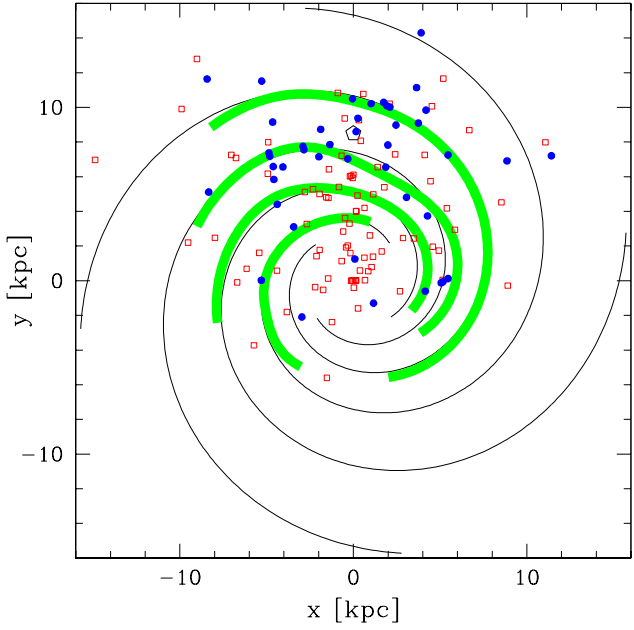}}
  \caption{Face-on view of the Galaxy -- distribution of low mass
  (open squares) and  high mass (filled circles) X-ray binaries. The
  origin of the coordinate is at the Galactic Centre. The Sun is
  located at x=0, y=8.5 (marked by the pentagon). The
  thin  solid line shows logarithmic 4-armed (m=4) spiral model  with
  pitch angle of $12\degr$ (e.g.~\citet{vallee:95}). The thick solid
  lines show the spiral model of the Galaxy based on optical and radio
  observation of the giant HII regions (\citet{georgelin:76},
  \citet{taylor:93}). The fact that the majority of sources is
  located at $y>0$ is due to the flux limited nature of the ASM sample
  and incompleteness of the optical identifications/distance
  measurements at the large distances from the Sun (see discussion in
  the text).}
  \label{fig:faceon}
\end{figure}

\begin{figure}[tb]
  \resizebox{\hsize}{!}{\includegraphics{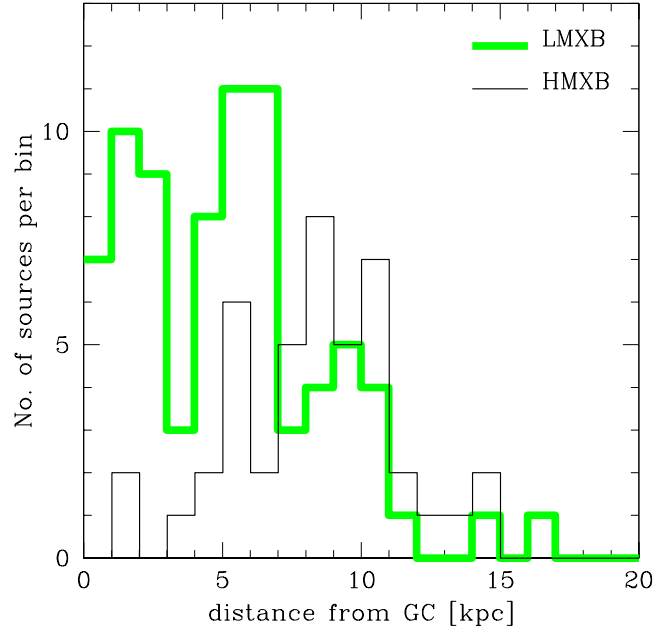}}
  \caption{Radial distributions of high mass (solid histogram) and low
  mass (thick grey histogram) X-ray binaries. The projected distance is
  defined as $\sqrt{x^2 + y^2}$, where $x$ and $y$ are Cartesian
  coordinates in the Galactic plane, see Fig. \ref{fig:faceon}.
  Note that the plotted distributions are not corrected for the volume
  of cylindrical shells ($\propto r$).}
  \label{fig:r_dist}
\end{figure}

In order to study the {\em spatial} distribution of X-ray
binaries we collected source distances from the literature. We
found distances for 140 X-ray binaries from the ASM
sample. For X-ray binaries with an average flux above
the ASM completeness limit, used for constructing the luminosity
functions in Sect. \ref{sec:lumf}, distances were determined for
all but 8 sources. In cases when the published distance estimates
disagree significantly we used the least model dependent estimates or
their average. For the compilation of the source distances see
http://www.mpa-garching.mpg.de/$\sim$grimm/. The spatial distribution
of X-ray binaries in various projections is shown in
Fig. \ref{fig:faceon}--\ref{fig:z_dist}.

\begin{table}[bt]
\caption{The parameters of the standard Galaxy model.}
\newcolumntype{Y}{>{\centering\arraybackslash}X}
\begin{tabularx}{\linewidth}{|c|Y|c|c|}
\hline
parameter & meaning &\multicolumn{2}{c|}{value}\\
\cline{3-4}
         &                            & HMXB   & LMXB \\
\cline{3-4}
\hline
$q$      & oblateness of bulge        & -- & 0.6     \\
$\gamma$ & --                         & -- & 1.8     \\
$R_e$    & scale length of spheroid   & -- & 2.8 kpc \\
$b$      & --                         & -- & 7.669   \\
$r_0$    & scale length of bulge      & -- & 1 kpc   \\
$r_t$    & truncation radius of bulge & -- & 1.9 kpc \\
$r_d$    & scale length of disk       & 3.5 kpc& 3.5 kpc \\
$r_z$    & vertical scale of disk     & 150 pc & 410 pc \\
$r_m$    & inner disk cut-off         & 6.5 kpc& 6.5 kpc\\
$R_{\text{mass}}$ & mass ratios Disk:Bulge:Spheroid& 1:0:0 & 2:1:0.8\\
\hline
\end{tabularx}
\label{tab:para}
\end{table}

\subsection{The Galaxy model}
\label{sec:spat_model}
As a starting point in constructing the spatial distribution of
X-ray binaries we employ the standard three component model of the
stellar mass distribution in the Galaxy \citep{bahc-son:80},
consisting of bulge, disk and spheroid. The parameterisation of
bulge and disk is taken from~\citet{dehnen:98} and for the spheroid we
take the model of~\citet{bahc-son:80}:
\begin{equation}
  \rho_{Bulge} = \rho_{0, Bulge} \cdot (\frac{\sqrt{r^{2} +
                 \frac{z^{2}}{q^{2}}}}{r_0})^{-\gamma} \cdot
                 \exp(-\frac{r^{2} +\frac{z^{2}}{q^{2}}}{r_t^2})
\label{eq:bulge}
\end{equation}
\begin{equation}
  \rho_{Disk} = \rho_{0, Disk} \cdot
  \exp(-\frac{r_m}{r}-\frac{r}{r_d} - \frac{|z|}{r_z})
\label{eq:disk}
\end{equation}
\begin{equation}
  \rho_{Sphere} = \rho_{0, Sphere} \cdot \frac{\exp(-b \cdot
  (\frac{R}{R_{e}})^{1/4})}{(\frac{R}{R_{e}})^{7/8}},
\label{eq:sphere}
\end{equation}
where $\rho_{0, Bulge}$, $\rho_{0, Disk}$ and $\rho_{0, Sphere}$ are
the normalisations, $r$ is the distance in the plane from the galactic
centre, $z$ is the distance perpendicular to the galactic plane, and
$R$ is the distance from the galactic centre in spherical
coordinates. All distances are in kiloparsec.
Meaning and values for other parameters are given in Table
\ref{tab:para}. 

In the standard Galaxy model the mass ratios of the components
are about 2:1:0.3 for disk:bulge:spheroid. These numbers follow from
the model using normalisations for the disk, $\rho_{0, Disk} = 0.05 
\text{M}_{\odot} \text{pc}^{-3}$, and spheroid population, $\rho_{0,
Sphere} = 1/500 \cdot  \rho_{0, Disk}$, observed in the vicinity of
the Sun ~\citep{zombeck:90} and a bulge mass of about $\sim 1.3 \cdot
10^{10}$ M$_{\odot}$~\citep{dwek:95}. All these masses refer to 
baryonic mass in the stars. 

All three components of the standard Galaxy model were used to
construct the spatial distribution of LMXB. The spheroid
component with appropriately adjusted normalisation was used to
account for the population of globular cluster sources.  Based on the
observed distribution and theoretical expectation that HMXBs trace the
star forming regions in the Galaxy, only the disk component was used
for the spatial distribution of HMXBs.

Several parameters, namely vertical scale height of the disk and
relative normalisation of the spheroid for the LMXBs, can be
determined directly from our sample of X-ray binaries. For these
parameters we used the best fit values inferred by the data. For the
rest of the parameters we accepted standard values for the stellar
mass distribution in the Galaxy. The final set of the parameters is
summarised in Table \ref{tab:para}. 

The disk component of the standard Galaxy model was modified in order
to account for the Galactic spiral structure. The description of the
spiral arms is based on the model of ~\citet{georgelin:76} derived
from the distribution of HII regions. To include it into our Galaxy
model we used the FORTRAN code provided by ~\citet{taylor:93}. The
spiral arms computed in this way are shown in Fig. \ref{fig:faceon} by
thick grey lines. This empirical model is close but not identical to a
4 arm logarithmic spiral with pitch angle of $12^{\circ}$
(e.g. \citet{vallee:95}) shown in Fig. \ref{fig:faceon} by thin solid
lines.

In the following two subsections we discuss spatial distribution of
HMXBs and LMXBs in more detail.

\subsection{High mass X-ray binaries}
The angular distribution of HMXBs in Fig. \ref{fig:gal_ang} shows
signatures of the Galactic spiral structure. These signatures are 
clearly seen in the distribution of sources over galactic longitude
which shows maxima approximately consistent with directions towards
tangential points of the spiral arms. No significant peak in the
direction to the Galactic centre is present. The signatures of the
spiral structure become more evident in the 3-D distribution of the
smaller sample of sources for which distance measurements are
available, Figs. \ref{fig:faceon},\ref{fig:r_dist}. The radial
distribution (Fig. \ref{fig:r_dist}) shows pronounced peaks at the
locations of the major spiral arms and is similar to that of primary
tracers of the Galactic spiral structure --  giant HII regions
(e.g. \citet{downes:80}) and warm molecular clouds
(e.g. \citet{solomon:85}). In particular, the central $\sim 3-4$ kpc
region of the Galaxy is almost void of HMXB well in accordance with
the radial distribution of the giant HII regions and warm CO
clouds. This appears to correspond to the interior of the 4-kpc
molecular ring.

\begin{figure*}[tb]
  \resizebox{0.5\hsize}{!}{\includegraphics{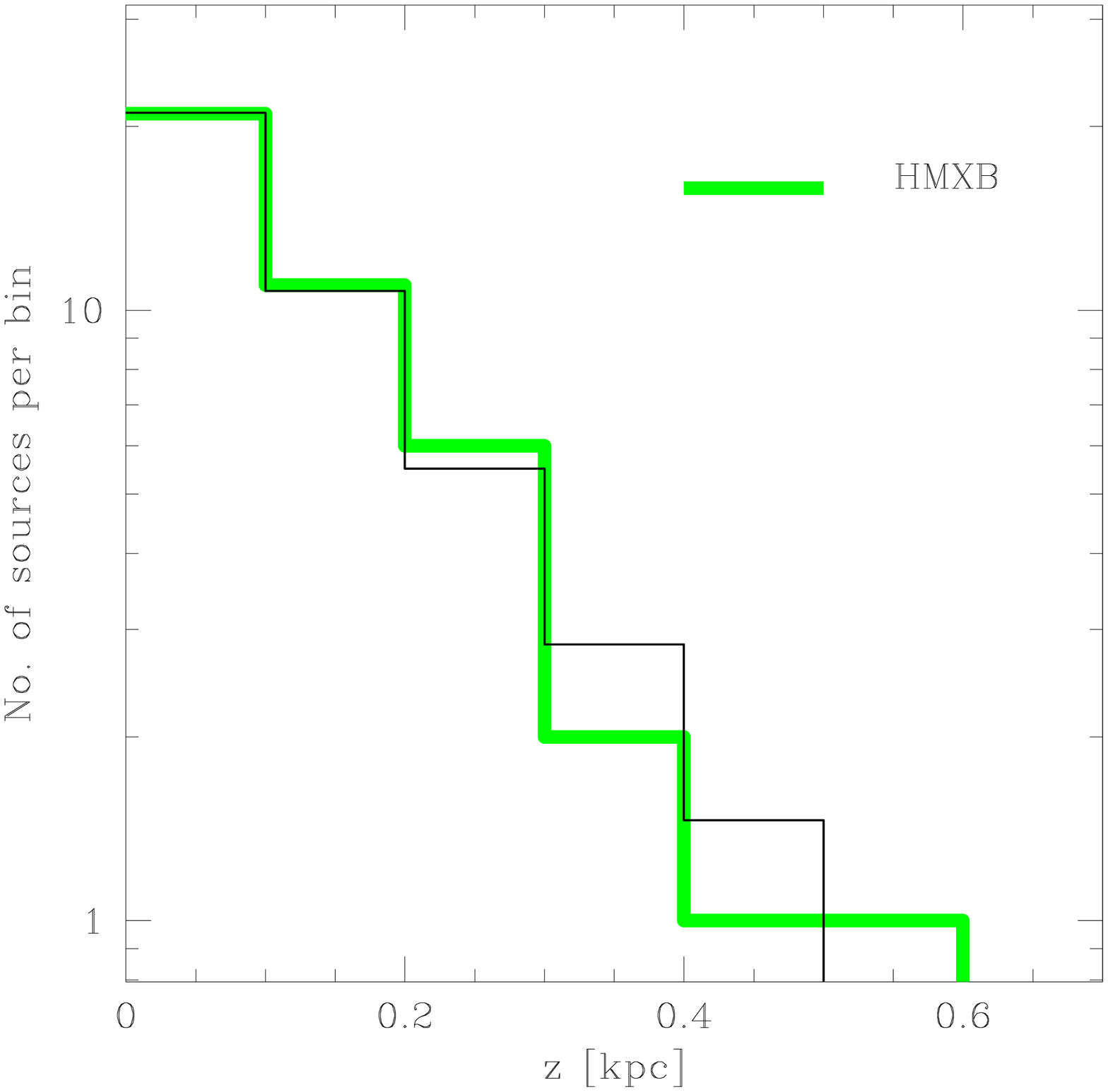}}
  \resizebox{0.5\hsize}{!}{\includegraphics{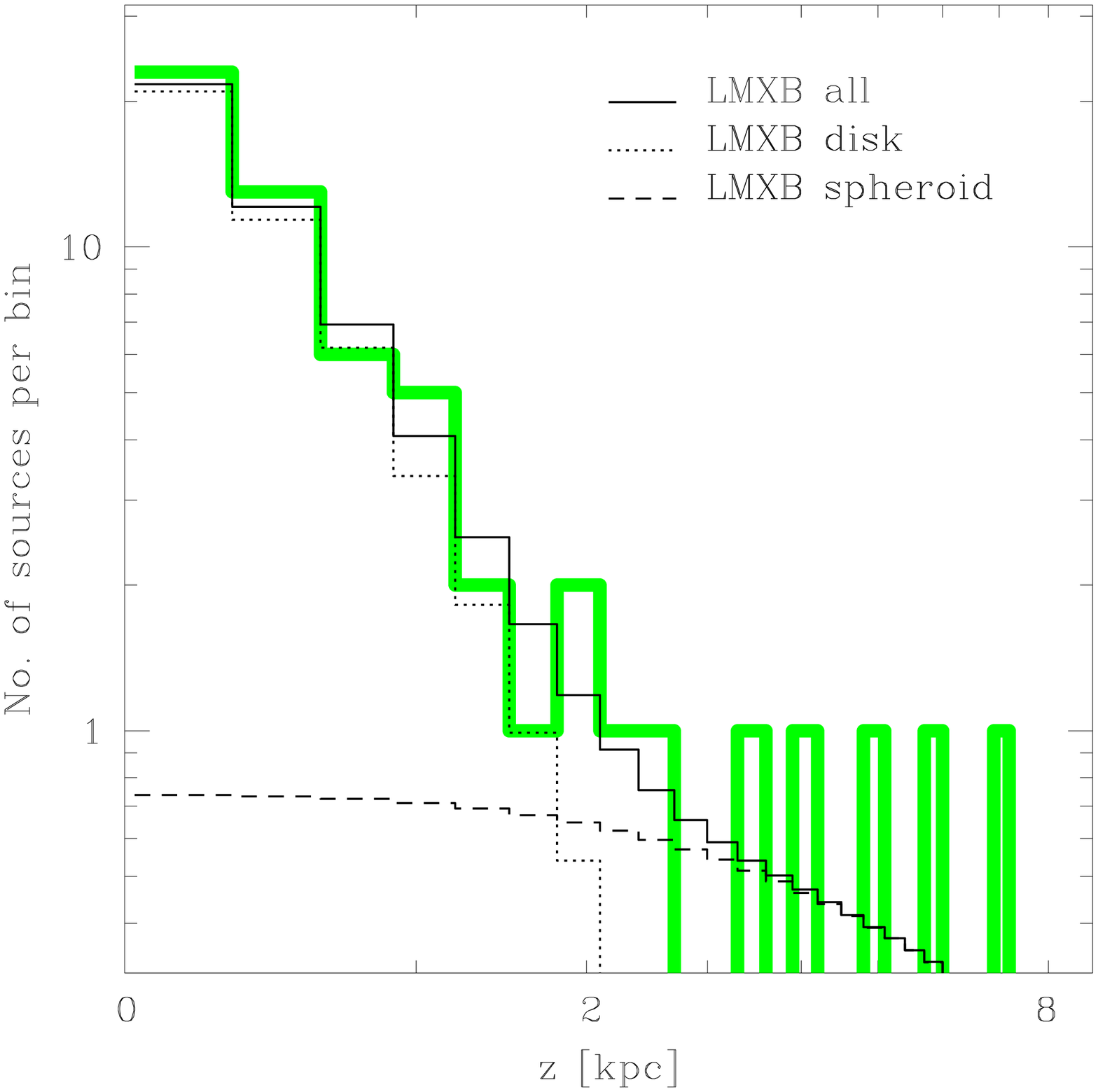}}
  \caption{Vertical distributions of high mass (left panel) and low
  mass (right panel) X-ray binaries. The vertical distributions were
  summed over northern and southern galactic hemispheres. In the case
  of LMXBs only sources with $R > 3.5$ kpc were used, to exclude bulge
  sources. The thick grey solid lines show the observed distributions
  and the thin solid and dashed lines the expected distributions for
  an exponential disk with 150 pc scale height for HMXBs, and an
  exponential with scale height 410 pc and a 25\% contribution of the
  spheroid for LMXBs, respectively. For the assumed model see
  Eqs. (\ref{eq:disk}, \ref{eq:sphere}).}
  \label{fig:z_dist}
\end{figure*}

The vertical distribution of HMXBs is significantly more concentrated
towards the Galactic Plane and sufficiently well described by
a simple exponential with a scale height of 150 pc as shown in the
left panel of Fig. \ref{fig:z_dist}.

Based on theoretical expectations and on the data shown in Fig.
\ref{fig:asm_sky},\ref{fig:gal_ang},\ref{fig:r_dist},\ref{fig:z_dist} 
we included only the disk component in the volume density distribution 
HMXBs. It is clear however that a simple exponential disk is not a
good description for the radial distribution of HMXB. Therefore,
following \citet{dehnen:98} we assumed the disk density distribution
in the form given by Eq. (\ref{eq:disk}), where the first term in the
exponential allows for the central density depression. To describe the
observed central depression for HMXBs a rather large value of
$r_m\approx 6-7$ kpc is required (cf. $r_m = 4$ kpc
from~\citet{dehnen:98}). The spiral arms were assumed to have a
Gaussian density profile along the Galactic Plane:
\begin{equation}
  \rho_{Spiral} \propto \sum_{j=1}^{j=4}\exp(-(\frac{s_j}{w_a})^2),
\label{eq:spiral}
\end{equation}
where $w_a=600$ pc is the width of the spiral arm, and $s_j$
is the distance to the nearest point of the spiral arm $j$ projected
to the Galactic Plane:
\begin{equation}
s_j = \sqrt{(x-x_j^\prime)^2 + (y-y_j^\prime)^2}.
\end{equation}
In order to account for the spiral structure the disk density, 
Eq. (\ref{eq:disk}), was multiplied by $\rho_{Spiral}$:
\begin{equation}
  \rho^{HMXB}_{Disk}\propto \rho_{Disk} \cdot \rho_{Spiral}
  \label{eq:hmxb_disk}
\end{equation}

\subsection{Low mass X-ray binaries}
\label{sec:lmxb_distr}

Contrary to HMXB, the angular distribution of LMXBs is strongly peaked
in direction to the Galactic centre and declines gradually along the
Galactic plane, see Fig. \ref{fig:gal_ang}. The central $\sim 2$ kpc
region is densely populated with Galactic Bulge LMXB sources and
contains $\sim 1/3$ of the LMXBs from our flux limited sample
(Fig. \ref{fig:r_dist}). A noticeable feature of the radial
distribution of LMXB is the pronounced minimum at $\sim 3-4$ kpc. This
minimum approximately coincides with the $\sim 1-3$ kpc gap in the
distribution of the molecular gas and the $\sim 2.2$ kpc minimum in
the density of infrared light distribution in the Galaxy
\citep{binney:97} and probably separates bulge sources from the disk
population. Similar to HMXBs, the signatures of the spiral structure
might be present in the radial distribution although they are less
pronounced.

The vertical distribution outside the bulge
(Fig. \ref{fig:z_dist}) is significantly broader than that of HMXBs
and includes a number of sources at high galactic $z$. A formal fit to
the observed distribution with an exponential law results in a large
scale height of $950 \pm 130$ pc, which is close to the value of 710
pc obtained by~\citet{vanparadijs:95} for NS LMXBs. However, due to
presence of a tail of sources at $|z|>1.5-2$ kpc, the observed
z-distributions cannot be 
adequately described by a simple exponential law. As only three out
of nine sources at $|z|>2$ kpc are located in globular clusters, 
this tail of high-z sources cannot be solely due to the globular
cluster component. A possible mechanism -- a kick received by a
compact object during the SN explosion, was considered e.g. by
~\citet{vanparadijs:95}. The relatively small number of high-z sources
does not allow one to determine the shape of their distribution based
on the data only. In order to account for the high-z sources and the
LMXB sources in globular clusters we chose to include in the
spatial distribution of LMXBs the spheroid component described by a de
Vaucouleurs profile  (Eq. (\ref{eq:sphere})). Note that a de Vaucouleurs
profile correctly represents the distribution of globular
clusters. The overall vertical distribution can be adequately
represented by a sum of an exponential law with a scale height of
$410^{+100}_{-80}$ pc and a de Vaucouleurs profile with the
parameters given in Table \ref{tab:para}. The spheroid component
represented by the de Vaucouleurs profile contains a $\sim 25\%$ of
the total number of LMXBs. Note, that this number is by a factor of
$\sim$ 2--3 larger than the mass fraction of the stellar spheroid in
the standard Galaxy model.  The enhanced fraction of the spheroid
component is generally consistent with the fact, that the number of
X-ray sources per unit mass is $\sim 100$ times higher in the globular
clusters than in the Galactic disk and 12 out of 104 LMXBs in our
sample are globular cluster sources.

The angular resolution of the ASM instrument does not permit to study
in detail the very central region of the Galaxy which is characterised
by the highest volume and surface density of X-ray binaries. Based on
GRANAT/ART-P data having significantly better sensitivity and angular
resolution, \citet{grebenev:96} showed that the distribution of
the  surface density of X-ray binaries in the central $8\degr \times 8
\degr$ of the Galaxy is consistent with the stellar mass distribution
in the Galactic Bulge.

To conclude, our model of the volume density distribution of LMXBs
includes all three components of the standard model of the Galaxy:
bulge, disk and spheroid with the disk-to-spheroid mass ratio
decreased to 4:1. Similarly to HMXBs, $r_m\approx 6-7$ kpc is required
to describe the central density suppression of the disk
population. The modulation of the disk component by the spiral pattern
at the 20\% level was also included:
\begin{equation}
  \rho^{\text{LMXB}}_{Disk}\propto \rho_{Disk} \cdot (1 + 0.2 \cdot
  \rho_{Spiral})
\end{equation}
where $\rho_{Disk}$ is given by Eq. (\ref{eq:disk}) and
$\rho_{Spiral}$ -- by Eq. (\ref{eq:spiral}).

\subsection{Completeness of the sample of the distance measurements.}
\label{sec:dist_distr}

\begin{figure}[tb]
  \resizebox{\hsize}{!}{\includegraphics{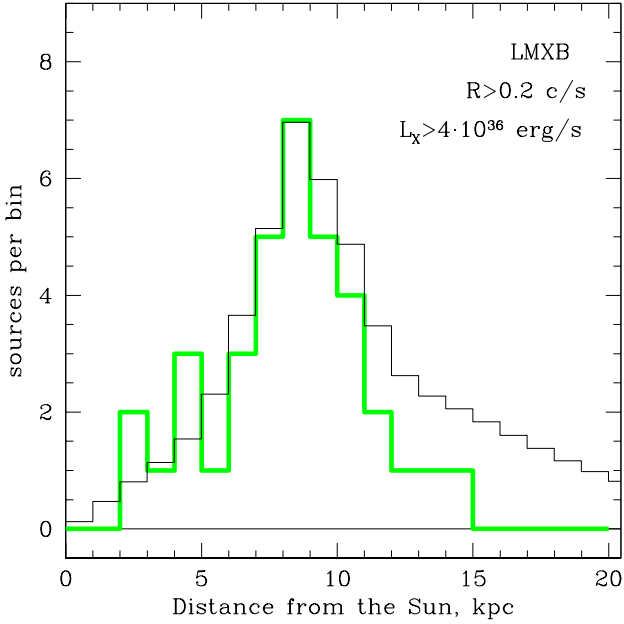}}
  \caption{Distribution of the LMXB sources over distance from the Sun
  (thick grey histogram). Only sources with luminosity $L_X>4\cdot
  10^{36}$ erg s$^{-1}$  are plotted. Given the ASM completeness flux
  limit of 0.2 cnts s$^{-1}$, sources with $L_X>4\cdot 10^{36}$ erg
  s$^{-1}$ should be visible from the distance of upto $\approx 20$
  kpc. The thin solid histograms shows the expected distribution of
  the sources in the model constructed in
  Sect. \ref{sec:lmxb_distr}. Deviation of the observed distribution
  from the prediction becomes visible at the distance $> 10-15$ kpc.}
  \label{fig:dist_distr}
\end{figure}

The fact that the majority of the sources in Fig. \ref{fig:faceon} is
located at  $y>0$ is related to the flux limited nature of the ASM
sample (obviously it is easier to observe weak sources located closer
to the Sun) and to the incompleteness of the available distance
measurements (more difficult to measure the distance to a more distant
source). The 3-D distribution of X-ray binaries enables one to check
the latter effect.

Plotted in Fig. \ref{fig:dist_distr} is the distribution of LMXB
sources with luminosities $L_X > 4\cdot 10^{36}$ erg s$^{-1}$ over the
distance from the Sun. For the ASM completeness flux limit of 0.2
cnts s$^{-1}$, sources with $L_X > 4 \cdot 10^{36}$ erg s$^{-1}$
should be visible up to a distance of $\approx 20$ kpc. However,
comparison with the expected distribution computed using the LMXB
volume density distribution constructed in Sect. \ref{sec:lmxb_distr}
shows an increasing deficiency of sources at distances $\ga 10-15$
kpc. In total $\sim 14$ sources in the distance range of 10-20 kpc are
``missing''. These ``missing'' sources should be hidden among the
$\sim 20$ unclassified sources in the ASM catalogue for which no
optical identification/distance determinations are available.

Recent observations by \citet{kuijken:01} lend support to this
interpretation. They measured proper motions of blue and red giants
in direction to the Galactic centre. The red giants, concentrated in
the Galactic bulge, have a velocity dispersion in Galactic
coordinates, $b^{II}$ versus $l^{II}$, symmetric around zero. However,
blue giants, located in the disk, have a velocity dispersion
asymmetric around zero with respect to $l^{II}$ which means that there
is a net motion of the observed blue giants in one
direction. Interpreting this as the motion of the disk around the
Galactic centre, it also means that there is a deficit of the observed
blue giants on the far side of the Galaxy
(c.f. Fig. \ref{fig:dist_distr}).

This comparison (Fig. \ref{fig:dist_distr}) shows that our sample 
of optical identifications/distance measurements for LMXB sources is
complete up to a distance of $\sim 10$ kpc. The significantly smaller
number of HMXBs above the ASM completeness flux limit did not permit
us to perform a similar analysis for HMXB sources. However, one might 
expect that due to the higher luminosity of the optical companion the
limiting distance for HMXBs is not smaller than for LMXBs. We
therefore accepted a value of $D_{max}=10$ kpc as a maximum source
distance for the luminosity function calculation for both types of
sources described in the next section.

\begin{figure}[tb]
  \resizebox{\hsize}{!}{\includegraphics{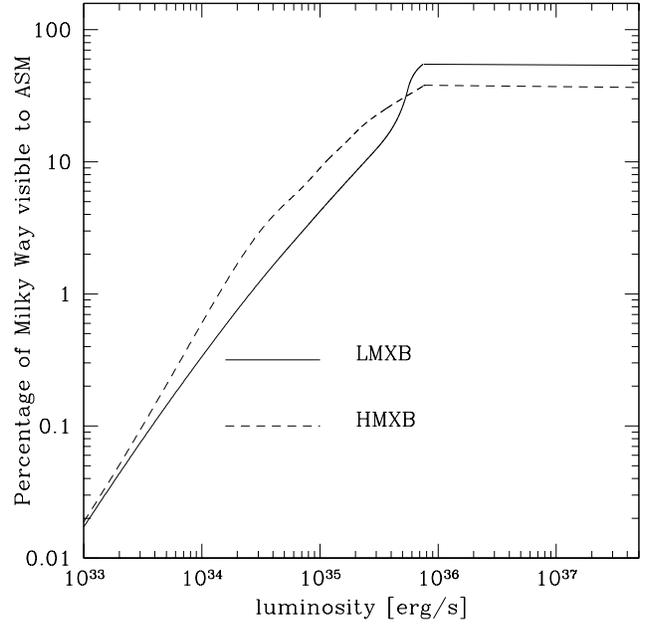}}
  \caption{Fraction of the mass of the Galaxy visible to ASM with
  account for the selection criteria described in the text as a
  function of source luminosity.}
  \label{fig:mass_frac}
\end{figure}

%
%
\section{Luminosity function}
\label{sec:lumf}
\begin{figure*}[tb]
  \resizebox{0.5\hsize}{!}{\includegraphics{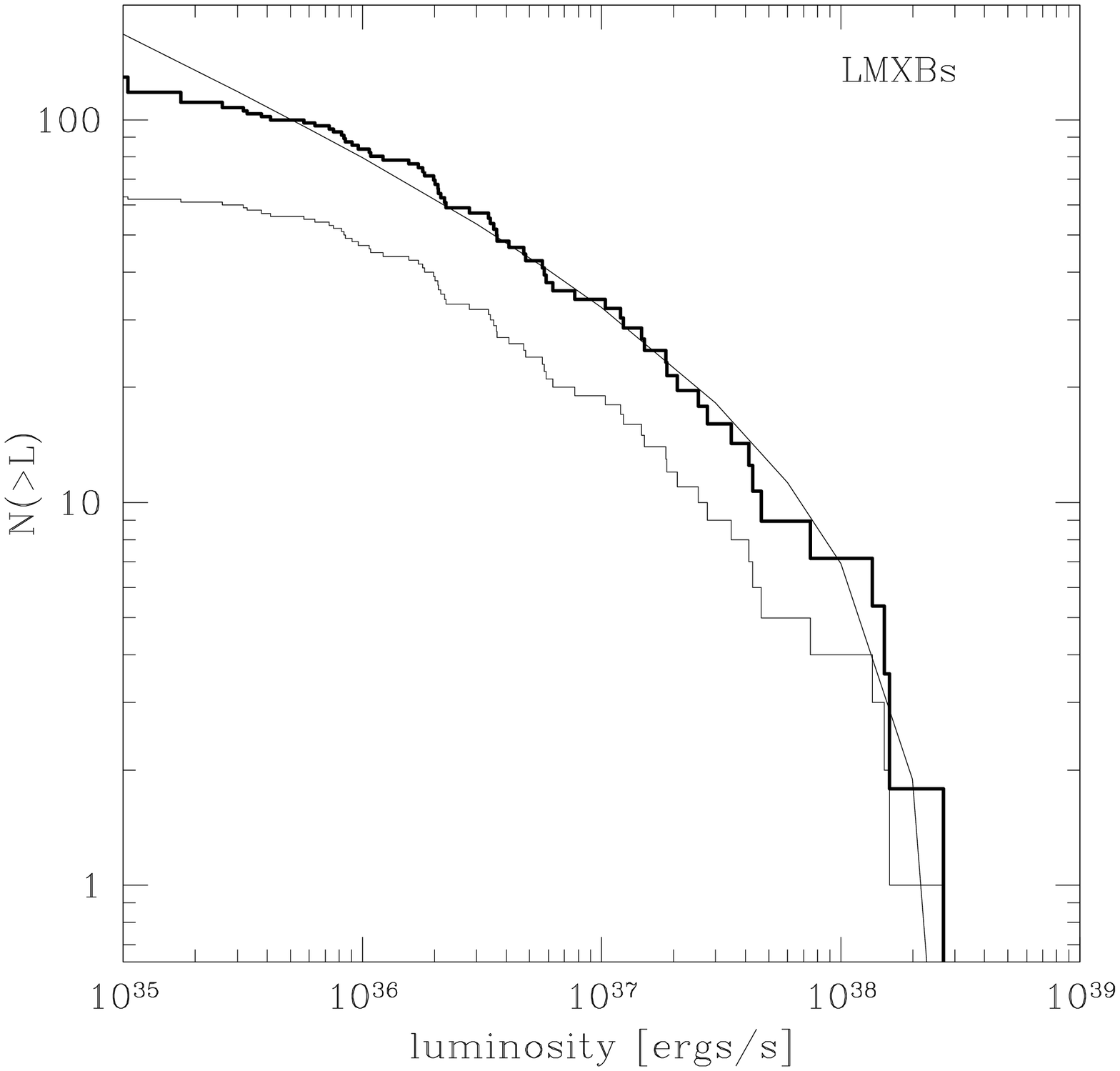}}
  \resizebox{0.5\hsize}{!}{\includegraphics{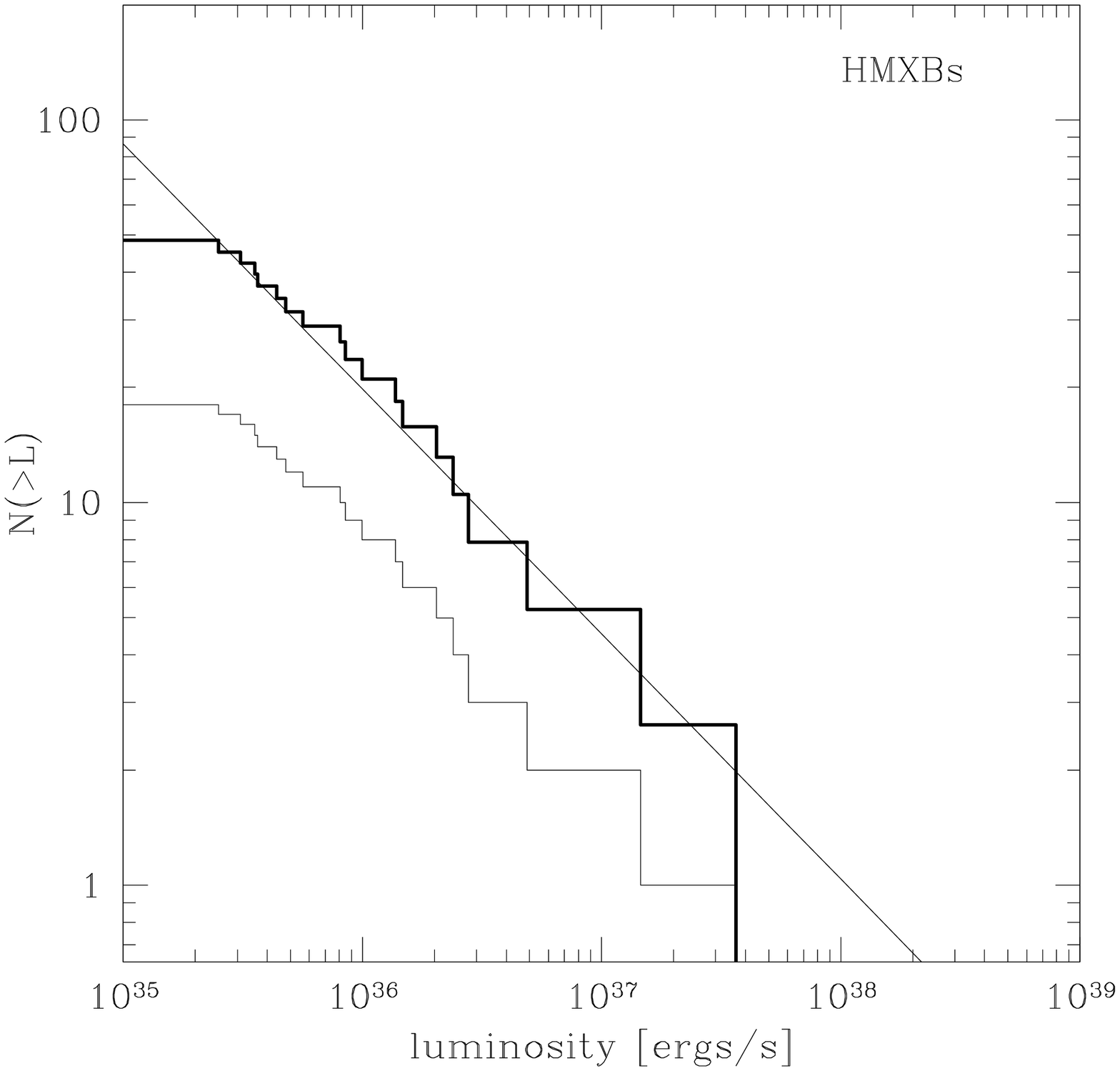}}
  \caption{The apparent (thin histogram) and volume corrected (thick
  histogram) cumulative luminosity function for LMXBs and HMXBs. The
  solid lines are the best fits to the data.} 
  \label{fig:cor_lumi}
\end{figure*}

\begin{figure*}[tb]
  \resizebox{0.5\hsize}{!}{\includegraphics{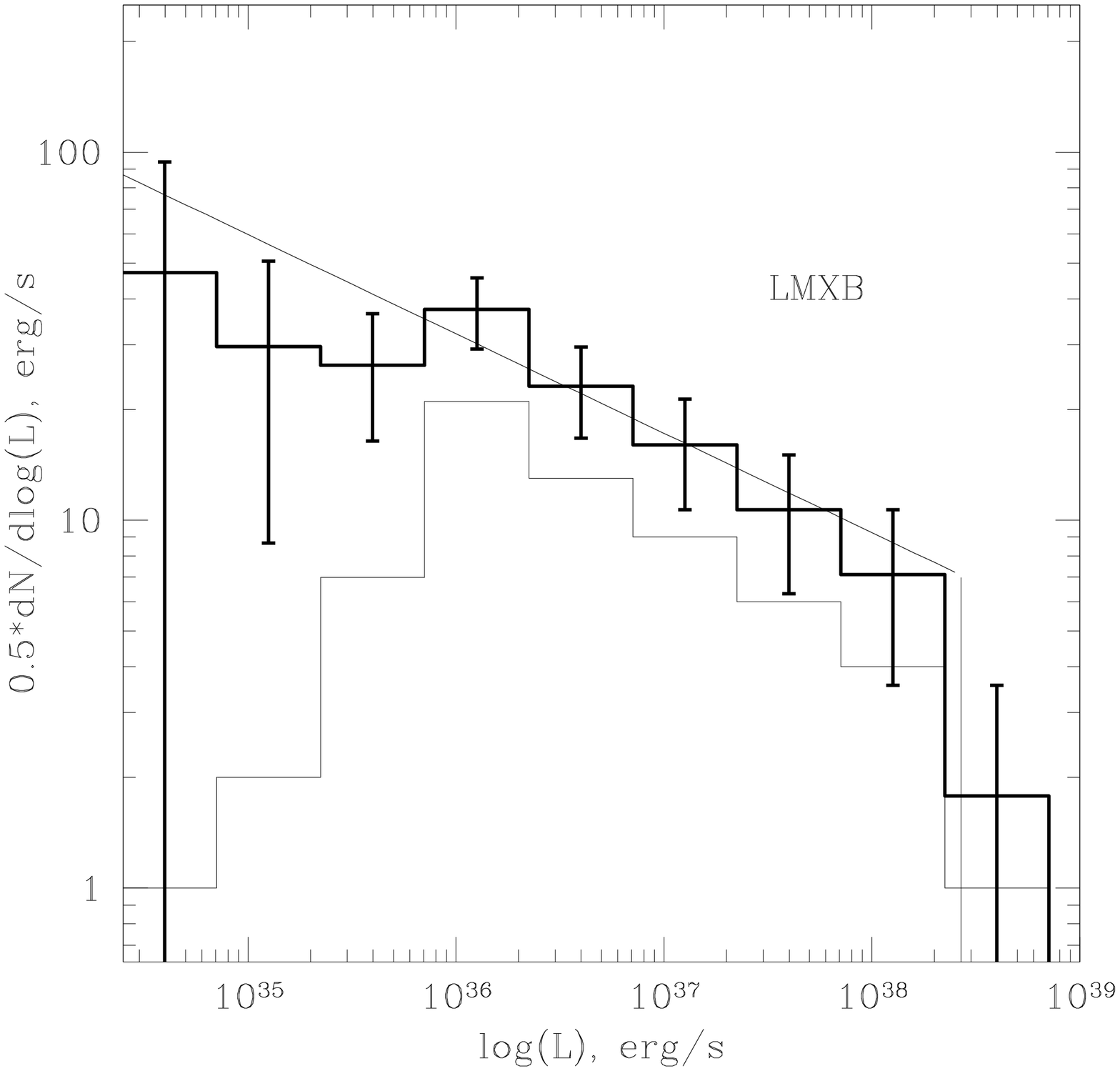}}
  \resizebox{0.5\hsize}{!}{\includegraphics{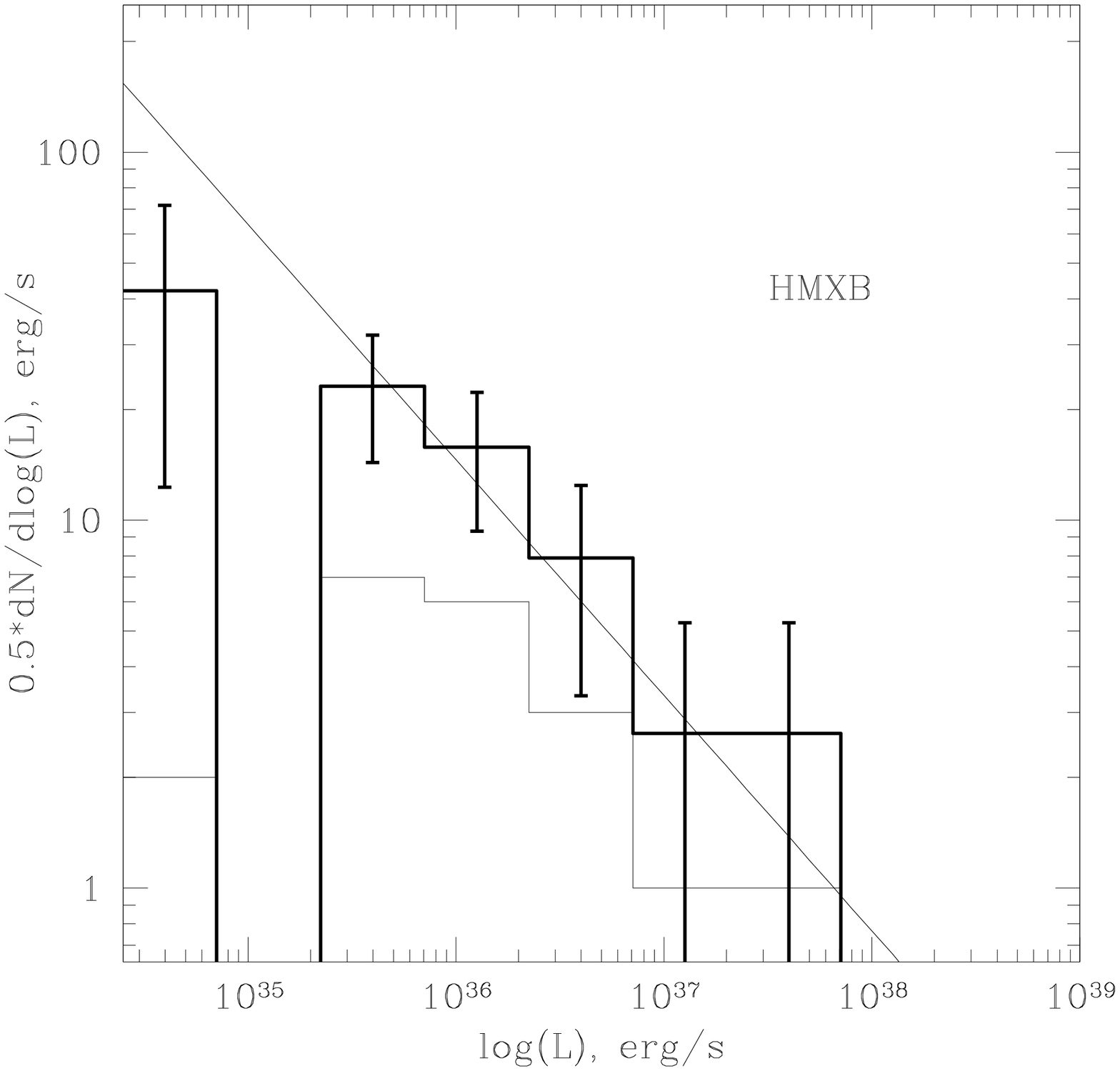}}
  \caption{The apparent (thin histogram) and volume corrected (thick
  histogram) differential luminosity function for LMXBs and HMXBs
  binned into bins with logarithmic with of 0.5. The
  solid lines are the best fits to the cumulative distributions. The
  fall-over of the apparent distributions below $\sim 10^{36}$ erg
  s$^{-1}$ are due to the flux limited nature of the ASM sample (see
  Fig. \ref{fig:mass_frac})}  
  \label{fig:cor_lumi_dif}
\end{figure*}

Due to the flux limited nature of the ASM sample and incompleteness of
the optical identifications/distance measurements beyond $\sim 10$
kpc, the {\em apparent} luminosity function which can be derived
straightforwardly  from the ASM flux measurements and the source  
distances (thin line histograms in Fig. \ref{fig:cor_lumi} and
\ref{fig:cor_lumi_dif}) needs to be corrected for the fraction of the
Galaxy observable by ASM. This correction can be performed using the
model of the spatial distribution of X-ray binaries constructed in
the previous section:
\begin{equation}
  \frac{dN}{dL} = \left( \frac{dN}{dL} \right )_{\text{obs}}\times 
\frac{M(<D(L))}{M_{tot}}
\label{eq:lumf_diff}
\end{equation}
where $\frac{dN}{dL}$ is the true luminosity function, 
$\left( \frac{dN}{dL} \right )_{\text{obs}}$ -- apparent luminosity
function constructed using ASM flux measurements and the source
distances, $M(<D)$ -- mass of the Galaxy inside distance $D$ from the
Sun computed using the volume density distributions for HMXB and LMXB
sources from the Sect. \ref{sec:spat}, $M_{tot}$ -- total mass of
the Galaxy, $D(L)$ is defined by:
\begin{equation}
  D(L)={\rm min} \left( \frac{L}{\sqrt{4\pi F_{lim}}},~ D_{max} \right)
\label{eq:lumf}
\end{equation}
where $F_{lim}$ is the limiting (minimum) flux and $D_{max}$ -- the
maximum distance from the Sun of the sources used for constructing the
luminosity function. As discussed in the previous sections we accepted
the following selection criteria: $F_{lim}=0.2$ cnts s$^{-1}$ $\approx
6.4 \cdot 10^{-11}$ erg s$^{-1}$ cm$^{-2}$, i.e. equal to the
completeness flux limit of the ASM catalogue, and $D_{max}=10$ kpc --
a completeness limit of distance measurements estimated in
Sect. \ref{sec:dist_distr}.

Obviously, for a given flux limit $F_{lim}$ the mass fraction of the
Galaxy $\frac{M(<D(L))}{M_{tot}}$ is a decreasing function of the
source luminosity as shown in Fig. \ref{fig:mass_frac}. For the ASM
sensitivity/completeness limit of $\approx 6.4 \cdot 10^{-11}$
erg s$^{-1}$ cm$^{-2}$ the entire volume inside $D_{max}=10$ kpc from
the Sun is observable down to a luminosity of $\approx 10^{36}$ erg
s$^{-1}$ (the flat part of the curves in Fig. \ref{fig:mass_frac})
below which the mass fraction of the observable part of the Galaxy
begins to decrease. As the spatial distributions of HMXB and LMXB
sources differ significantly, the volume correction and the luminosity
function were calculated separately for HMXBs and LMXBs.
The apparent and volume corrected (true) cumulative luminosity
functions are presented in Fig. \ref{fig:cor_lumi}.
Fig. \ref{fig:cor_lumi_dif} shows the corresponding differential
distributions binned logarithmically over luminosity.

The cumulative luminosity function of HMXBs (Fig. \ref{fig:cor_lumi},
right panel) does not seem to contradict to a power law distribution
down to a luminosity of $\sim 2\cdot 10^{35}$ erg s$^{-1}$ with some 
indication of flattening at lower luminosity. However, limited
sensitivity of ASM and correspondingly large values of the 
correction factor (Fig. \ref{fig:mass_frac}) at low luminosities do
not allow one to draw a definite conclusion regarding the shape of the
luminosity function at these low luminosities (see comparison with
ASCA source counts in Sec. \ref{sec:lowlum}).
We therefore fitted the luminosity function of HMXBs in the $L>2\cdot
10^{35}$ erg s$^{-1}$ range with a power law distribution. Using a
Maximum-Likelihood method the best fit  parameters are:
\begin{equation}
  N(>L) = 20 \cdot (\frac{L}{10^{36} \text{erg s$^{-1}$}})^{-0.64 \pm
  0.15}
\label{eq:lumf_hmxb}
\end{equation}
where $L$ is the source luminosity in erg s$^{-1}$ and $N(>L)$ --
total number of sources on the sky with luminosity greater than $L$.

The shape of both cumulative and differential luminosity function for
LMXBs (Figs. \ref{fig:cor_lumi}, \ref{fig:cor_lumi_dif}, left panels)
indicates the presence of a high luminosity cut-off. We fitted the
unbinned cumulative distribution with the functional form
\begin{equation}
  N(>L) = A \cdot (L^{-\alpha} - L_{max}^{-\alpha}).
\label{eq:lumf_lmxb_1}
\end{equation}
corresponding to a power law differential luminosity function with a
sharp cut-off at $L_{max}$.  The value of the cutoff was set equal to 
to $2.7 \cdot 10^{38}$ erg s$^{-1}$ which corresponds to the
luminosity of the most luminous source within 10 kpc, \object{Sco
X-1}. The best fit values of other parameters are:
\begin{equation}
  N(>L) = 105 \cdot ((\frac{L}{10^{36} \text{erg s$^{-1}$}})^{-0.26
  \pm 0.08} - 270^{-0.26}).
\label{eq:lumf_lmxb}
\end{equation}

Note that the smaller number of sources and the steeper slope of
luminosity function make the HMXB data insensitive to a high
luminosity cut-off above $\sim {\rm few}\times 10^{36}$ erg s$^{-1}$.

\subsection{Effect of the Galaxy model on the luminosity function}

From Eq. (\ref{eq:lumf_diff}) it is clear that the luminosity
function depends on the spatial distribution of XRBs in the Galaxy. As
discussed above, using the distance measurements available, we were
able to determine some of the parameters of their distribution. But
the data are not sufficient to determine the entire distribution
unambiguously. Thus we had to assume a spatial distribution of XRBs
in the Galaxy. In order to investigate the effect of the adopted
spatial distribution of X-ray sources on the derived luminosity
function we varied our model and computed the respective luminosity
functions.

\begin{figure*}[tb]
  \resizebox{0.5\hsize}{!}{\includegraphics{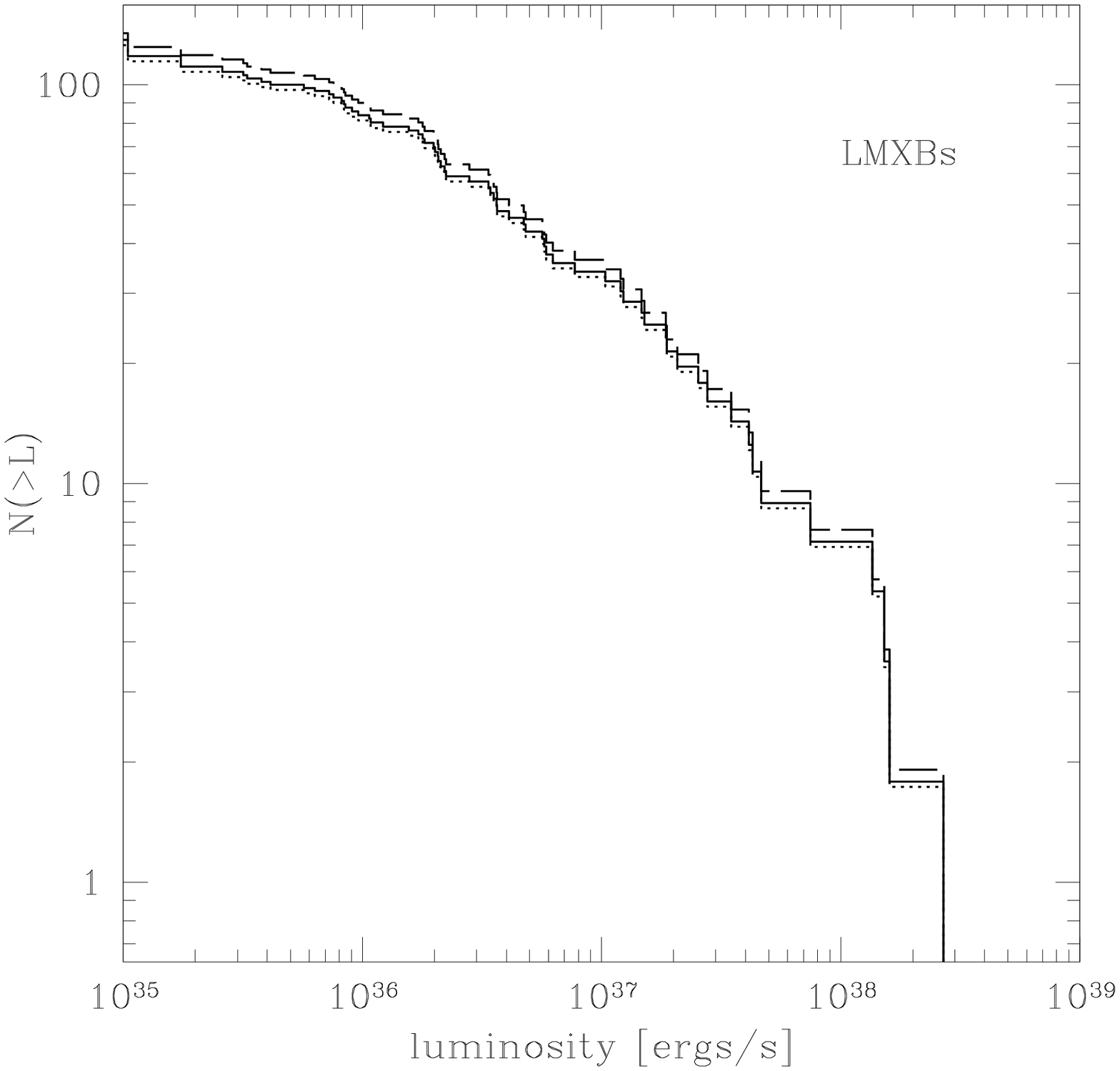}}
  \resizebox{0.5\hsize}{!}{\includegraphics{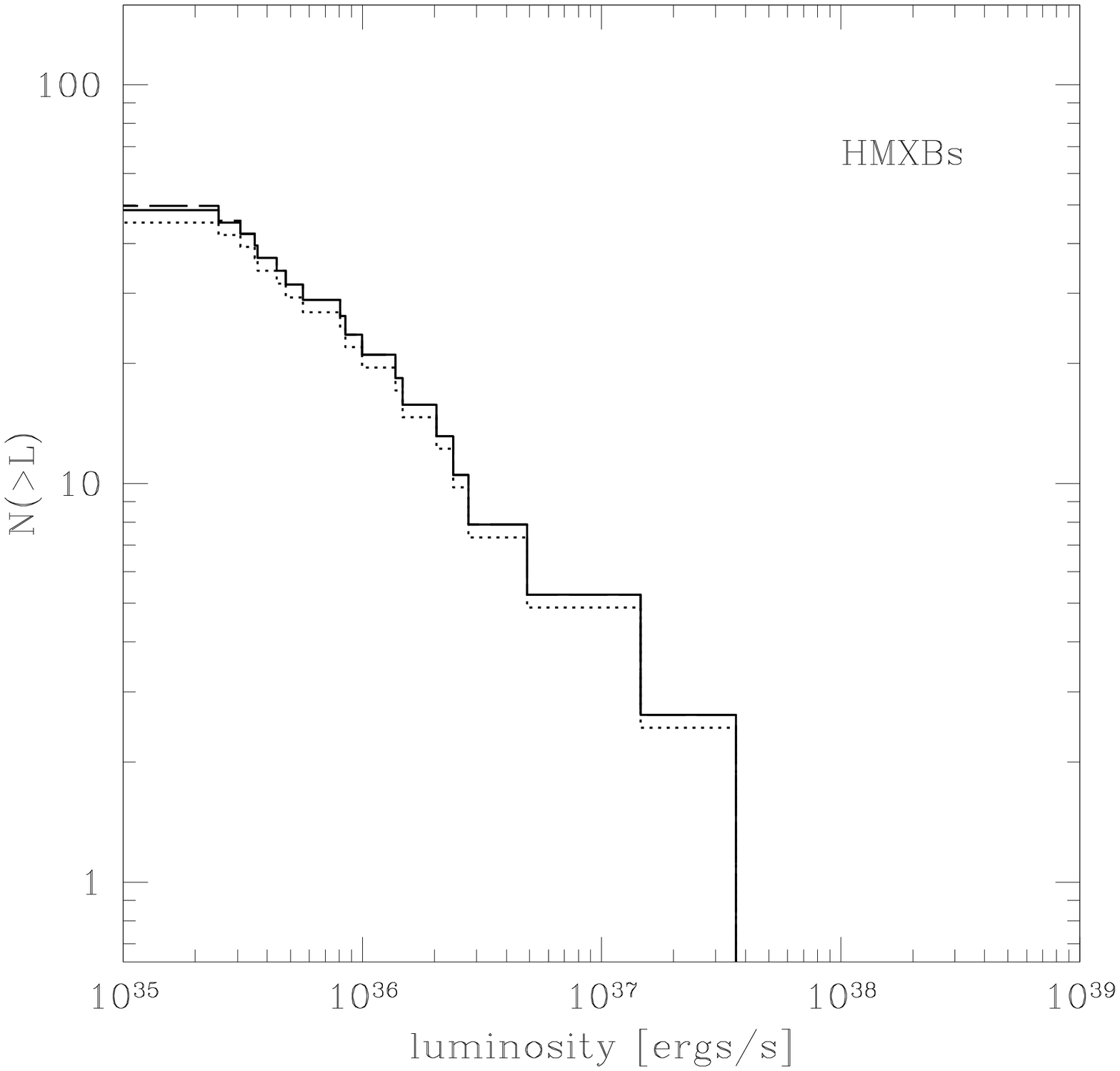}}
  \caption{Dependence of the luminosity function on the adopted
  model of the spatial distribution of XRBs. The figures show the
  luminosity functions of LMXBs (left panel) and HMXBs (right panel)
  for three different Galaxy models. The solid, dotted and dashed
  lines in both panels correspond to the models A, B and C.}
  \label{fig:model_comp}
\end{figure*}

For our analysis we used three different distributions for LMXBs and
HMXBs. In the case of HMXBs, only the disk component was included in
each of the three distributions. The modulation of the disk
distribution by the spiral pattern, when present, was 100\% for HMXB
and 20\% for LMXB. The models are:
\begin{itemize}
  \item Model A: Our primary model constructed in Sec.
    \ref{sec:spat} and used to derive the luminosity function
    above (shown as a solid histogram in Fig. \ref{fig:model_comp}).\\
  \item Model B: The same as the model A, except that the inner
    cut-off of the disk was set to $r_m=4$ kpc in accordance with the
    result of \citet{dehnen:98}(dotted histogram in
    Fig. \ref{fig:model_comp}).\\
  \item Model C: The spheroid component is the same as in Model A. The
    disk radial distribution is without the inner cut-off, i.e. $r_m=0$
    and without modulation by the spiral structure. No bulge component
    is included for either LMXBs or HMXBs. The resulting density
    distribution is similar to that derived by \citet{vanparadijs:95}
    for NS LMXBs (dashed histogram in Fig. \ref{fig:model_comp}).\\
\end{itemize}

The resulting luminosity function for each of the three models are
shown in Fig. \ref{fig:model_comp}. It is obvious that there is no
strong dependence of the luminosity function on the mass
distribution. The slopes vary in the range from $1.28$ -- $1.30$ for
LMXBs and $1.64$ -- $1.72$ for HMXBs. The total number of sources
varies from 88 to 90 for LMXBs and  from 21 to 26 for HMXBs. It is
worth noting that the spiral pattern is no significant factor in the
determination of the luminosity function of HMXBs although the spatial
distribution shows clear signs of them.

\subsection{Total X-ray luminosity of Galactic X-ray binaries}

The total luminosity of all X-ray binaries in the Galaxy is calculated
in the following way. Down to a luminosity of $10^{36}$ erg s$^{-1}$
we sum the measured luminosities of the individual sources to obtain a
more precise number. For the lower luminosities that contribute only a
small fraction to the total luminosity we use the analytical
description of the luminosity function given by
Eqs. (\ref{eq:lumf_hmxb}) and (\ref{eq:lumf_lmxb}).

\begin{table}[tb]
\caption{List of the most luminous LMXB sources contributing
  $\approx 90\%$ to the integrated luminosity of LMXBs in the 2--10
  keV band, averaged over 1996--2000. The 12 most luminous sources
  contribute $\approx 80\%$ of the integrated luminosity.}
  \begin{tabularx}{\linewidth}{|l|ccc|c|l|}
  \hline
  Source & \multicolumn{3}{c|}{$L_X$ [$10^{38}$erg s$^{-1}$]}& dist.& Ref.\\
         & avg. & min.$^{(a)}$ & max.$^{(a)}$ &  [kpc] &\\
  \hline
  \object{Cir X-1}      & 4.4 & 0.3  & 10  & 10.9 & 1\\
  \object{GRS 1915+105} & 3.7 & 1    & 11  & 12.5 & 2\\
  \object{Sco X-1}      & 2.7 & 2    & 4.5 & 2.8  & 3\\
  \object{Cyg X-2}      & 1.8 & 0.9  & 3.4 & 11.3 & 1,4--7\\
  \object{GX 349+2}     & 1.6 & 1.1  & 2.7 & 9.2  & 1,7,8\\
  \object{GX 17+2}      & 1.5 & 1.1  & 2.4 & 9.5  & 1,7,9,\\
&&&&&10\\
  \object{GX 5-1}       & 1.4 & 1      & 1.8 & 7.2  & 1,7\\
  \object{GX 340+0}     & 1.3 & 0.9    & 1.8 & 11.0 & 1,7\\
  \object{GX 9+1}       & 0.75 &0.5    & 1.0 & 7.2  & 9\\
  \object{NGC 6624}     & 0.47 &0.15   & 0.8 & 8.0  & 10,11\\
  \object{Ser X-1}      & 0.43 &0.26   & 0.6 & 8.4  & 12\\
  \object{GX 13+1}      & 0.41 & 0.25  & 0.6 & 7.0  & 13\\
  \object{X 1735-444}   & 0.35 & 0.2   & 0.6 & 9.2  & 1\\
  \object{XTE J1550-564}& 0.35 &0.005  & 2.1 & 5.3  & 14\\
  \object{KS 1731-260}  & 0.28 & 0.06  & 0.6 & 8.5  & 15--17\\
  \object{X 1705-440}   & 0.25 &$<$0.04& 0.6 & 7.4  & 18\\
  \object{X 1624-490}   & 0.24 &$<$0.13& 0.4 & 13.5 & 9\\
  \hline
  \end{tabularx}
  $^{(a)}$ min. and max. luminosity were estimated by eye from the 1
  day averaged light curves.\\
  References for the distances: (1) -- \citet{vanparadijs:95},
(2) -- \citet{mirabel:94}, (3) -- \citet{bradshaw:99},  
(4) -- \citet{orosz:99}, (5) -- \citet{cowley:79}, (6) --
\citet{smale:98}, (7) -- \citet{penninx:89}, (8) --
\citet{wachter:96}, (9) -- \citet{christian:97},
(10) -- \citet{djorgovski:93}, (11) -- \citet{webbink:85},
(12) -- \citet{ebisuzaki:84}, (13) -- \citet{bandyopadhyay:99},
(14) -- \citet{orosz:01},
(15) -- \citet{barret:98}, (16) -- \citet{smith:97}, (17) --
\citet{sunyaev:90},  (18) -- \citet{haberl:95}
\label{tab:bright_l}
\end{table}

\begin{table}[tb]
\caption{List of the most luminous HMXB sources that contribute
  $\approx 40\%$ to the integrated luminosity of HMXBs in the 2--10
  keV band, averaged over 1996--2000.}
\begin{tabularx}{1.01\linewidth}{|l|ccc|c|l|} 
  \hline
  Source & \multicolumn{3}{c|}{$L_X$ [$10^{38}$erg s$^{-1}$]}& dist.& Ref.\\
         & avg. & min.$^{(a)}$ & max.$^{(a)}$ &  [kpc] &\\
  \hline
  \object{Cyg X-3}   & 0.5   & 0.08   & 1.4  & 9.0  &1\\
  \object{Cen X-3}   & 0.15  &$<$0.03 & 0.7  & 9.0  &2--5\\
  \object{Cyg X-1}   & 0.05  &  0.02  & 0.17 & 2.1  &6\\
  \object{X 1657-415}& 0.043 &$<$0.02 & 0.22 & 11.0 &7\\
  \object{V4641 Sgr} & 0.028 &$<$0.02 & 7.3  & 9.9  &8\\
  \hline
  \object{GX 301-2}      & 0.02  &$<0.005$& 0.4 & 5.3  &9\\
  \object{XTE J1855-024}& 0.015 &$<$0.01& 0.11 & 10.0 &10\\
  \object{X1538-522}    & 0.014 &$<0.008$& 0.08 & 6.4  &11\\
  \object{GS1843+009}   & 0.01  &$<0.007$& 0.11 & 10.0 &12\\
  \object{X1908+075}    &0.008&$<0.006$& 0.05 & 6.4  &13,\\
                        &     &        &      &      &14\\
  \hline
  \end{tabularx}
  $^{(a)}$ min. and max. luminosity were estimated by eye from the 1
  day averaged light curves.\\
References for the distances: (1) -- \citet{predehl:00}, (2) --
\citet{krzeminski:74}, (3) -- \citet{hutchings:79}, (4) --
\citet{motch:97},(5) -- \citet{bahcall:78}, (6) -- \citet{massey:95},
(7) -- \citet{chakrabarty:93}, (8) -- \citet{orosz:00}, (9) --
\citet{kaper:95}, (10) -- \citet{corbet:99}, (11) --
\citet{reynolds:92}, (12) -- \citet{israel:01}, (13) --
\citet{wen:00}, (14) -- \citet{vanparadijs:95} 
\label{tab:bright_h}
\end{table}

The integrated luminosity of HMXBs and LMXBs in the
2--10 keV ASM band calculated in such way are $\approx 2 \cdot 10^{38}$
erg s$^{-1}$ and  $\approx 2.5 \cdot 10^{39}$ ergs s$^{-1}$,
respectively. Note that these numbers refer to the luminosity {\it
averaged} over the period from 1996--2000. The variability of
individual sources or  an outburst of a bright transient can change
the luminosity by a factor of up to $\sim 2-3$. Due to the shallow
slopes of the luminosity functions the integrated X-ray emission of
the Milky Way is dominated by the $\sim 5-10$ most luminous sources
(see Table \ref{tab:bright_l} and \ref{tab:bright_h}). The maximum and
minimum values for the luminosities were estimated by eye from the 1
day averaged light curves. The values in the tables therefore differ
from the values in Table \ref{tab:edd_source}.

Normalised to the star formation rate which is about 4
M$_{\odot}$ yr$^{-1}$ in the Milky Way~\citep{mckee:97} galactic HMXBs
emit about $\sim 5\cdot 10^{37}$ erg s$^{-1}$/(M$_{\odot}$
yr$^{-1}$). The luminosity of LMXBs normalised to the stellar mass is
about $\sim 5 \cdot 10^{28}$erg s$^{-1}$ M$^{-1}_{\odot}$, assuming a
stellar mass of the Galaxy of about $5 \cdot 10^{10}$ M$_{\odot}$.

The contribution of Be X-ray binaries from the ASM sample to the
integrated luminosity of HMXBs is $\sim$ 5\%.

Note that poor knowledge of the shape of the luminosity function at
low luminosities, $L \la 10^{35}$ erg s$^{-1}$ should not influence
the total luminosity considerably unless the luminosity function
steepens significantly at these low luminosities (see
Sec. \ref{sec:lowlum}).

The total number of X-ray binaries above $2 \cdot 10^{35}$ erg
s$^{-1}$ obtained from the luminosity functions is about $\sim 190$ of
which $\sim 55$ are HMXBs and $\sim 135$ -- LMXBs.

\subsection{Luminosity function and $\dot{M}$ distribution of X-ray
binaries} 

The X-ray luminosity function is obviously related to the
distribution of X-ray binary systems over the mass loss rate of
the secondary, $\dot{M}$. The simplest assumption would be that both
distributions  have the same slope in the range corresponding to
luminosities of $\sim (0.01-1) L_{Edd}$. At larger luminosities,
$L\ga L_{Edd}$, the luminosity function has a break or cut-off, well
in accordance with theoretical expectation, that the luminosity due to
accretion cannot exceed the Eddington luminosity of the primary star
by a large factor (see discussion in Sect. \ref{sec:eddington}).
The donor star in a binary system, on the other hand,  ``does not
know'' about the Eddington critical luminosity, therefore the
distribution of binary systems over the mass loss rate of the
secondary, $\dot{M}$, is not expected to break near the Eddington
value for the compact object. Thus the distribution of binary systems
over $\dot{M}$ is expected to continue with the same slope well beyond
the Eddington value.

Extremely super-Eddington values of the mass accretion rate
$\dot{M}$ can result in quenching of the X-ray source and/or its
obscuration by the matter expelled from the system by radiation
pressure \citep{shakura:73}. This would lead to the appearance of a
peculiar object, dim in X-rays and extremely bright in the
optical and UV band -- similar to \object{SS 433} or the recent fast
transient \object{V4641 Sgr} at the peak of its optical outburst.
Such objects would emit only a negligible fraction in the X-ray
band and would contribute to the lower luminosity end of the XRB
luminosity function.

For moderately super-Eddington values of $\dot{M}\la 10-100
\dot{M}_{Edd}$, however, one might expect the appearance of a near- or
slightly super-Eddington source, therefore all such systems are
expected to cluster near $L_{Edd}$. For a given slope of the
luminosity function the number of such sources can be easily
estimated. For the observed parameters of the LMXB luminosity function
(slope $=1.3$, 42 sources with $36.5<\log(L_X)<38$) and assuming that
the $\dot{M}$ distribution continues with the same slope $=1.3$, the
total number of sources with $\dot{M}$ corresponding to the
range of luminosities of $10^{38}-10^{39}$ and $10^{39}-10^{40}$ erg
s$^{-1}$ is $\approx 10$ and $\approx 6$ correspondingly ($\approx 7$
sources are expected to have $\dot{M}$ corresponding to $L>10^{40}$
erg s$^{-1}$). These estimates are in disagreement with the actually
observed number of sources with $L\ga10^{38}$ erg s$^{-1}$, which is
equal to 8. In order to reconcile the expected number of sources near
$L_{Edd}$ with the observations, a slope of the $\dot{M}$ distribution
of $\ga 1.35-1.40$ is required which is somewhat steeper than the
observed value of $\sim 1.3$. We note that the slope of $\sim 1.35$ is
within $\sim 1\sigma$ of the the observed value.

Finally, there are several effects that can suppress the number of the
low luminosity sources, i.e. make the luminosity function flatter than
the $\dot{M}$ distribution. The most obvious and important are
discussed below.
\begin{itemize}
  \item In the case of HMXBs
  the magnetosphere of the strongly magnetised, rapidly rotating
  neutron star can prevent the accretion at low $\dot{M}$ via the
  propeller effect \citep{illarionov:75}.

  \item Be-systems are characterised by regular
  outbursts corresponding to the passage of the neutron star through
  the equatorial stellar wind. Therefore for such sources the true
  value of the $\dot{M}$ in the binary system is measured  by the peak
  luminosity during the outbursts whereas  the long term averaged
  luminosity, used to construct the luminosity function, can give a
  significantly underestimated value.

  \item A common property of LMXBs, containing both neutron stars and
  black hole, is the presence of relativistic jets which might carry
  away a sizable fraction of the energy of accretion
  \citep{mirabel:99}. The presence of jets correlates with the X-ray
  spectral state: the jets are absent (and hence the true accretion
  efficiency is higher) in the soft spectral state corresponding to
  higher values of $\dot{M}$. The jets exist only in the hard spectral
  state \citep{fender:01}, thus decreasing the accretion efficiency
  at lower $\dot{M}$.

  \item In the case of black hole binaries an ADAF can form at low
  accretion rate in which case the accretion efficiency is
  proportional to $\dot{M}$ and the X-ray luminosity scales as $L
  \propto \dot{M}^2$ \citep{narayan:95}.

  \item At sufficiently low accretion rates a source becomes a
  transient with a recurrence time varying from $\sim 1$ to $\ga 50$
  years \citep{white:84}. This would decrease the number of low and
  intermediate luminosity sources in the luminosity function
  constructed on the several years baseline.
\end{itemize}

The number of luminous X-ray binaries in the Milky Way is
insufficient to study the shape of the luminosity function near
$L_{Edd}$ in detail. On the other hand within next several years
CHANDRA X-ray observatory will study compact sources in a large
number of nearby, $d\la 50$ Mpc galaxies and the total number of the
X-ray binaries detected in other galaxies can easily reach several
hundred or thousand. In this context it might be interesting to
construct a combined luminosity function of X-ray binaries in our
and other galaxies to study its exact shape at the high luminosity
end and search for a possible excess of sources near $L_{Edd}$.

\begin{figure}[tb]
  \resizebox{\hsize}{!}{\includegraphics{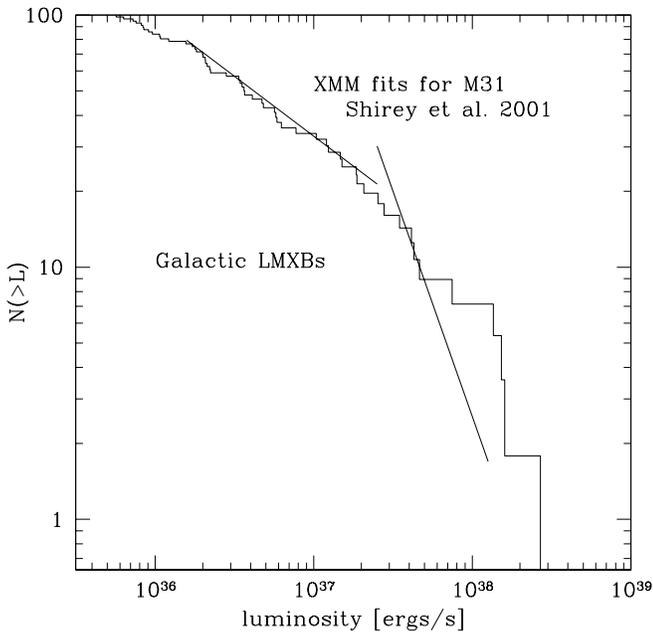}}
  \caption{Cumulative luminosity function of Galactic LMXBs and also
  the best fit values for the XMM-Newton observation of M 31 by
  \citet{shirey:00}}
  \label{fig:m31}
\end{figure}

\begin{figure*}[bt]
  \resizebox{0.5\hsize}{!}{\includegraphics{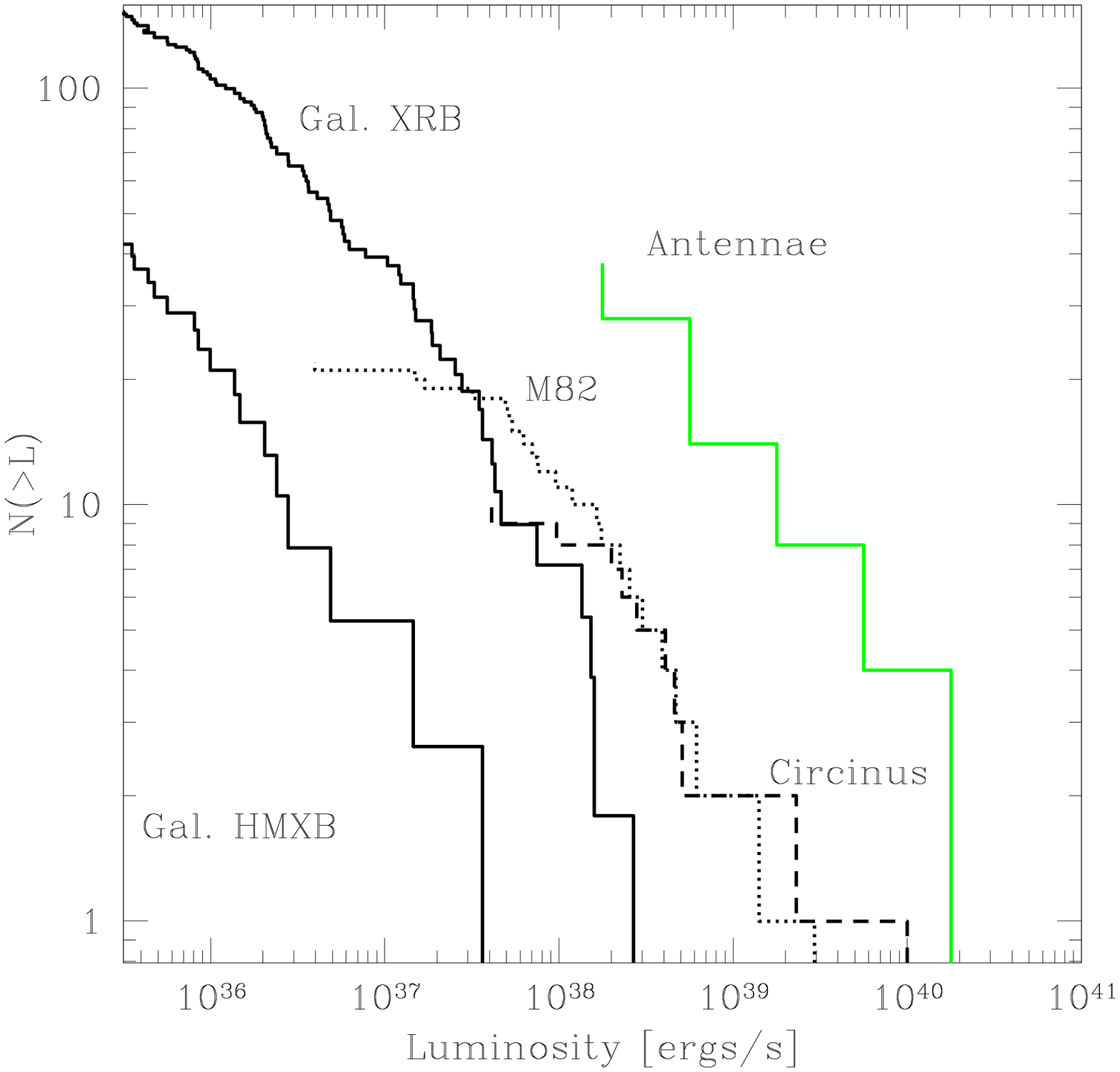}}
  \resizebox{0.5\hsize}{!}{\includegraphics{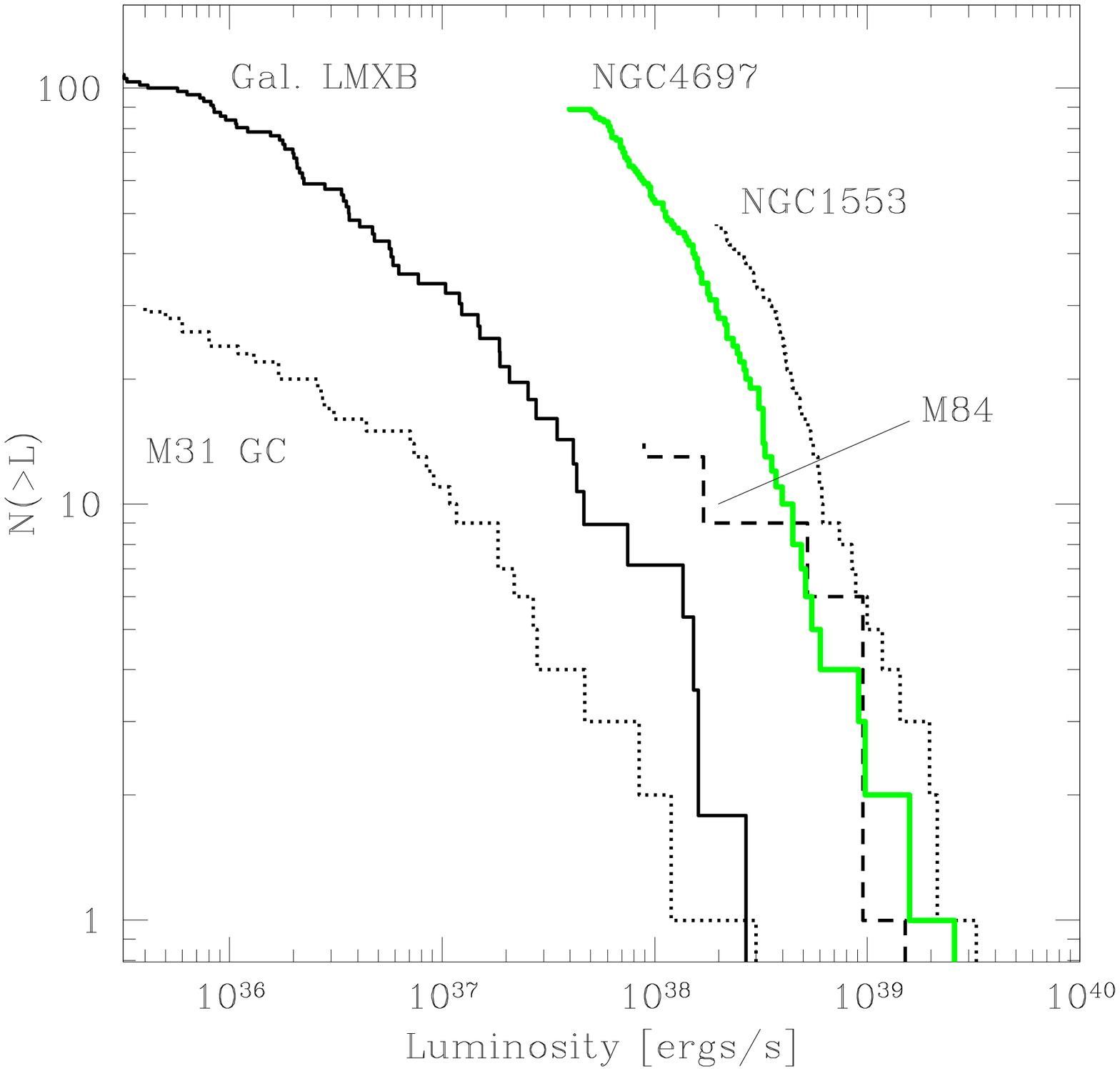}}
  \caption{Cumulative luminosity functions of galaxies observed with
  CHANDRA. The left panel shows actively star forming spiral galaxies
  that include \object{NGC 4038/39} and \object{M 82} which are
  supposed to be dominated by HMXBs. For comparison the luminosity
  functions of Galactic X-ray binaries and HMXBs alone are shown. The
  right panel shows elliptical galaxies including the SO galaxy
  \object{NGC 1553}. For comparison the luminosity function of
  Galactic LMXBs is shown.}
  \label{fig:galaxies}
\end{figure*}

%
%
\section{Comparison with nearby galaxies}

The total luminosity of X-ray binaries in the Milky Way, 
$\sim 2-3 \cdot 10^{39}$ erg s$^{-1}$ in the 2--10 keV band, agrees
sufficiently well with observations of \object{M 31}, for which GINGA
has found a luminosity of $5 \cdot 10^{39}$ ergs s$^{-1}$ between 2-20
keV \citep{makishima:89}.

Recently XMM-Newton observed the inner 30$^{\prime}$ region of
\object{M 31} \citep{shirey:00}. In total 116 sources were detected
above the limiting luminosity of $6 \cdot 10^{35}$ erg s$^{-1}$ in the
0.3--12 keV energy range, assuming a distance of 760 kpc. 
\citet{shirey:00} distinguish between two luminosity ranges, $36.2 <
\log(L_X) < 37.4$, for which the best fit slope is $-0.47 \pm 0.03$,
and $37.4 < \log(L_X) < 38.1$ where the best fit slope is $-1.79 \pm
0.26$. At the distance of 760 kpc 30$^{\prime}$ correspond to $\approx
6.6$ kpc therefore these data should be compared with the luminosity
function of Galactic LMXBs, assuming that similarly to the Milky Way
the inner part of \object{M31} is populated mainly with LMXBs. The two
luminosity functions are plotted in Fig. \ref{fig:m31}. Although the
general shapes of the luminosity functions of LMXBs in the Milky Way
and in \object{M 31} are similar, it is obvious that one can not be
obtained from the other by a shift along the vertical axis as one
would expect if the luminosity function was simply proportional to the
mass of the host galaxy.

CHANDRA observations have produced luminosity functions of compact
sources in a number of nearby galaxies, including ellipticals:
\object{NGC 4697} \citep{sarazin:00}, \object{M 84}
\citep{finoguenov:01} and \object{NGC 1553} \citep{blanton:01}, 
spirals: \object{M 81} \citep{tennant:01}, \object{Circinus}
\citep{smith:01},  \object{M31} \citep{garcia:00} 
and starburst galaxies: \object{NGC 4038/39} (Antennae)
\citep{fabbiano:01} and \object{M 82} \citep{griffiths:00}.
The luminosity functions of the compact sources in these galaxies are
compared to that of the Milky Way in Fig. \ref{fig:galaxies}. The left
panel in Fig. \ref{fig:galaxies} shows spirals and starbursts which
are expected to have a higher fraction of HMXBs due to higher star
formation rates. These are compared with the luminosity functions of
HMXBs and all X-ray binaries in the Milky Way. The right panel in
Fig. \ref{fig:galaxies} shows elliptical galaxies along with the
luminosity function of Galactic LMXBs. 

As the example of our Milky Way shows, X-ray binaries in globular
clusters play an important role in determination and understanding the
properties of the population. It is also well known that globular
cluster systems are quite different for early- and late-type galaxies,
in terms of number per galaxy luminosity \citep{harris:79} as well as
depend on the environment of the host galaxy \citep{bridges:90}. Taken
together this shows the need for a closer study of X-ray binaries in
globular clusters -- ideally they should be treated separately, when
studying the luminosity function of LMXB sources. Unfortunately only
for few galaxies there are observations which allow the separation of
globular cluster X-ray sources, e.g. \object{M 31}
\citep{distefano:02} and \object{NGC 1399} \citep{angelini:01}.  We
therefore decided to ignore in the present study the possible effects
of the globular cluster sources on the overall luminosity function.

Comparing the HMXB luminosity function in our and nearby star forming
galaxies we could check the proportionality of the HMXB luminosity to
star forming rate. There might be several additional factors involved
including chemical abundance of the particular galaxy. For example,
the HMXB sources in LMC and SMC appear to be significantly more
luminous than the HMXB sources in our Galaxy, even though the star
formation rates are comparable. Especially interesting is the
case of the Antennae galaxies where the difference from the Galactic
HMXB luminosity function is extremely impressive. It seems that it can
not be explained simply by the difference in the star formation rate,
which is about 20 times higher \citep{neff:00} whereas the number of
X-ray sources is a factor of more than 50 higher. This example shows
that the knowledge of the HMXB luminosity function seems to be
insufficient to measure the star formation rate in galaxies and to
estimate the distances to them with acceptable precision.

CHANDRA observations are also opening an important possibility to
check the proportionality of LMXB luminosity functions to the mass of
the parent galaxies.

%
%
\section{Low luminosity sources}
\label{sec:lowlum}
\subsection{Extension of \lns~towards lower fluxes}
Since the sensitivity of ASM is limited to relatively high flux
sources it is interesting to investigate the behaviour of the
\lns~at lower fluxes. Note that, given the slope observed by ASM (1.2
and 1.61 for LMXBs and HMXBs), the \lns~ distribution should flatten
at low fluxes since the total number of sources in the Galaxy is
finite.

\begin{figure}[b]
  \resizebox{\hsize}{!}{\includegraphics{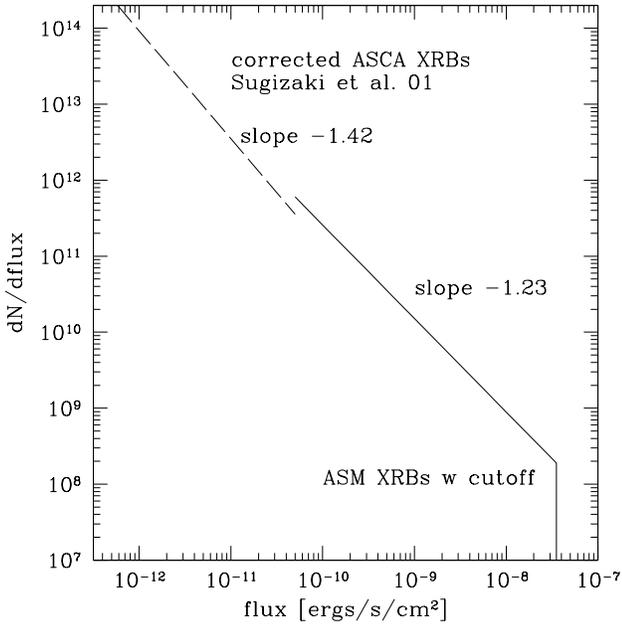}}
  \caption{Comparison of the differential \lns~relation for Galactic
  X-ray binaries obtained by ASM (solid line with break) and by ASCA
  Galactic Ridge Survey (dashed line). The ASCA number--flux relation
  was multiplied by an approximate correction factor accounting for
  the difference in the sky coverage of the ASM and ASCA surveys (see
  text for details).} 
  \label{fig:asca_comp}
\end{figure}

In order to study the low flux regime below the ASM completeness limit
of $\approx 6.4 \cdot 10^{-11}$ erg s$^{-1}$ cm$^{-2}$, we use ASCA
data from the Galactic Ridge Survey \citep{sugizaki:01} covering
$\approx 40$ square degrees with the limiting sensitivity of $\sim 3
\cdot 10^{-13}$ erg s$^{-1}$ cm$^{-2}$. Since most of the sources in
the ASCA survey are unidentified we followed the criterion suggested
by \citet{sugizaki:01} in order to discriminate X-ray binary
candidates from other sources: that X-ray binary candidates have
either a spectral photon index $\Gamma < 1$, or a spectral photon
index $\Gamma < 3$ and a column density $N_H < 0.8 \cdot 10^{22}$
cm$^{-2}$. Excluding otherwise identified sources with these spectral
properties there remain 28 sources. We fit the \lns~of the selected
sources with the procedure similar to that used for ASM sources,
modified to account for the flux dependent sky coverage of the ASCA
survey (Fig. 7 in \citet{sugizaki:01}). The resulting \lns~is:
\begin{eqnarray}
  N(>S) = 9.4 \cdot 10^{-5} \cdot S^{-0.42 \pm 0.08}
\end{eqnarray}
where $S$ is flux in units of $10^{-12}$ erg s$^{-1}$ cm$^{-2}$.
To compare ASCA data with an extrapolation of the ASM number--flux
relation one needs to account for the difference in their sky coverage
($|l|\la 40^{\circ}$ and $|b|\la 0.3^{\circ}$ for ASCA survey and
entire sky for the ASM data). An approximate value of the correction
factor can be estimated as the fraction of the mass of the Milky Way
covered by the ASCA Galactic  Ridge Survey with account for its
sensitivity and the particular pattern of its sky coverage (Fig. 1 in
\citet{sugizaki:01}). The mass fraction was calculated using the
Galaxy model described in Sect. \ref{sec:spat} and equals to $\sim$
1:21. Converting the cumulative \lns~to differential \lns~for ASCA
X-ray binary candidates and all ASM X-ray binaries and multiplying the
resulting ASCA \lns~by 21 we obtain the result shown in
Fig. \ref{fig:asca_comp}.

It is obvious that the agreement between ASM and
ASCA data is sufficiently good. The slopes are different at the 
$\sim 2 \sigma$ level. On the other hand since the sources are all
unidentified and their distances unknown it is not possible to
distinguish between high and low mass X-ray binaries which have
different slopes of their \lns~distributions in the ASM
sample. Indeed, due to the small range in Galactic latitude $b^{II}$
covered by the ASCA survey and due to the fact that HMXBs have a 3
times smaller vertical scale height (cf. Sec. \ref{sec:spat}), the
ratio of HMXBs to LMXBs should be different for the ASCA and ASM
samples. The fraction of HMXBs, having steeper \lns, should be larger
in the ASCA sample and thus the resulting \lns~should be somewhat
steeper. We conclude that the data of the ASCA Galactic Ridge Survey
indicate that there are no significant deviations in the \lns~from the
extrapolations of the ASM data down to the sensitivity limit of the
ASCA survey of $\sim 5 \cdot 10^{-13}$ erg s$^{-1}$ cm$^{-2}$.

\subsection{Low luminosity end of X-ray binary luminosity function}

Knowledge of the \lns~observed by ASCA and the spatial distribution of
sources in the Galaxy gives a possibility to constrain the low
luminosity end of the luminosity function. If the luminosity function
observed with ASM continues to lower luminosities then it should be
possible to reproduce the \lns~observed by ASCA according to the
formula
\begin{equation}
  N(>S) = \int_{L_{min}}^{L_{max}}\frac{dN}{dL} \cdot
  \frac{M(<r)_{\text{ASCA}}}{M_{total}}dL,
\label{eq:art_lns}
\end{equation}
with
\begin{equation}
  r = \sqrt{\frac{L}{4\pi \cdot S}}.
\end{equation}
where $N(>S)$ is the number of sources with a flux higher than $S$
observed by ASCA, $\frac{dN}{dL}$ is the differential luminosity
function, and $\frac{M(<r)_{\text{ASCA}}}{M_{total}}$ is the fraction
of mass within a radius $r$ from the Earth within the field of view of
the ASCA survey, $L_{max}$ is the high luminosity cut-off of the
luminosity function (Eqs. (\ref{eq:lumf_hmxb}) and
(\ref{eq:lumf_lmxb})). The $L_{min}$ is the low luminosity cut-off of
the luminosity function below which it is assumed to be equal to
zero. This quantity characterises roughly the luminosity level at
which the luminosity function deviates significantly from the
extrapolation of the ASM power law.

\begin{figure}[t]
  \resizebox{\hsize}{!}{\includegraphics{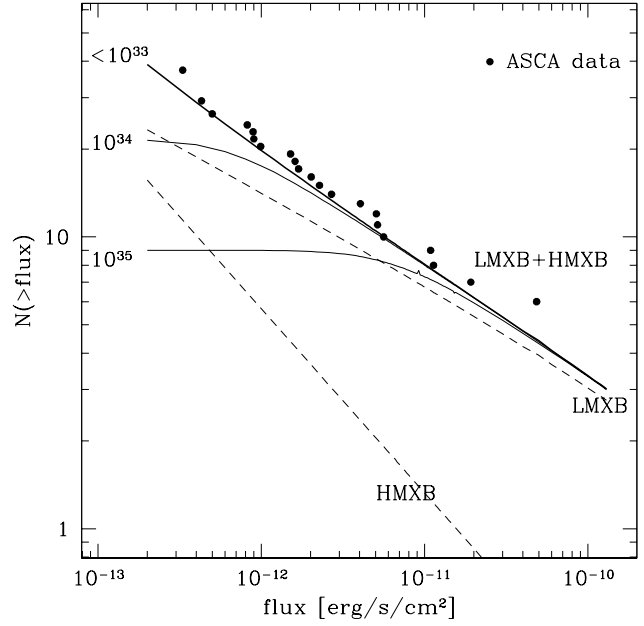}}
  \caption{Comparison of the number-flux relation observed in the ASCA
  Galactic Ridge Survey (points) and the predicted number--flux
  relation based on the extrapolation of the ASM luminosity function
  to low luminosities (lines). The vertical axis shows the number of
  sources in the entire field of the ASCA survey. The ASCA number-flux
  relation was corrected for the flux dependent sky coverage (Fig. 7
  in \citet{sugizaki:01}). The predicted number--flux relations were
  computed according to Eq. (\ref{eq:art_lns}) using the extrapolation
  of the ASM luminosity functions and the volume density distributions
  of X-ray binaries described in Sect. \ref{sec:spat}. The thick
  solid lines show the combined \lns~ of LMXBs and HMXBs for different
  values of the low luminosity cut-off. The thin dashed lines show
  the contributions of LMXBs and HMXBs separately for the case without
  cut-off.}
\label{fig:art_lnls}
\end{figure}

The predicted \lns~calculated from Eq. (\ref{eq:art_lns}) is compared
with the \lns~of X-ray binary candidates from the ASCA survey in
Fig. \ref{fig:art_lnls}. In plotting the ASCA data (solid circles) we
added five bright sources located in the ASCA field of view that were
excluded from the final catalogue in \citet{sugizaki:01} and corrected
for the flux dependent sky coverage of the ASCA survey (Fig. 7 in
\citet{sugizaki:01}). The predicted \lns~ was calculated according to
Eq. (\ref{eq:art_lns}) separately for HMXB and LMXB using the
extrapolation of the respective ASM luminosity functions. The mass
integral $M(<r)$ in Eq. (\ref{eq:art_lns}) was calculated taking
approximately  into account the actual pattern of ASCA pointings and
using the volume density distributions constructed in
Sect. \ref{sec:spat}. The predicted combined \lns~ of HMXB and LMXB
sources is shown in Fig. \ref{fig:art_lnls} by the thick solid lines
for different values of the low luminosity cut-off $L_{min}$. The thin
solid and dashed lines show the contributions of HMXBs and LMXBs
respectively for the case without low luminosity cut-off.

It is clear from Fig. \ref{fig:art_lnls} that the predicted
number--flux relation of X-ray binaries agrees with the ASCA data
very well. Given the volume density distributions of X-ray binaries
in the Galaxy, the low flux end of the ASCA \lns~is sensitive to
sources with luminosities of $\sim 10^{34}$ erg s$^{-1}$. The good
agreement with the predicted \lns~distribution implies that the data
do not require a low luminosity cut-off of the luminosity function
down to $\sim 10^{34}$ erg s$^{-1}$.

\begin{figure*}[htb]
  \resizebox{0.5\hsize}{!}{\includegraphics{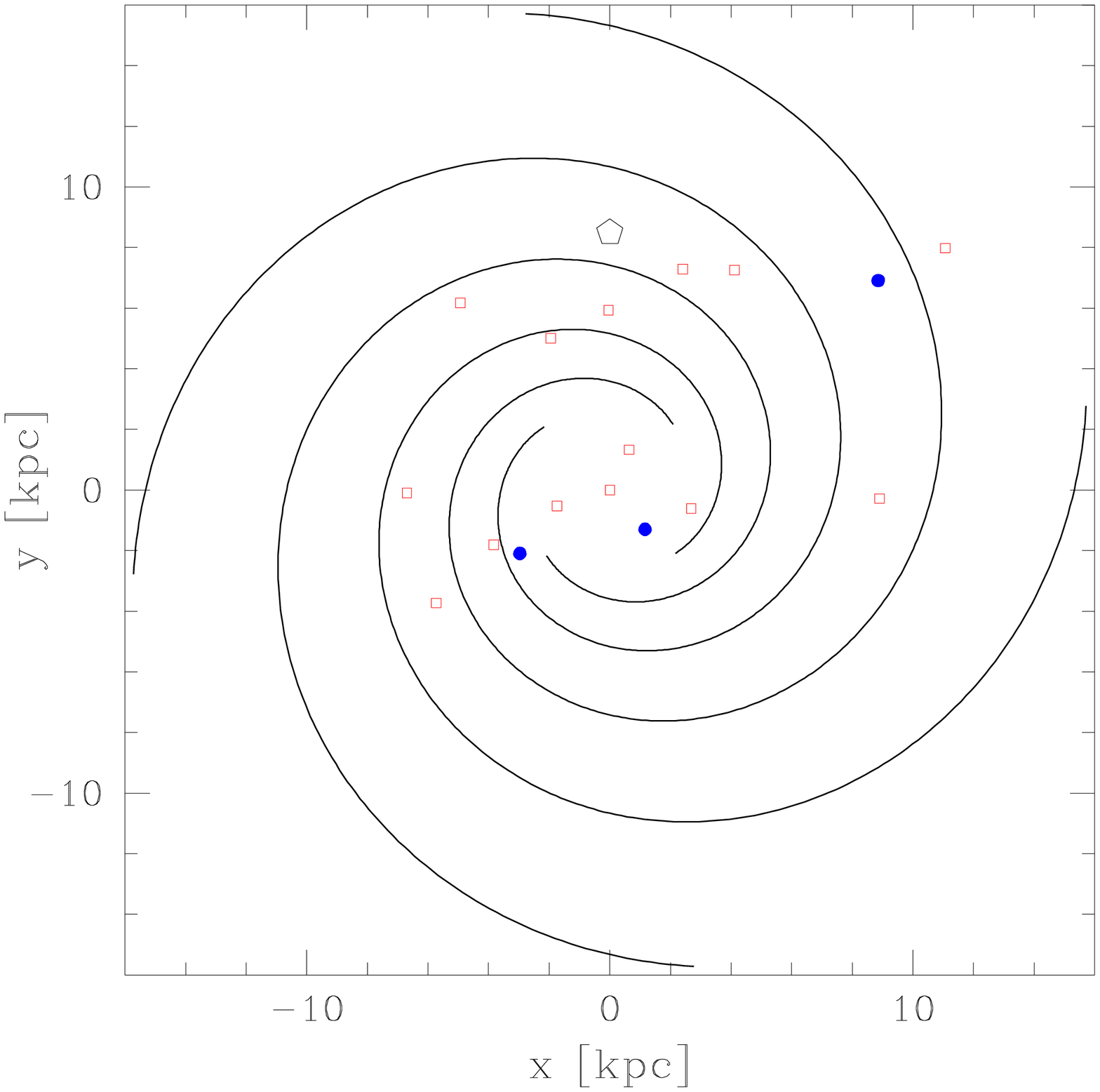}}
  \resizebox{0.5\hsize}{!}{\includegraphics{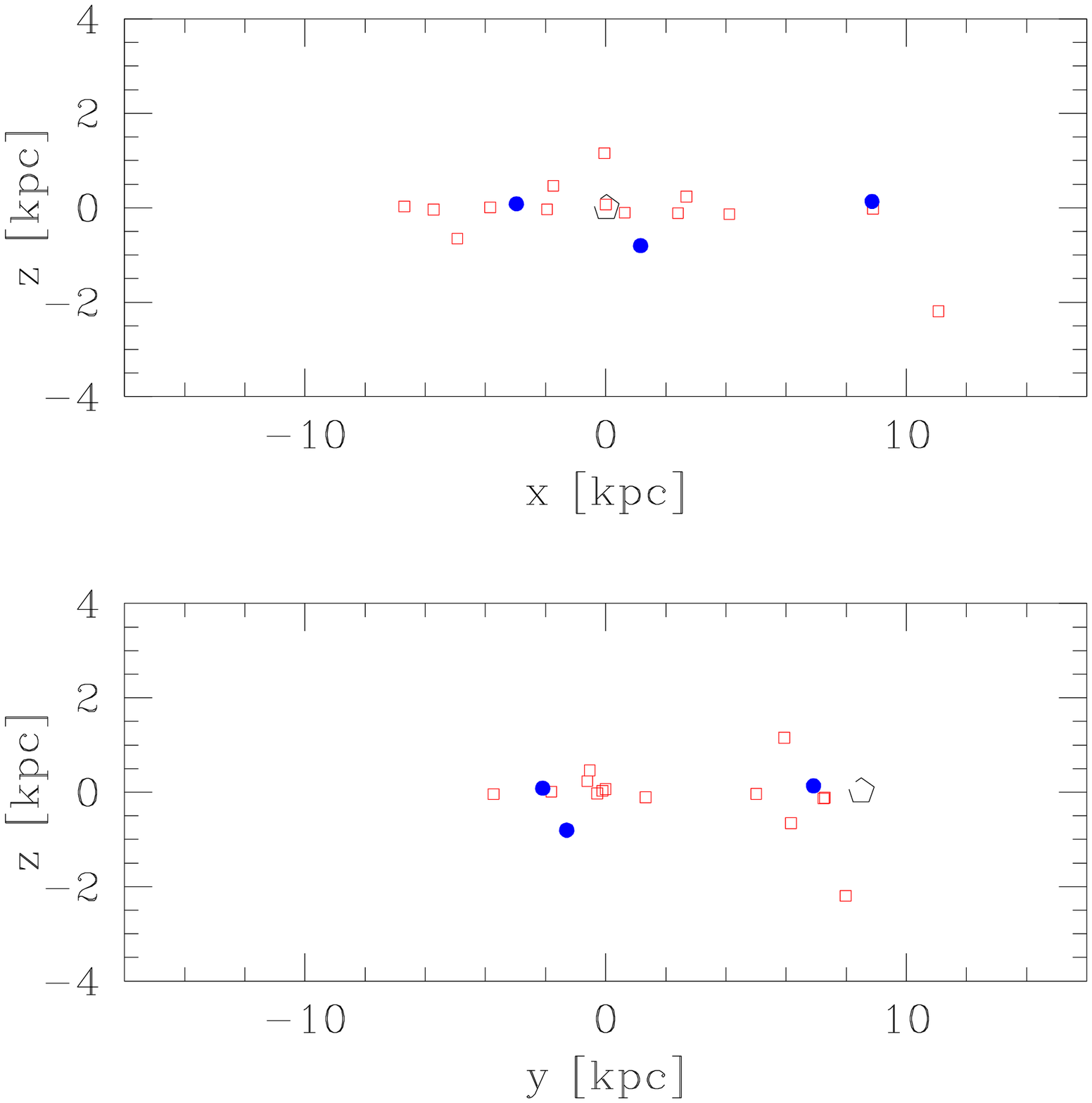}}
  \caption{The spatial distribution of Galactic X-ray binaries  that
  have shown episodes of Eddington or super-Eddington luminosity for a
  1.4  M$_{\odot}$ neutron star. The coordinate system is the same as
  in Fig. \ref{fig:faceon}. Filled circles indicate HMXBs, open
  squares indicate LMXBs. Note that fact that the majority of the
  sources are located at $y>0$ reflects the flux limited nature of the
  ASM sample.}
  \label{fig:xrb_bright}
\end{figure*}

\subsection{Young objects in star forming regions}
Recent observations
with the CHANDRA X-ray observatory of the Orion Nebula cluster allow
one to estimate the contribution to the X-ray emission from young
objects in the  star forming regions. \citet{schulz:00} observed the
Orion Trapezium region and found 111 sources above the sensitivity
threshold of $6.6 \cdot 10^{28}$ erg s$^{-1}$, assuming a distance of
440 pc. The total luminosity of their sample is about $5.6 \cdot
10^{32}$ erg s$^{-1}$. This luminosity is dominated by the brightest
source in the Orion Nebula cluster, \object{$\theta^1$ Ori C}, which
provides about $1.8 \cdot 10^{32}$ erg s$^{-1}$. Extrapolating this
result to the whole Orion Nebula Cluster in which CHANDRA observed
about 1000 sources we obtain a total luminosity of the star cluster of
about $4 \cdot 10^{33}$ erg s$^{-1}$, counting the luminosity of
\object{$\theta^1$ Ori C} only once and multiplying the rest by 10,
assuming the luminosity function of the Trapezium region is
representative for the whole Orion Nebula cluster. To estimate the
X-ray luminosity of all star forming regions in the Galaxy one can
proceed in two ways. Taking the mass of the molecular gas in the Orion
cluster to be $\sim 10^{5} $M$_{\odot}$ \citep{maddalena:86}, and the
total mass of the molecular gas in the Galaxy to be $\sim 10^{9}
$M$_{\odot}$ \citep{williams:97}, the total luminosity is $\sim 4
\cdot 10^{37}$ erg s$^{-1}$. On the other hand one can use the star
formation rate in the Orion Nebula cluster and the Galaxy as the
determining factor. Taking the SFR in Orion to be $ \ge 10^{-4}$
M$_{\odot}$ yr$^{-1}$ \citep{hillenbrand:97}, and the SFR in the
Galaxy to be 4 M$_{\odot}$ yr$^{-1}$ \citep{mckee:97}, the total
luminosity of young objects in the star forming regions in the Galaxy
is $\la 1.6 \cdot 10^{38}$ erg s$^{-1}$. Taking into account that the
latter value is an upper limit, both numbers agree sufficiently
well. Therefore star forming regions contribute less than $\sim$ few
per cent  to the integrated X-ray emission of the Galaxy but $\sim
20\%$ or more to the luminosity of HMXBs in the energy range from 2-10
keV. On the other hand the spectrum of young stellar objects is much
softer than the spectrum of X-ray binaries. 

%
%
\section{High luminosity sources}
\label{sec:eddington}

In recent months the CHANDRA X-ray observatory was able to resolve
single X-ray sources in other galaxies that appear to radiate at or
above the Eddington limit for a 1.4 M$_{\odot}$ neutron star,
i.e. $\sim 2 \cdot 10^{38}$ erg s$^{-1}$. Similar behaviour is also
observed in Galactic X-ray binaries by ASM. The slightly different
spectral band used in these CHANDRA observations, usually 0.3--10 keV
compared to 2--10 keV for ASM, does not lead to significant
differences in luminosity.

Table \ref{tab:edd_source} lists the sources which were observed
either by ASM or some other instrument to emit at or above the
Eddington limit for a 1.4 M$_{\odot}$ neutron star. The spatial
distribution of these sources is shown in Fig. \ref{fig:xrb_bright}
and can be compared to the distributions of the brightest sources
observed by CHANDRA in other galaxies.

There are several reasons why sources can emit super-Eddington
luminosity:
\begin{itemize}
  \item For accreting black holes in high state radiation is coming
  from the quasi-flat accretion disk where electron scattering gives
  the main contribution to the opacity. Under these conditions the
  radiation is emitted according to
  \begin{equation}
    f(\mu) = (1 + 2.08\mu)\mu
  \end{equation}
  where $\mu = \cos(i)$ where $i$ is the inclination angle. It is easy
  to show that the radiation flux perpendicular to the plane of the
  disk exceeds the average value by 3 times (see ~\citet{shakura:73}
  for discussion).

  \item Some of the normal stars entering the X-ray binary phase are
  strongly evolved and have an unusual chemical abundance, e.g. if a
  He-enriched star supplies matter the Eddington luminosity is twice
  higher than for hydrogen plasma due to the change in cross-section
  per nucleus.

  Just these two factors permit to surpass the classical Eddington
  limit by a factor of $\sim$6.
  \item The star supplying material to the neutron star or black hole
  ``does not know'' about the existence of the Eddington luminosity
  limit due to accretion. Therefore some part of the matter will
  outflow forming a supercritical disk. In the approach
  of~\citet{shakura:73} it is possible to gain a factor of 
  $\ln(\dot{m}) \approx 3-5$ for $\dot{m} >> 1$ with $\dot{m} =
  \frac{\dot{M}}{\dot{M_{Edd}}}$.
  \citet{paczynski:80} and ~\citet{abramowicz:88} constructed the
  solution of slim disks which also permits luminosities higher than
  the Eddington luminosity.

  \item Many X-ray binaries show from time to time the acceleration of
  powerful jets~\citep{mirabel:99}. These relativistic jets might
  produce strongly beamed X-ray emission with flux strongly exceeding
  the average and Eddington critical value for isotropic sources. See
  also the discussion by \citet{koerding:01}, ~\citet{fabrika:00}
  and~\citet{king:01}. 

  \item In the case of accretion on to a neutron star with strong
  magnetic field the accretion columns form near the surface of the
  neutron star in the polar regions. Such columns can have a
  super-Eddington luminosity, because photons are emitted
  perpendicular to the axis of the accretion column and the light
  pressure force is balanced by magnetic field \citep{basko:76}.

  \item In Z-sources (luminous accreting neutron stars with low
  magnetic field) the boundary layer width expands rapidly with
  increasing accretion rate reaching several star radii
  \citep{popham:01}. This quasi-flat continuation of accretion disk
  might also have super-Eddington luminosity of the type of the slim
  disk.

\end{itemize}

%
%
\section{Summary}
We studied the population of X-ray binaries in the Milky Way. 
\begin{itemize}

  \item In good agreement with theoretical expectations and earlier
  results \citep{vanparadijs:95,white:96,koyama:90,nagase:89} we found
  significant differences in the spatial (3-D) distribution of high
  and low mass X-ray binaries. HMXBs are more concentrated towards the
  Galactic Plane with a vertical scale height of 150 pc, tend to avoid
  the Galactic Bulge and central $\sim 3-4$ kpc of the Galaxy and show
  clear signatures of the spiral structure. The distribution of
  LMXB sources, on the contrary, peaks strongly at the Galactic Bulge
  and shows a pronounced minimum at $\sim 3-4$ kpc. Some signatures of
  the Galactic spiral structure are also present. The vertical
  distribution of LMXB sources is significantly broader, with a
  scale height of 410 pc.

  \item We constructed the {\em long-term averaged} \lns~distribution
  of high and low mass X-ray binaries in the 2--10 keV energy range
  using the data of the ASM instrument aboard RXTE from 1996-2000 to
  the limiting sensitivity of $\approx 6.4 \cdot 10^{-11}$
  erg s$^{-1}$ cm$^{-2}$. The \lns~distribution of HMXBs is well
  described by a simple power law with a slope of the differential
  distribution of  $1.61_{+0.12}^{-0.14}$ down to a flux limit of
  $\approx 6.4 \cdot 10^{-11}$ erg s$^{-1}$ cm$^{-2}$. The
  differential \lns~distribution of LMXBs has a slope of $-1.2 \pm
  0.06$ and requires a high-flux cutoff at $\sim 110$ ASM cnts
  s$^{-1}$, $\approx 3.5\cdot 10^{-8}$ erg s$^{-1}$ cm$^{-2}$. A
  comparison with data of the ASCA Galactic Ridge Survey
  \citep{sugizaki:01} which covered $\sim 40$ square degrees with
  $\sim 100$ times better sensitivity did not reveal any evidence of
  significant departures of the \lns~from an extrapolation of the ASM
  data down to $\approx 5 \cdot 10^{-13}$ erg s$^{-1}$ cm$^{-2}$.

  \item Using the source distances available and assuming a model for
  the volume density distribution we constructed luminosity functions
  for HMXBs and LMXBs in the 2--10 keV energy range. The sensitivity
  limit of the ASM catalogue allows one to study the XRB luminosity
  functions down to a luminosity of $\sim 2\cdot 10^{35}$ erg
  s$^{-1}$. The differential luminosity functions can be described by
  a power law with slopes of 1.64 and 1.27 for HMXBs and LMXBs
  respectively. For LMXB sources a cut-off at $\sim 2.7 \cdot 10^{38}$
  erg s$^{-1}$ is required. The HMXB data are insufficient to detect a
  high luminosity cut-off above $\sim{\rm few} \times 10^{36}$ erg
  s$^{-1}$. A comparison with the data of ASCA Galactic Ridge Survey
  did not find evidence for significant departures from these power
  laws down to luminosities of $\sim 10^{34}$ erg s$^{-1}$.

  \item The complete catalogue of our sample of X-ray binaries is
  available at http://www.mpa-garching.mpg.de/$\sim$grimm/. Properties
  of the brightest sources are summarised in Tables
  \ref{tab:bright_l}, \ref{tab:bright_h}, \ref{tab:edd_source}. 

  \item The integrated luminosity of X-ray binaries in the Milky
  Way in the 2--10 keV band averaged over 1996--2000 is $\sim 2-3
  \cdot 10^{39}$ erg s$^{-1}$ to which LMXB sources contribute 
  $\sim 90\%$. Normalised to the stellar mass and the star formation
  rate the integrated luminosity of LMXBs ($\sim 2.5 \cdot 10^{39}$
  erg s$^{-1}$) and HMXBs ($\sim 2 \cdot 10^{38}$ erg s$^{-1}$)
  correspond to $\sim 5 \cdot 10^{28}$ erg s$^{-1}$ M$^{-1}_{\odot}$
  and $\sim 5\cdot 10^{37}$ erg s$^{-1}$/(M$_{\odot}$ yr$^{-1}$),
  respectively. The total number of the X-ray binaries brighter than
  $2 \cdot 10^{35}$ erg s$^{-1}$ is $\sim 190$ of which $\sim 55$ are
  high mass and $\sim 135$ are low mass binaries. Extrapolating the
  luminosity functions towards low luminosities we estimate the total
  number of the X-ray binaries brighter than $10^{34}$ erg s$^{-1}$ as
  $\sim 705$ ($\sim 325$ LMXB and $\sim 380$ HMXB sources). These
  estimates might be subject to the uncertainty of a factor of $\sim
  2$ due to insufficient knowledge of the spatial distribution of
  X-ray binaries in the Galaxy.

  \item Due to the shallow slope of the luminosity function the
  integrated X-ray emission of the Milky Way is dominated by $\sim
  5-10$ brightest sources. Variability of individual sources or an
  outburst of a bright transient source can increase the integrated
  luminosity of the Milky Way by as much as a factor of $\sim
  2$.

  \item We found that at least 16 sources in the Galaxy showed
  episodes of super-Eddington luminosity for a 1.4 M$_{\odot}$ neutron
  star. We plotted the distribution of these sources across the Galaxy
  in various projections, which can be used to compare with the recent
  CHANDRA and XMM-Newton images of the nearby galaxies.
\end{itemize}

\section*{acknowledgements}
This research has made use of data and results provided by the
ASM/RXTE teams at MIT and at the RXTE SOF and GOF at NASAs GSFC. Data
were obtained through the High Energy Astrophysics Science Archive
Research Center Online Service, provided by the NASA/Goddard Space
Flight Center.
We also want to thank the referee for helpful remarks on the paper.
\addcontentsline{toc}{chapter}{Literaturverzeichnis}
   \bibliographystyle{aa}
   \bibliography{paper}

\begin{thebibliography}{115}
\expandafter\ifx\csname natexlab\endcsname\relax\def\natexlab#1{#1}\fi

\bibitem[{Abramowicz {et~al.}(1988)Abramowicz, Czerny, Lasota, \&
  Szuszkiewicz}]{abramowicz:88}
Abramowicz, M., Czerny, B., Lasota, J., \& Szuszkiewicz, E. 1988, ApJ, 332, 646

\bibitem[{{Angelini} {et~al.}(2001){Angelini}, {Loewenstein}, \&
  {Mushotzky}}]{angelini:01}
{Angelini}, L., {Loewenstein}, M., \& {Mushotzky}, R.~F. 2001, \apjl, 557, L35

\bibitem[{Bahcall(1978)}]{bahcall:78}
Bahcall, J. 1978, in Proceedings of the International School of Physics "Enrico
  Fermi", Vol.~65, Physics and astrophysics of neutron stars and black holes,
  ed. R.~Giacconi \& R.~Ruffini (Italian Phys. Soc.), 63

\bibitem[{Bahcall \& Soneira(1980)}]{bahc-son:80}
Bahcall, J.~N. \& Soneira, R.~M. 1980, ApJS, 44, 73

\bibitem[{{Bandyopadhyay} {et~al.}(1999){Bandyopadhyay}, {Shahbaz}, {Charles},
  \& {Naylor}}]{bandyopadhyay:99}
{Bandyopadhyay}, R.~M., {Shahbaz}, T., {Charles}, P.~A., \& {Naylor}, T. 1999,
  MNRAS, 306, 417

\bibitem[{{Barret} {et~al.}(1996){Barret}, {McClintock}, \&
  {Grindlay}}]{barret:97}
{Barret}, D., {McClintock}, J.~E., \& {Grindlay}, J.~E. 1996, ApJ, 473, 963

\bibitem[{{Barret} {et~al.}(1998){Barret}, {Motch}, \& {Predehl}}]{barret:98}
{Barret}, D., {Motch}, C., \& {Predehl}, P. 1998, A\&A, 329, 965

\bibitem[{{Basko} \& {Sunyaev}(1976)}]{basko:76}
{Basko}, M.~M. \& {Sunyaev}, R.~A. 1976, MNRAS, 175, 395

\bibitem[{Binney {et~al.}(1997)Binney, Gerhard, \& Spergel}]{binney:97}
Binney, J., Gerhard, O., \& Spergel, D. 1997, MNRAS, 288, 365

\bibitem[{Blanton {et~al.}(2001)Blanton, Sarazin, \& Irwin}]{blanton:01}
Blanton, E.~L., Sarazin, C.~L., \& Irwin, J.~A. 2001, ApJ, 552, 106

\bibitem[{{Bradshaw} {et~al.}(1999){Bradshaw}, {Fomalont}, \&
  {Geldzahler}}]{bradshaw:99}
{Bradshaw}, C.~F., {Fomalont}, E.~B., \& {Geldzahler}, B.~J. 1999, ApJ, 512,
  L121

\bibitem[{Brandt {et~al.}(1996)Brandt, Rotschild, \& Swank}]{brandt:96}
Brandt, H., Rotschild, R., \& Swank, J. 1996, in Memorie della Societa
  Astronomica Italiana, Vol.~67, 593

\bibitem[{{Bridges} \& {Hanes}(1990)}]{bridges:90}
{Bridges}, T.~J. \& {Hanes}, D.~A. 1990, \aj, 99, 1100

\bibitem[{{Chakrabarty} {et~al.}(1993){Chakrabarty}, {Grunsfeld}, {Prince},
  {Bildsten}, {Finger}, {Wilson}, {Fishman}, {Meegan}, \&
  {Paciesas}}]{chakrabarty:93}
{Chakrabarty}, D., {Grunsfeld}, J.~M., {Prince}, T.~A., {et~al.} 1993, ApJ,
  403, L33

\bibitem[{{Chevalier} \& {Ilovaisky}(1990)}]{chevalier:90}
{Chevalier}, C. \& {Ilovaisky}, S.~A. 1990, A\&A, 238, 163

\bibitem[{{Christian} \& {Swank}(1997)}]{christian:97}
{Christian}, D.~J. \& {Swank}, J.~H. 1997, ApJS, 109, 177

\bibitem[{{Corbet} {et~al.}(1999){Corbet}, {Marshall}, {Peele}, \&
  {Takeshima}}]{corbet:99}
{Corbet}, R.~H.~D., {Marshall}, F.~E., {Peele}, A.~G., \& {Takeshima}, T. 1999,
  ApJ, 517, 956

\bibitem[{{Cowley} {et~al.}(1979){Cowley}, {Crampton}, \&
  {Hutchings}}]{cowley:79}
{Cowley}, A.~P., {Crampton}, D., \& {Hutchings}, J.~B. 1979, ApJ, 231, 539

\bibitem[{Dehnen \& Binney(1998)}]{dehnen:98}
Dehnen, W. \& Binney, J. 1998, MNRAS, 294, 429

\bibitem[{{Di Stefano} {et~al.}(2002){Di Stefano}, {Kong}, {Garcia}, {Barmby},
  {Greiner}, {Murray}, \& {Primini}}]{distefano:02}
{Di Stefano}, R., {Kong}, A.~K.~H., {Garcia}, M.~R., {et~al.} 2002, \apj, 570,
  618

\bibitem[{Djorgovski(1993)}]{djorgovski:93}
Djorgovski, S. 1993, in ASP Conf. Ser. 50: Structure and Dynamics of Globular
  Clusters, 373

\bibitem[{{Downes} {et~al.}(1980){Downes}, {Wilson}, {Bieging}, \&
  {Wink}}]{downes:80}
{Downes}, D., {Wilson}, T.~L., {Bieging}, J., \& {Wink}, J. 1980, A\&AS, 40,
  379

\bibitem[{Dwek {et~al.}(1995)Dwek, Arendt, Hauser, Kelsall, Lisse, Moseley,
  Silverberg, Sodroski, \& Weiland}]{dwek:95}
Dwek, E., Arendt, R., Hauser, M., {et~al.} 1995, ApJ, 445, 716

\bibitem[{{Ebisuzaki} {et~al.}(1984){Ebisuzaki}, {Sugimoto}, \&
  {Hanawa}}]{ebisuzaki:84}
{Ebisuzaki}, T., {Sugimoto}, D., \& {Hanawa}, T. 1984, PASJ, 36, 551

\bibitem[{Englmaier \& Gerhard(1999)}]{englmaier:99}
Englmaier, P. \& Gerhard, O. 1999, MNRAS, 304, 512

\bibitem[{Fabbiano {et~al.}(2001)Fabbiano, Zezas, \& Murray}]{fabbiano:01}
Fabbiano, G., Zezas, A., \& Murray, S. 2001, ApJ, 554, 1035

\bibitem[{Fabrika \& Mescheryakov(2000)}]{fabrika:00}
Fabrika, S. \& Mescheryakov, A. 2000, in IAU Symposium, Vol. 205, E105

\bibitem[{{Fender}(2001)}]{fender:01}
{Fender}, R.~P. 2001, \mnras, 322, 31

\bibitem[{Finoguenov \& Jones(2001)}]{finoguenov:01}
Finoguenov, A. \& Jones, C. 2001, ApJL, 547, 107

\bibitem[{Forman {et~al.}(1978)Forman, Jones, Cominsky, Julien, Murray, Peters,
  Tananbaum, \& Giacconi}]{forman:78}
Forman, W., Jones, C., Cominsky, L., {et~al.} 1978, ApJS, 38, 357

\bibitem[{{Garcia} {et~al.}(2000){Garcia}, {Murray}, {Primini}, {Forman}, \&
  {Jones}}]{garcia:00}
{Garcia}, M.~R., {Murray}, S.~S., {Primini}, F.~A., {Forman}, W.~R., \&
  {Jones}, C. 2000, in IAU Symposium, Vol. 205

\bibitem[{Georgelin \& Georgelin(1976)}]{georgelin:76}
Georgelin, Y. \& Georgelin, Y. 1976, A\&A, 49, 57

\bibitem[{{Grebenev} {et~al.}(1996){Grebenev}, {Pavlinsky}, \&
  {Sunyaev}}]{grebenev:96}
{Grebenev}, S.~A., {Pavlinsky}, M.~N., \& {Sunyaev}, R. 1996, in
  R\"ontgenstrahlung from the Universe, 141

\bibitem[{Greiner {et~al.}(2001)Greiner, Cuby, McCaughrean, Castro-Tirado, \&
  Mennickent}]{greiner:01}
Greiner, J., Cuby, J., McCaughrean, M., Castro-Tirado, A., \& Mennickent, R.
  2001, submitted to A\&A

\bibitem[{{Greiner} {et~al.}(1994){Greiner}, {Hasinger}, {Molendi}, \&
  {Ebisawa}}]{greiner:94}
{Greiner}, J., {Hasinger}, G., {Molendi}, S., \& {Ebisawa}, K. 1994, A\&A, 285,
  509

\bibitem[{Griffiths {et~al.}(2000)Griffiths, Ptak, Feigelson, Garmire,
  Townsley, Brandt, Sambruna, \& Bregman}]{griffiths:00}
Griffiths, R., Ptak, A., Feigelson, E., {et~al.} 2000, Science, 290, 1325

\bibitem[{{Haberl} \& {Titarchuk}(1995)}]{haberl:95}
{Haberl}, F. \& {Titarchuk}, L. 1995, A\&A, 299, 414

\bibitem[{{Harris} \& {Racine}(1979)}]{harris:79}
{Harris}, W.~E. \& {Racine}, R. 1979, \araa, 17, 241

\bibitem[{Hillenbrand(1997)}]{hillenbrand:97}
Hillenbrand, L.~A. 1997, AJ, 113, 1733

\bibitem[{{Hutchings} {et~al.}(1979){Hutchings}, {Cowley}, {Crampton}, {van
  Paradus}, \& {White}}]{hutchings:79}
{Hutchings}, J.~B., {Cowley}, A.~P., {Crampton}, D., {van Paradus}, J., \&
  {White}, N.~E. 1979, ApJ, 229, 1079

\bibitem[{Illarionov \& Sunyaev(1975)}]{illarionov:75}
Illarionov, A. \& Sunyaev, R. 1975, A\&A, 39, 185

\bibitem[{Israel {et~al.}(2001)Israel, Negueruela, Campana, Covino, Paola,
  Maxwell, Norton, Speziali, Verrecchia, \& Stella}]{israel:01}
Israel, G., Negueruela, I., Campana, S., {et~al.} 2001, A\&A, 371, 1018

\bibitem[{{Kaper} {et~al.}(1995){Kaper}, {Lamers}, {Ruymaekers}, {van den
  Heuvel}, \& {Zuidervijk}}]{kaper:95}
{Kaper}, L., {Lamers}, H.~J.~G.~L.~M., {Ruymaekers}, E., {van den Heuvel},
  E.~P.~J., \& {Zuidervijk}, E.~J. 1995, A\&A, 300, 446

\bibitem[{King {et~al.}(2001)King, Davies, Ward, Fabbiano, \& Elvis}]{king:01}
King, A., Davies, M., Ward, M., Fabbiano, G., \& Elvis, M. 2001, ApJ, 552, L109

\bibitem[{{King}(1993)}]{king:93}
{King}, A.~R. 1993, MNRAS, 260, L5

\bibitem[{{Kitamoto} {et~al.}(1992){Kitamoto}, {Tsunemi}, {Miyamoto}, \&
  {Hayashida}}]{kitamoto:92}
{Kitamoto}, S., {Tsunemi}, H., {Miyamoto}, S., \& {Hayashida}, K. 1992, ApJ,
  394, 609

\bibitem[{{Koerding} {et~al.}(2001){Koerding}, {Falcke}, {Markoff}, \&
  {Fender}}]{koerding:01}
{Koerding}, E., {Falcke}, H., {Markoff}, S., \& {Fender}, R. 2001, in
  Astronomische Gesellschaft Meeting Abstracts, Vol.~18, 176

\bibitem[{{Koyama} {et~al.}(1990){Koyama}, {Kawada}, {Kunieda}, {Tawara}, \&
  {Takeuchi}}]{koyama:90}
{Koyama}, K., {Kawada}, M., {Kunieda}, H., {Tawara}, Y., \& {Takeuchi}, Y.
  1990, \nat, 343, 148

\bibitem[{{Krzeminski}(1974)}]{krzeminski:74}
{Krzeminski}, W. 1974, ApJ, 192, L135

\bibitem[{{Kuijken} \& {Rich}(2001)}]{kuijken:01}
{Kuijken}, K. \& {Rich}, R.~M. 2001, in American Astronomical Society Meeting,
  Vol. 199, 9113

\bibitem[{{Lamb} {et~al.}(1980){Lamb}, {Markert}, {Hartman}, {Thompson}, \&
  {Bignami}}]{lamb:80}
{Lamb}, R.~C., {Markert}, T.~H., {Hartman}, R.~C., {Thompson}, D.~J., \&
  {Bignami}, G.~F. 1980, \apj, 239, 651

\bibitem[{Levine {et~al.}(1996)Levine, Bradt, Cui, Jernigan, Morgan, Remillard,
  Shirley, \& Smith}]{levine:96}
Levine, A.~M., Bradt, H., Cui, W., {et~al.} 1996, ApJ, 469, L33

\bibitem[{{Liu} {et~al.}(2000){Liu}, {van Paradijs}, \& {van den
  Heuvel}}]{liu:00}
{Liu}, Q.~Z., {van Paradijs}, J., \& {van den Heuvel}, E.~P.~J. 2000, A\&AS,
  147, 25

\bibitem[{{Liu} {et~al.}(2001){Liu}, {van Paradijs}, \& {van den
  Heuvel}}]{liu:01}
---. 2001, A\&A, 368, 1021

\bibitem[{Lochner \& Remillard(1997)}]{xte-asm}
Lochner, J. \& Remillard, R. 1997, The XTE All Sky Monitor Data Products,
  http://heasarc.gsfc.nasa.gov/docs/xte/ asm\_products\_guide.html

\bibitem[{Maddalena {et~al.}(1986)Maddalena, Morris, Moscowitz, \&
  Thaddeus}]{maddalena:86}
Maddalena, R., Morris, M., Moscowitz, J., \& Thaddeus, P. 1986, ApJ, 303, 375

\bibitem[{Makishima {et~al.}(1989)Makishima, Hayashida, Inoue, Koyama, Takano,
  Tanaka, Yoshida, Turner, Thomas, Stewart, Williams, Awaki, \&
  Tawara}]{makishima:89}
Makishima, K., Hayashida, K., Inoue, H., {et~al.} 1989, PASJ, 41, 697

\bibitem[{Markert {et~al.}(1979)Markert, Winkler, Laird, Clark, Hearn, Sprott,
  Li, Bradt, Lewin, \& Schnopper}]{markert:79}
Markert, T., Winkler, P., Laird, F., {et~al.} 1979, ApJS, 39, 573

\bibitem[{{Martin} {et~al.}(1995){Martin}, {Casares}, {Charles}, {van der
  Hooft}, \& {van Paradijs}}]{martin:95}
{Martin}, A.~C., {Casares}, J., {Charles}, P.~A., {van der Hooft}, F., \& {van
  Paradijs}, J. 1995, \mnras, 274, L46

\bibitem[{{Massey} {et~al.}(1995){Massey}, {Johnson}, \&
  {Degioia-Eastwood}}]{massey:95}
{Massey}, P., {Johnson}, K.~E., \& {Degioia-Eastwood}, K. 1995, ApJ, 454, 151

\bibitem[{Matilsky {et~al.}(1973)Matilsky, Gursky, Kellogg, Tananbaum, Murray,
  \& Giacconi}]{matilsky:73}
Matilsky, T., Gursky, H., Kellogg, E., {et~al.} 1973, ApJ, 181, 753

\bibitem[{McKee \& Williams(1997)}]{mckee:97}
McKee, C.~F. \& Williams, J.~P. 1997, ApJ, 476, 144

\bibitem[{Mirabel \& Rodriguez(1994)}]{mirabel:94}
Mirabel, I. \& Rodriguez, L. 1994, Nature, 371, 46

\bibitem[{Mirabel \& Rodr{\'i}guez(1999)}]{mirabel:99}
Mirabel, I. \& Rodr{\'i}guez, L. 1999, ARA\&A, 37, 409

\bibitem[{Motch {et~al.}(1997)Motch, Haberl, Dennerl, Pakull, \&
  Janot-Pacheco}]{motch:97}
Motch, C., Haberl, F., Dennerl, K., Pakull, M., \& Janot-Pacheco, E. 1997,
  A\&A, 323, 853

\bibitem[{Murdoch \& Crawford(1973)}]{murdoch:73}
Murdoch, H.~S. \& Crawford, D.~F. 1973, ApJ, 183, 1

\bibitem[{{Nagase}(1989)}]{nagase:89}
{Nagase}, F. 1989, in Two Topics in X-Ray Astronomy, Volume 1: X Ray Binaries.
  Volume 2: AGN and the X Ray Background, 45--55

\bibitem[{{Nakamura} {et~al.}(1989){Nakamura}, {Dotani}, {Inoue}, {Mitsuda},
  {Tanaka}, \& {Matsuoka}}]{nakamura:89}
{Nakamura}, N., {Dotani}, T., {Inoue}, H., {et~al.} 1989, PASJ, 41, 617

\bibitem[{Narayan \& Yi(1995)}]{narayan:95}
Narayan, R. \& Yi, I. 1995, ApJ, 452, 710

\bibitem[{{Neff} \& {Ulvestad}(2000)}]{neff:00}
{Neff}, S.~G. \& {Ulvestad}, J.~S. 2000, AJ, 120, 670

\bibitem[{{Nishiuchi} {et~al.}(1999){Nishiuchi}, {Koyama}, {Maeda}, {Asai},
  {Dotani}, {Inoue}, {Mitsuda}, {Nagase}, {Ueda}, \&
  {Kouveliotou}}]{nishiuchi:99}
{Nishiuchi}, M., {Koyama}, K., {Maeda}, Y., {et~al.} 1999, ApJ, 517, 436

\bibitem[{Ogasaka {et~al.}(1998)Ogasaka, Kii, Ueda, Takahashi, Yamada, Inoue,
  Ishisaki, Ohta, Yamada, Makishima, Miyaji, \& Hasinger}]{ogasaka:98}
Ogasaka, Y., Kii, T., Ueda, Y., {et~al.} 1998, AN, 319, 43

\bibitem[{{Orosz} {et~al.}(1996){Orosz}, {Bailyn}, {McClintock}, \&
  {Remillard}}]{orosz:96}
{Orosz}, J.~A., {Bailyn}, C.~D., {McClintock}, J.~E., \& {Remillard}, R.~A.
  1996, ApJ, 468, 380

\bibitem[{{Orosz} {et~al.}(2002){Orosz}, {Groot}, {van der Klis}, {McClintock},
  {Garcia}, {Zhao}, {Jain}, {Bailyn}, \& {Remillard}}]{orosz:02}
{Orosz}, J.~A., {Groot}, P.~J., {van der Klis}, M., {et~al.} 2002, \apj, 568,
  845

\bibitem[{{Orosz} \& {Kuulkers}(1999)}]{orosz:99}
{Orosz}, J.~A. \& {Kuulkers}, E. 1999, MNRAS, 305, 132

\bibitem[{{Orosz} {et~al.}(2000){Orosz}, {Kuulkers}, {van der Klis},
  {McClintock}, {Garcia}, {Callanan}, {Jain}, {Bailyn}, \&
  {Remillard}}]{orosz:00}
{Orosz}, J.~A., {Kuulkers}, E., {van der Klis}, M., {et~al.} 2000, in American
  Astronomical Society Meeting, Vol. 197, 8320

\bibitem[{Paczynsky \& Wiita(1980)}]{paczynski:80}
Paczynsky, B. \& Wiita, P. 1980, A\&A, 88, 23

\bibitem[{Penninx(1989)}]{penninx:89}
Penninx, W. 1989, in X-ray astronomy, ed. J.~Hunt \& B.~Battrick, Proceedings
  of the 23rd ESLAB Symposium No. ESA SP-296 (ESA), 185

\bibitem[{Piccinotti {et~al.}(1982)Piccinotti, Mushotzky, Boldt, Holt,
  Marshall, Serlemitsos, \& Shafer}]{piccinotti:82}
Piccinotti, G., Mushotzky, R., Boldt, E., {et~al.} 1982, ApJ, 253, 485

\bibitem[{{Popham} \& {Sunyaev}(2001)}]{popham:01}
{Popham}, R. \& {Sunyaev}, R. 2001, ApJ, 547, 355

\bibitem[{{Predehl} {et~al.}(2000){Predehl}, {Burwitz}, {Paerels}, \& {Tr{\"
  u}mper}}]{predehl:00}
{Predehl}, P., {Burwitz}, V., {Paerels}, F., \& {Tr{\" u}mper}, J. 2000, A\&A,
  357, L25

\bibitem[{{Reynolds} {et~al.}(1992){Reynolds}, {Bell}, \&
  {Hilditch}}]{reynolds:92}
{Reynolds}, A.~P., {Bell}, S.~A., \& {Hilditch}, R.~W. 1992, MNRAS, 256, 631

\bibitem[{Ritter \& Kolb(1998)}]{ritter:98}
Ritter, H. \& Kolb, U. 1998, A\&AS, 129, 83

\bibitem[{Sarazin {et~al.}(2000)Sarazin, Irwin, \& Bregman}]{sarazin:00}
Sarazin, C., Irwin, J., \& Bregman, J. 2000, ApJ, 544, L101

\bibitem[{{Sazonov} {et~al.}(1997){Sazonov}, {Sunyaev}, \& {Lund}}]{sazonov:97}
{Sazonov}, S.~Y., {Sunyaev}, R.~A., \& {Lund}, N. 1997, Pis ma Astronomicheskii
  Zhurnal, 23, 326

\bibitem[{Schulz {et~al.}(2001)Schulz, Canizares, Huenemoerder, Kastner,
  Taylor, \& Bergstrom}]{schulz:00}
Schulz, N., Canizares, C., Huenemoerder, D., {et~al.} 2001, ApJ, 549, 441

\bibitem[{Shakura \& Sunyaev(1973)}]{shakura:73}
Shakura, N. \& Sunyaev, R. 1973, A\&A, 24, 337

\bibitem[{Shirey {et~al.}(2001)Shirey, Soria, Borozdin, Osborne, Tiengo,
  Guainazzi, Hayter, Palombara, Mason, Molendi, Paerels, Pietsch, Priedhorsky,
  Read, Watson, \& West}]{shirey:00}
Shirey, R., Soria, R., Borozdin, K., {et~al.} 2001, A\&A, 365, L195

\bibitem[{Simonson(1976)}]{simonson:76}
Simonson, S.~C. 1976, A\&A, 46, 261

\bibitem[{{Smale}(1998)}]{smale:98}
{Smale}, A.~P. 1998, ApJ, 498, L141

\bibitem[{{Smith} {et~al.}(1997){Smith}, {Morgan}, \& {Bradt}}]{smith:97}
{Smith}, D.~A., {Morgan}, E.~H., \& {Bradt}, H. 1997, ApJ, 479, L137

\bibitem[{Smith \& Wilson(2001)}]{smith:01}
Smith, D.~S. \& Wilson, A.~S. 2001, ApJ, 557, 180

\bibitem[{Solomon {et~al.}(1985)Solomon, Sanders, \& Rivolo}]{solomon:85}
Solomon, P., Sanders, D., \& Rivolo, A. 1985, ApJL, 292, 19

\bibitem[{Sugizaki {et~al.}(2001)Sugizaki, Mitsude, Kaneda, Matsuzaki,
  Yamauchi, \& Koyama}]{sugizaki:01}
Sugizaki, M., Mitsude, K., Kaneda, H., {et~al.} 2001, ApJS, 134, 77

\bibitem[{Sunyaev(1990)}]{sunyaev:90}
Sunyaev, R. 1990, IAUC, 5104, 1

\bibitem[{{Sunyaev} \& {Revnivtsev}(2000)}]{sunyaev:00}
{Sunyaev}, R. \& {Revnivtsev}, M. 2000, A\&A, 358, 617

\bibitem[{Tanaka(1992)}]{tanaka:92}
Tanaka, Y. 1992, in X-ray binaries and recycled pulsars, ed. E.~van~den Heuvel
  \& S.~Rappaport, Vol. C 377 (Kluwer), 37

\bibitem[{Taylor \& Cordes(1993)}]{taylor:93}
Taylor, J. \& Cordes, J. 1993, ApJ, 411, 674

\bibitem[{Tennant {et~al.}(2001)Tennant, Wu, Ghosh, Kolodziejczak, \&
  Swartz}]{tennant:01}
Tennant, A.~F., Wu, K., Ghosh, K.~K., Kolodziejczak, J.~J., \& Swartz, D.~A.
  2001, ApJ, 549, L43

\bibitem[{{Tsunemi} {et~al.}(1989){Tsunemi}, {Kitamoto}, {Okamura}, \&
  {Roussel-Dupre}}]{tsunemi:89}
{Tsunemi}, H., {Kitamoto}, S., {Okamura}, S., \& {Roussel-Dupre}, D. 1989, ApJ,
  337, L81

\bibitem[{Vall\'ee(1995)}]{vallee:95}
Vall\'ee, J. 1995, ApJ, 454, 119

\bibitem[{van Paradijs(1994)}]{paradijs:94}
van Paradijs, J. 1994, Cambridge Astrophysics Series, Vol.~26, X-ray binaries
  (Cambridge University Press), 536

\bibitem[{{van Paradijs} \& {White}(1995)}]{vanparadijs:95}
{van Paradijs}, J. \& {White}, N. 1995, ApJ, 447, L33

\bibitem[{{Verbunt}(1996)}]{verbunt:96}
{Verbunt}, F. 1996, in NATO ASIC Proc. 477: Evolutionary Processes in Binary
  Stars, 201

\bibitem[{{Wachter} \& {Margon}(1996)}]{wachter:96}
{Wachter}, S. \& {Margon}, B. 1996, AJ, 112, 2684

\bibitem[{Warwick {et~al.}(1981)Warwick, Marshall, Fraser, Watson, Lawrence,
  Page, Pounds, Ricketts, Sims, \& Smith}]{warwick:81a}
Warwick, R., Marshall, N., Fraser, G., {et~al.} 1981, MNRAS, 197, 865

\bibitem[{{Watson} {et~al.}(1978){Watson}, {Ricketts}, \&
  {Griffiths}}]{watson:78}
{Watson}, M.~G., {Ricketts}, M.~J., \& {Griffiths}, R.~E. 1978, \apjl, 221, L69

\bibitem[{{Webbink}(1985)}]{webbink:85}
{Webbink}, R.~F. 1985, in IAU Symp. 113: Dynamics of Star Clusters, Vol. 113,
  541

\bibitem[{{Wen} {et~al.}(2000){Wen}, {Remillard}, \& {Bradt}}]{wen:00}
{Wen}, L., {Remillard}, R.~A., \& {Bradt}, H.~V. 2000, ApJ, 532, 1119

\bibitem[{White {et~al.}(1984)White, Kaluzienski, \& Swank}]{white:84}
White, N., Kaluzienski, J., \& Swank, J. 1984, in High Energy Transients in
  Astrophysics, ed. S.~Woosley, AIP Conference Proceedings No. 115 (New York:
  American Institute of Physics), 31--48

\bibitem[{{White} {et~al.}(1980){White}, {Becker}, {Pravdo}, {Boldt}, {Holt},
  \& {Serlemitsos}}]{white:80}
{White}, N.~E., {Becker}, R.~H., {Pravdo}, S.~H., {et~al.} 1980, \apj, 239, 655

\bibitem[{{White} \& {van Paradijs}(1996)}]{white:96}
{White}, N.~E. \& {van Paradijs}, J. 1996, ApJ, 473, L25

\bibitem[{Williams \& McKee(1997)}]{williams:97}
Williams, J.~P. \& McKee, C.~F. 1997, ApJ, 476, 166

\bibitem[{Zezas {et~al.}(2001)Zezas, Fabbiano, Ward, Prestwich, \&
  Murray}]{zezas:01}
Zezas, A., Fabbiano, G., Ward, M., Prestwich, A., \& Murray, S. 2001, in
  American Astronomical Society Meeting, Vol. 198, 5011

\bibitem[{Zombeck(1990)}]{zombeck:90}
Zombeck, M.~V. 1990, Handbook of Space Astronomy and Astrophysics, 2nd edn.
  (Cambridge University Press)

\end{thebibliography}
\clearpage

\end{document}